\documentclass[11pt,a4paper,english,nofootinbib,,superscriptaddress]{revtex4}
\usepackage{lmodern}

\usepackage[T1]{fontenc}
\usepackage[latin9]{inputenc}
\usepackage{babel}
\usepackage{color}

\usepackage{amsmath}
\usepackage{graphicx}
\usepackage{amssymb}
\usepackage{esint}
\usepackage[unicode=true, pdfusetitle,
 bookmarks=true,bookmarksnumbered=false,bookmarksopen=false,
 breaklinks=false,pdfborder={0 0 1},backref=false,colorlinks=false]
 {hyperref}
\usepackage{amsmath}
\usepackage{amssymb}
\def\b{\begin{equation}}
\def\e{\end{equation}}
\def\K{{\cal K}}

\def\K{{\cal K}}

\def\K{{\cal K}}

\def\1{\mbox{I\hspace{-.15em}1}}

\def\Z{\mbox{Z\hspace{-.3em}Z}}

\def\R{{\rm I\hspace{-.15em}R}}

\def\C{\hspace{3pt}{\rm l\hspace{-.47em}C}}

\def\b{\begin{equation}}
\def\e{\end{equation}}
\def\bee{\begin{enumerate}}
\def\eee{\end{enumerate}}

\newcommand{\lbl}[1]{\label{eq: #1}}

\def\Z{\mbox{Z\hspace{-.3em}Z}}

\def\R{{\rm I\hspace{-.15em}R}}

\def\C{\hspace{3pt}{\rm l\hspace{-.47em}C}}

\def\tr{\mathop{\rm tr}\nolimits}

\def\bd{\begin{displaystyle}}
\def\ed{\end{displaystyle}}
\def\ba{\begin{array}}
\def\ea{\end{array}}

\def\bee{\begin{enumerate}}
\def\eee{\end{enumerate}}

\def\bes{\begin{eqnarray*}}
\def\ees{\end{eqnarray*}}
\def\be{\begin{eqnarray}}
\def\ee{\end{eqnarray}}

\makeatletter
\@ifundefined{textcolor}{}
{%
 \definecolor{BLACK}{gray}{0}
 \definecolor{WHITE}{gray}{1}
 \definecolor{RED}{rgb}{1,0,0}
 \definecolor{GREEN}{rgb}{0,1,0}
 \definecolor{BLUE}{rgb}{0,0,1}
 \definecolor{CYAN}{cmyk}{1,0,0,0}
 \definecolor{MAGENTA}{cmyk}{0,1,0,0}
 \definecolor{YELLOW}{cmyk}{0,0,1,0}
 }

\usepackage{latexsym}\usepackage{bm}

\makeatother

\begin{document}

\title{ Quantum Field Theory in de Sitter Universe:\\ Ambient Space Formalism}

\author{M.V. Takook}

\email{takook@razi.ac.ir, mtakook@yahoo.com}

\affiliation{Department of Physics, Razi University,
Kermanshah, Iran}
\affiliation{Department of Physics,
Science and Research branch, \\Islamic Azad University, Tehran,
Iran}

\date{\today}

\begin{abstract}

Quantum field theory in the $4$-dimensional de Sitter space-time is constructed in the ambient space formalism in a rigorous mathematical framework. This work is based on the group representation theory and the analyticity of the complexified pseudo-Riemannian manifolds. The unitary irreducible representations of de Sitter group and their corresponding Hilbert spaces are reformulated in the ambient space formalism. Defining the creation and annihilation operators, quantum field operators and their corresponding analytic two-point functions for various spin fields have been constructed. The various spin massless fields can be constructed in terms of the massless conformally coupled scalar field in this formalism. Then the quantum massless minimally coupled scalar field operator, for the first time, is also constructed on Bunch-Davies vacuum state which preserve the analyticity. We show that the massless fields with $s \geq 3$  cannot propagate in de Sitter ambient space formalism. The massless gauge invariant field equations for $s=1, \frac{3}{2}, 2$ are studied. The gauge spin-$\frac{3}{2}$ fields satisfy the Grassmannian algebra, and hence provoke one to couple them with the gauge spin-$2$ field and the super-algebra is naturally appeared.

\end{abstract}
\maketitle

\vspace{0.5cm}
{\it Proposed PACS numbers}: 04.62.+v, 03.70+k, 11.10.Cd, 98.80.H
\vspace{1.5cm}


\tableofcontents
\newpage

\section{Introduction}

The quantum field theory in $4$-dimensional de Sitter space-time along the logical lines from the first principles is presented. The following principles are used as the axioms in our construction: 
\begin{itemize}
\item (A) As it is indicated by the observation, our universe can be well approximated by the de Sitter space-time with its symmetrical group $SO(1,4)$. 
\item (B) The quantum fields actually are fundamental objects and their corresponding free field operators must be transformed by the unitary irreducible representation of the de Sitter group. 
\item (C) The interaction between these fields are governed by the gauge principle (gauge theory)!
\item (D) The conformal symmetry is preserved in the early universe.  
\end{itemize}
The de Sitter (dS) space-time can be considered as a $4$-dimensional hyperboloid embedded in a $5$-dimensional Minkowskian space-time. The ambient space formalism is used in this paper. This formalism allows us to reformulate the quantum field theory (QFT)
in a rigorous mathematical framework, based
on the analyticity of the complexified pseudo-Riemannian manifold and the group representation theory, similar to the QFT in the Minkowskian space-time. The unitary irreducible representations (UIR) of dS group were completed and finalized by Takahashi \cite{tho,new,dix,tak}. The analyticity of the complexified dS space-time had been studied by Bros et al \cite{brgamo,brmo,brmo03,ta97}. In what follows, we combine these two subjects to construct the field equations (or the Lagrangian), the quantum field operators, the quantum states and the two-point functions in dS space-time for various spin fields. The construction of QFT in dS space-time, (free field quantization, interaction
fields or gauge theory, super-symmetry and super-gravity) is
necessary for gaining a better understanding of the evolution of our universe and the quantum effects of the gravitational field. Here the field quantization and the gauge theory are reformulated in the ambient space formalism.

First, the field equations are obtained by using the second order Casimir operator of the dS group and then the UIRs of the dS group and their corresponding Hilbert spaces are reformulated in ambient space formalism. The UIRs are classified by the eigenvalues of the Casimir operators of the dS group. The eigenvalues are written in terms of the two parameters $j$ and $p$, respectively playing the role of the spin $s$ and mass $m$ in Poincar\'e group or Minkowskian space-time in the null curvature limit. The representations of dS group are expressed by three series: principal, complementary and discrete series  \cite{tho,new,dix,tak,lip,vikl}. 

The quantum free fields are divided here into the three distinguishable types: massive fields, massless fields and auxiliary fields. A field is called massive when it propagates inside the dS light-cone and corresponds to the massive Poincar\'e fields in the null curvature limit. The massive field operators transform by the principal series representation of dS group \cite{brgamo,gata,taazba,berotata}. On the other hand, the massless fields propagate on the dS light-cone and consequently, they possess an additional symmetry, namely, the conformal symmetry. They correspond to the massless Poincar\'e field at the flat space-time limit \cite{gagarota,taro12}. The auxiliary fields do not have any counterpart in the null curvature limit, but they appear in the indecomposable representations of the massless fields and also in the conformal invariance of the theory \cite{gagarota}.

The free field operators, which correspond to the principal and complementary series representations and discrete series with $j \neq p $ can be constructed by using the principle (B). The field operators in these cases transform by the UIR of the dS group, and by defining the creation and annihilation operators on the corresponding Hilbert spaces, the quantum free field operators can be explicitly calculated similar to the Minkowskian space-time, presented by Weinberg \cite{wei}. The UIR of dS group can exactly address the quantum states or the vectors in Hilbert space and then the creation operator and vacuum states are defined from this quantum states. For obtaining a well-defined field operator in this ambient space-time formalism, it must be constructed in terms of dS plane wave on the complex dS space-time \cite{brgamo,brmo}. Then the analytic two-point function can be calculated directly from the complex dS plane wave and the vacuum states, up to normalization constant. Normalization constant can be fixed by imposing the local Hadamard condition which select the Bunch-Davies vacuum state \cite{brmo}. The Wightman two-point function is the boundary value of this analytic two-point function \cite{brmo}.

The massless conformally coupled scalar fields correspond to the complementary series representation with $j=0$ and $p=0$ or $p=1$ \cite{brgamo,brmo,ta97}, where these two values of $p$ are unitary equivalent \cite{tak}. The massless spinor fields correspond to the discrete series representation with $j=p=\frac{1}{2}$ which is also conformally invariant \cite{tak,ta97,bagamota}. The procedure of defining the field operators for these fields are also similar to the massive case. For all of them, the fields operators are defined as a map on the Fock space:
$$ \mbox{Field Operators}: \;{\cal F}({\cal H}) \longrightarrow {\cal F}({\cal H}),$$ where is constructed by the corresponding Hilbert spaces.

The massless minimally coupled scalar fields correspond to the value $j=0$ and $p=2$ which can not correspond with an UIR of dS group. Previously we constructed the minimally coupled scalar field operator in Krein space which transforms by an indecomposable representation of dS group \cite{gareta}. But it breaks the analyticity. Here using the ambient space formalism, we construct the quantum field operator on Bunch-Davies vacuum state which transforms by a representation of dS group and satisfies the analyticity properties. It can be written in terms of the massless conformally coupled scalar fields.

For the other free massless fields, which correspond to $j=p\geq 1$ \cite{ta97,gagarota,taro12}, the Hilbert spaces and consequently the quantum states cannot be defined uniquely due to the appearance of a gauge invariance \cite{tak}. For defining the quantum field operator, one must fix the gauge invariant. The action of the creation operator on the Hilbert spaces results in the states which are out of the Hilbert spaces. Therefore, the field operators act on vector spaces which are constructed by an indecomposable representation of dS group. For such fields the massless quantum states can be expressed in three subspaces:
$$ {\cal M}=V_1 \oplus V_2 \oplus V_3, $$ 
where $V_1 $ and $V_3$ are the space of the gauge dependent states and the pure gauge states, respectively. The massless physical states appear in the vector space $V_2/V_3\equiv {\cal H}$, which is the Hilbert space constructed by the corresponding UIR of dS group. This is known as the Gupta-Bleuler triplets \cite{anla,la,gareta}. In these cases, the massless field operators are defined as a map on the Fock space which are constructed on the vector space ${\cal M}$:
$$ \mbox{Massless Field Operators}: \; {\cal F}({\cal M}) \longrightarrow {\cal F}({\cal M}).$$
The structure of unphysical states $V_3$ and $V_1$ is obtained by using the gauge invariant transformation and the gauge invariant field equation, respectively. The gauge invariant field equation and the gauge invariant transformation are obtained by using the Casimir operator of dS group and dS algebra. The physical states (correspond to the Hilbert spaces) are associated with the UIR of the discrete series representation of dS group with $1\leq j=p < 3$. The vector field  $(j=p=1)$ and the vector-spinor field $(j=p=\frac{3}{2})$ were considered in a previously papers \cite{gagarota,paenta}. The linear conformal quantum gravity $(j=p=2)$ will be discussed. The interactions between the various fields are also written utilizing the gauge principle or the local transformation.

In dS ambient space formalism the tensor (-spinor) fields are homogeneous functions of degree $\lambda$ which can be written in terms of a polarization tensor (-spinor) and the dS plane wave \cite{brgamo,brmo,ta97,gata,gagata,gagarota}, $$ \Psi_{\alpha_1...\alpha_l}(x)=D_{\alpha_1...\alpha_l}(x,\partial,\lambda) (x.\xi)^{\lambda}.$$
The coordinate system $x^\alpha$ is a five-vector in ambient space notation whereas the $\xi^\alpha$ is a null five-vector which in the null curvature limit becomes the energy-momentum four-vector $k^\mu=(k^0, \vec{k})$. For the principal series the homogeneous degree $\lambda$ in the null curvature limit has a relation with the mass in Minkowskian space-time. For the discrete series with $p\geq 3$, the homogeneous degree is positive ($\lambda\geq 0 )$ and the dS plane wave cannot be defined properly since the plane wave solution has the singularity in the limit  $x\longrightarrow \infty $. Therefore the massless fields with $j=p\geq 3$ cannot propagate in dS space. From the principle (D), one can deduce that at the early universe the massless fields with $j=p< 3$ may exist. After the symmetry breaking, such fields get mass. 

Here the three types of massless gauge fields or gauge potential are considered; spin-$1$ vector fields $(K_\alpha\;, \;j=p=1)$, spin-$\frac{3}{2} $ vector-spinor fields $(\Psi_\alpha\;, \;j=p=\frac{3}{2})$ and spin-$2$ field $(j=p=2)$. For a spin-$2$ field there are two possibilities for construction the field operator: a rank-$2$ symmetric tensor field $({\cal H}_{\alpha\beta})$ or a rank-$3$ mix-symmetric tensor field $({\cal K}_{\alpha\beta\gamma}^M)$. 

The massless gauge vector fields can be associated to the electromagnetic, weak and strong nuclear forces in the framework of the abelian and non-abelian gauge theory. Here the quantization of these fields will be reformulated in the ambient space formalism. The massless spin-$2$ field with a rank-$2$ symmetric tensor field in Dirac six-cone or dS ambient space formalisms breaks the conformal transformation \cite{bifrhe,tatafa,tata} and for preserving the
conformal transformation, this field must be represented by a rank-$3$ mix-symmetric tensor field ${\cal K}_{\alpha\beta\gamma}^M(x)$ \cite{bifrhe,tatafa,tata}. This gauge potential can be associated to the gravitational waves in the framework of the dS gauge gravity.

It is interesting to note that there exist other massless vector-spinor fields or gauge potentials which correspond to the discrete series representation with $j=p=\frac{3}{2}$. This gauge potential is spinor and also a Grassmannian function, which satisfy the anti-commutation relations. In the framework of the gauge theory, the infinitesimal generators, which are associated to the corresponding gauge group, $({\cal Q}_i)$, must be spinorial or Grassmannian functions and they satisfy the anti-commutation relations. The third principle (C) leads us to the super-algebra, where in this case the multiplication of two spinor-generators or the Grasmanian functions become the usual function or algebra and therefore, the spinor-generators cannot satisfy a closed super-algebra. We need the dS group generators $L_{\alpha\beta}$ to obtain a closed super-algebra \cite{parota},
$$ \{{\cal Q}_i,{\cal Q}_j\}=\left(S^{(\frac{1}{2})}_{\alpha\beta}\gamma^4
\gamma^2\right)_{ij}L^{\alpha\beta}.$$
The dS group generators, $L_{\alpha\beta}$, in the gauge gravity model may be coupled with the gauge potential ${\cal K}_{\alpha\beta\gamma}(x)$. In the framework of the gauge theory, the vector-spinor gauge field $\Psi_\alpha$ may be considered as a potential of a new force in the nature, but this gauge field must couple with the gauge potential ${\cal K}_{\alpha\beta\gamma}$ and consequently, $\Psi_\alpha$ may be considered as a new part of the gravitational field. It means that the gravitational field may be decomposed into three parts: the de Sitter background $\theta_{\alpha\beta}$, the gravitational waves ${\cal K}_{\alpha\beta\gamma}^M$ and $\Psi_\alpha$. 
 
The content of this paper is organized as follows. In the next section, the applied mathematical notations of the paper has been introduced, including the definition of the dS group, two independent Casimir operators and the UIR of the dS group. Section III is dedicated to defining the field equations by using the second order Casimir operator of dS group for various spin fields in the dS ambient space formalism in two different spaces: $x$-space and $\xi$-space. These two spaces play the similar role of space-time and energy-momentum in Minkowski space-time. By using the plane wave solution in dS ambient space formalism, we calculate the homogeneous degrees of various spin fields. This process is analogous to the first quantization in the framework of Minkowskian space-time.

The massive quantum field operators (or second quantization) are presented in section IV, where the massive free field operators are introduced. A unique vacuum state, Bunch-Davies or Hawking-Ellis vacuum state, is selected for these fields by using the local Hadamard condition. In section V, the quantum field operators for discrete series are studied. This section is divided into two parts:  $j\neq p$ and $j=p$. For $j\neq p$ the procedure of defining the field operator is exactly similar to the massive fields. These fields are auxiliary since they appear in the indecomposable representations of the quantum massless fields. For $j=p\geq 1$, we discus that one cannot define the creation operators in the Hilbert space because of the gauge invariant, the quantum state cannot be defined uniquely. In section VI, at first by using the Casimir operator and dS algebra, the gauge invariant field equations are obtained for the massless spin $1,\frac{3}{2}$ and $2$ fields, then we try to obtain these field equations by defining the gauge-covariant derivative in ambient space formalism and the gauge invariant Lagrangian. The gauge fixing terms are also presented and the gauge fixing field equations are obtained.
 
The massless quantum fields are introduced in section VII. The field operators of massless conformally coupled scalar fields and corresponding two-point functions are constructed on the Bunch-Davies vacuum state. In section VII-B, we introduce a magic identity in ambient space formalism, which permits us to write the massless minimally coupled scalar field in terms of the massless conformally coupled scalar field. Therefore, we define the field operator and the two-point function of a minimally coupled scalar field in terms of a conformally coupled scalar field and its two-point function. It means that a unique vacuum state, Bunch-Davies vacuum state, is used to construct the minimally coupled scalar field or equivalently for the linear gravity in dS space. Then the problem of the infra-red divergence of the linear gravity in dS ambient space formalism is completely solved in the Bunch-Davies vacuum state. The procedure preserves the analyticity. 

The massless spinor field can be defined similar to the massive field. In this case a new type of invariance has been introduced. The massless fields, with $j=1,\frac{3}{2},2$ are also constructed in section VII. Since the massless spin-$2$ rank-$2$ symmetric tensor field breaks the conformal invariance, the massless spin-$2$ rank-$3$ mix-symmetric tensor field is also studied. It has been shown that the two-point functions of all the massless fields can be written in terms of a polarization tensor (-spinor) and the two-point functions of the massless conformally coupled scalar field.

In section VIII, a brief review of the massless conformal field theory in dS ambient space formalism is presented. First, the Dirac $6$-cone formalism has been discussed which can be simply mapped on the dS ambient space notation, then the discrete series representation of the conformal group $SO(2,4)$ is introduced, so, the conformal gauge gravity based on the massless spin-$2$ rank-$3$ mixed-symmetric tensor fields has been discussed. The conclusions and an outlook for further investigation have been presented in section IX. In appendix,  the relation between our construction (ambient space formalism) with the intrinsic coordinates is studied and the structure of the maximally symmetric bi-tensors (-spinors) is introduced.


\setcounter{equation}{0}
\section{Notation}

The dS space-time can be identified with a 4-dimensional hyperboloid embedded in 5-dimensional Minkowskian space-time with the equation:
     \b \label{dSs} M_H=\{x \in \R^5| \; \; x \cdot x=\eta_{\alpha\beta} x^\alpha
 x^\beta =-H^{-2}\},\;\; \alpha,\beta=0,1,2,3,4, \e
where $\eta_{\alpha\beta}=$diag$(1,-1,-1,-1,-1)$ and $H$ is Hubble parameter. The dS metric is \b \label{dsmet}  ds^2=\eta_{\alpha\beta}dx^{\alpha}dx^{\beta}|_{x^2=-H^{-2}}=
g_{\mu\nu}^{dS}dX^{\mu}dX^{\nu},\;\; \mu=0,1,2,3,\e 
where $X^\mu$ is a 4 space-time intrinsic coordinates on dS hyperboloid. In this paper, we use the 5-dimensional Minkowskian space-time $x^\alpha$ with condition ($x\cdot x=-H^{-2}$) which is called the ambient space formalism. For simplicity from now on we take $H=1$ in some equations and been inserted again, whenever it was needed.

\subsection{de Sitter group}

The dS group is defined by: $$ SO(1,4)=\left\lbrace \Lambda \in GL(5,\R)| \;\; \det \Lambda=1,\;\; \Lambda \eta \Lambda^t= \eta \right\rbrace,$$ where $\Lambda^t$ is the transpose of $\Lambda$. The action of dS group on the intrinsic coordinate $X^\mu$ is non-linear, but on the ambient space coordinate $x^\alpha$ is linear:
$$ x'^\alpha=\Lambda^\alpha_{\;\;\beta} x^\beta, \;\;\;\Lambda \in SO(1,4) \Longrightarrow x\cdot x=x'\cdot x'=-H^{-2}. $$  The ambient space coordinate $x^\alpha$ can be defined by a $4\times 4$ matrix ${\bf X}$:
\b \label{mx44} {\bf X}=\left( \begin{array}{clcr} x^0I & \;\;{\bf x} \\ {\bf x}^\dag & x^0I \\    \end{array} \right),\e
$I$ is $2\times 2$ unit matrix and
\b \label{quat} {\bf x}=\left( \begin{array}{clcr} x^4+ix^3 &\;\; ix^1-x^2 \\ ix^1+x^2 & \;\;x^4-ix^3 \\    \end{array} \right).\e 
${\bf x}$ can be represented by a quaternion ${\bf x}\equiv(x^4, \vec x)$ and its norm is as fallow: 
$$|{\bf x}|=({\bf x}{\bf  \tilde{ x}})^{\frac{1}{2}}=\sqrt{(x^1)^2+(x^2)^2+(x^3)^2+(x^4)^2},$$
where ${\bf \tilde{ x}}\equiv(x^4,-\vec x)$ is quaternion conjugate of ${\bf x}$ \cite{tak,ta97}. In the matrix notation, we have ${\bf  \tilde{ x}}={\bf x}^\dag$ and the norm can be written as $|{\bf x}|^2=\frac{1}{2} \tr( {\bf x}{\bf  x}^\dag) $. The matrix ${\bf X}$ may be introduced in an alternative form, which is convenient for our considerations in this paper: 
\b \label{gamax} \not x =\eta_{\alpha\beta}\gamma^\alpha x^\beta={\bf X}\gamma^0=\left( \begin{array}{clcr} x^0I & \;\;-{\bf x} \\ {\bf  \tilde{ x}} & -x^0I \\    \end{array} \right),\;\;\; \not x\not x =x\cdot x \;\1,\;\;\frac{1}{4}\tr \left( \not x\not x\right) =x \cdot x  ,\e 
where $\1$ is a $4\times 4$ unit matrix and the five matrices $\gamma^{\alpha}$ satisfy the following conditions \cite{ta97,ta96,bagamota}:
$$\gamma^{\alpha}\gamma^{\beta}+\gamma^{\beta}\gamma^{\alpha}
=2\eta^{\alpha\beta}\qquad
\gamma^{\alpha\dagger}=\gamma^{0}\gamma^{\alpha}\gamma^{0}.$$ 
The following representation for $\gamma$ matrices is used in this paper \cite{ta97,ta96,bagamota}:
$$ \gamma^0=\left( \begin{array}{clcr} I & \;\;0 \\ 0 &-I \\ \end{array} \right)
      ,\;\;\;\gamma^4=\left( \begin{array}{clcr} 0 & I \\ -I &0 \\ \end{array} \right) , $$ \b \label{gammam}
   \gamma^1=\left( \begin{array}{clcr} 0 & i\sigma^1 \\ i\sigma^1 &0 \\
    \end{array} \right)
   ,\;\;\gamma^2=\left( \begin{array}{clcr} 0 & -i\sigma^2 \\ -i\sigma^2 &0 \\
      \end{array} \right)
   , \;\;\gamma^3=\left( \begin{array}{clcr} 0 & i\sigma^3 \\ i\sigma^3 &0 \\
      \end{array} \right),\e
where $\sigma^i$ $(i=1,2,3)$ are the Pauli matrices. In this representation, the matrix $\not x$ transforms by the group $Sp(2,2)$ as \cite{ta96}:
$$ \not x'=g\not x g^{-1}, \;\;\;g \in Sp(2,2),\;\;\tr \left( \not x\not x\right)=\tr \left( \not x'\not x'\right) \Longrightarrow x\cdot x=x'\cdot x'=-H^{-2}.$$
The group $Sp(2,2)$ is: \b \label{sp22} Sp(2,2)=\left\lbrace g=\left( \begin{array}{clcr} {\bf a} & {\bf b} \\ {\bf c} & {\bf d} \\    \end{array} \right), \;\; \det g=1 , \;\; \gamma^0 \tilde{ g}^t \gamma^0=g^{-1} \right\rbrace . \e 
The elements ${\bf a},{\bf b},{\bf c}$ and ${\bf d}$ are quaternions, $Sp(2,2)$ is the universal covering group of $SO(1,4)$ \cite{ta96,bagamota}:
 \b SO_0(1,4)  \approx Sp(2,2)/ \Z_2,\;\; \;\;\Lambda_{\alpha}^{\;\;\beta}\gamma^\alpha=g \gamma^\beta g^{-1} .\e 
For the $4 \times 4$ matrix $g$, we have $\tilde{ g}^t=g^\dag$. The ambient space coordinate, $x^\alpha$, under the action of the  group $Sp(2,2)$ transforms as \cite{tak}:
 \b \label{sp2tra} \left\lbrace \begin{array}{clcr} x_0'=(|{\bf a}|^2+|{\bf b}|^2)x_0 & +{\bf a}{\bf x}{\bf \tilde{b}}+{\bf b}{\bf \tilde{x}}{\bf \tilde{a}}, \\ {\bf x}'=({\bf a}{\bf \tilde{c}}+{\bf b}{\bf \tilde{d}})x_0+ & {\bf b}{\bf \tilde{x}}{\bf \tilde{c}}+{\bf a}{\bf x}{\bf \tilde{d} }. \\    \end{array} \right. \e
 
In this paper, two different types of homogeneous spaces are used to construct the UIR of the dS group: the quaternion ${\bf u}\equiv(u^4, \vec u)$ with norm $|{\bf u}|=1$ which is called three-sphere $S^3$ (or ${\bf u}$-space), the quaternion ${\bf q}\equiv(q^4, \vec q)$ with norm $|{\bf q}|<1$ which is called closed unit ball or for simplicity the unit ball $B$ (or ${\bf q}$-space). Their transformations under the group $Sp(2,2)$ are \cite{tak}:
$${\bf u}'= g\cdot {\bf u}=({\bf a}{\bf u}+{\bf b})({\bf c}{\bf u}+{\bf d})^{-1}, \;\;\; |{\bf u}' |=1,$$ \b \label{qt} {\bf q}'=g\cdot {\bf q}=({\bf a}{\bf q}+{\bf b})({\bf c}{\bf q}+{\bf d})^{-1},\;\;\; |{\bf q}'|<1.\e These two homogeneous spaces can be considered as the sub-spaces of the positive cone $C^+$, which is defined by $C^+=\left\lbrace \xi \in \R^5|\;\; \xi\cdot \xi=0,\;\; \xi^{0}>0 \right\rbrace$. Then in a unique way, the null five vector $\xi^\alpha=(\xi^0, \vec \xi, \xi^4)\in C^+$ can be written as:
\b \label{2orbit} \xi^\alpha_u \equiv (\xi^0, \xi^0 \; {\bf u}),\;|{\bf u}|=1;\;\;\; \xi^\alpha_B \equiv (\xi^0 , \xi^0\; \coth \kappa \; {\bf q}),\; \; |{\bf q}|=|\tanh \kappa|=r<1.\e 
Since $\xi\cdot \xi=0$, the $\xi^0$ is completely arbitrary from the mathematical point of view, {\it i.e.} $\xi^\alpha$ is scale invariant. $\xi^0$ transforms under the dS group as \cite{gasiyo}
\b  \label{xi0t}  \xi'^0=\xi^0 \left|{\bf c}{\bf u}+{\bf d}\right|^2  .\e
If we choose ${\bf q}=r{\bf u}$ with $r<1$, we obtain $$ \xi^\alpha_B = (\xi^0 , \xi^0\; \coth \kappa \; {\bf q})\equiv (\xi^0, \xi^0 \; {\bf u})=\xi^\alpha_{u}.$$
With this choice the unit ball $B$ may be considered as the compactified Minkowski space as a group manifold of the unity group $U(2)$ \cite{ja,ja2,ma77}. It may be considered also as the Shilov boundary of the bounded homogeneous complex domain $SU(2,2)/S(U(2)\times U(2))$ \cite{ja,ja2,ru72,ru73}.

In this notation, $x\cdot\xi$ can be written in the following form:
\b \label{x.xi} \not x \not \xi + \not \xi \not x=2\; x\cdot \xi\; \1 \Longrightarrow x\cdot\xi= \frac{1}{4} \tr \not x \not \xi ,\e and under the action of the dS group it is a scalar:
\b \label{x.xit}  x'\cdot \xi'  = x\cdot \xi,\;\;  \tr \not x' \not \xi'= \tr \not x \not \xi.\e The $\xi$-space plays the role of the energy-momentum $k^\mu$ in Minkowski space-time and it can be chosen for massive field as \cite{brgamo,ta97}
$$ \xi^\alpha_u \equiv \xi^0\left(1, \frac{\vec{k}}{k^0} , \; \frac{H \nu}{k^0}\right) ,\;\; k^0 \neq 0,\; \nu \neq 0,\;\;\; (k^0)^2- \vec{k}\cdot \vec{k}=(H\nu)^2, $$ 
$\nu$ is principal series parameter which is introduced in the next subsection. For massless field, $\xi$ is:
$$ \xi^\alpha_B = \xi^0 \left( 1, \; \frac{{\bf q}}{r}\right)\equiv \xi^0\left(1,  \frac{\vec{k}}{k^0} , \; \frac{H}{k^0}\right) ,\;\; k^0 \neq 0,\; \;\;\; (k^0)^2- \vec{k}\cdot \vec{k}=H^2.$$
In the null curvature limit, we exactly obtain  the massive and the massless energy momentum,  $(k^0)^2- \vec{k}\cdot \vec{k}=m^2$ and $(k^0)^2- \vec{k}\cdot\vec{k}=0$ respectively.

\subsection{Casimir operators}

The dS group has two Casimir operators: a second-order Casimir operator,
 \b \label{casi1} Q^{(1)}=-\frac{1}{2}L_{\alpha\beta}L^{\alpha\beta},\;\;  \alpha, \beta=0,1,2,3,4, \e and a fourth-order Casimir operator,
      \b Q^{(2)}=-W_\alpha W^\alpha\;\;,\;\;W_\alpha =\frac{1}{8}
      \epsilon_{\alpha\beta\gamma\delta\eta} L^{\beta\gamma}L^{\delta\eta},\e
where  $\epsilon_{\alpha\beta\gamma\delta\eta}$ is the usual anti-symmetric tensor in $\R^5$ and $L_{\alpha\beta}$ are the infinitesimal generators of dS group,  $L_{\alpha\beta}=M_{\alpha\beta}+S_{\alpha\beta}$. In the ambient space formalism, the orbital part, $M_{\alpha\beta}$, is
          \b \label{genm} M_{\alpha \beta}=-i(x_\alpha \partial_\beta-x_\beta
      \partial_\alpha)=-i(x_\alpha\partial^\top_\beta-x_\beta
        \partial^\top_\alpha),\e
$\partial^\top_\beta=\theta_\beta^{\;\;\alpha}\partial_\alpha$ and $\theta
_{\alpha\beta}=\eta_{\alpha\beta}+H^2x_\alpha x_\beta$ is the projection tensor on dS hyperboloid.
In order to calculate the action of the spinorial  part, $S_{\alpha\beta}$, on a tensor field or tensor-spinor field, one must
treat the integer and half-integer cases, separately. Integer spin
fields can be represented by the tensor fields of rank $l$,
 ${\cal K}_{\gamma_1...\gamma_l}(x)$, and the spinorial action reads as
\cite{frhe,gaha}:
       \b \label{gens} S_{\alpha \beta}^{(l)}{\cal K}_{\gamma_1......\gamma_l}=-i\sum^l_{i=1}
          \left(\eta_{\alpha\gamma_i}
        {\cal K}_{\gamma_1....(\gamma_i\rightarrow\beta).... \gamma_l}-\eta_{\beta\gamma_i}
          {\cal K}_{\gamma_1....(\gamma_i\rightarrow \alpha).... \gamma_l}\right),\e
where $(\gamma_i\rightarrow\beta)$ means $\gamma_i$ is replaced
by $\beta$. Half-integer spin fields with spin $s=l+\frac{1}{2}$
are represented by four component symmetric tensor-spinor field $\left[\Psi_{
\gamma_{1}..\gamma_{l} }(x)\right]^{i}$ with the spinor index $i=1,2,3,4$. In this case, the
spinorial part is
$$S_{\alpha\beta}^{(s)}=S_{\alpha\beta}^{(l)}+S_{\alpha\beta}^{(\frac{1}{2})}\qquad
\mbox{with}\qquad
S_{\alpha\beta}^{(\frac{1}{2})}=-\frac{i}{4}\left[\gamma_{\alpha},\gamma_{\beta}\right].$$ The
$S_{\alpha\beta}^{(\frac{1}{2})}$ acts only on the spinor index i.

For $j=l=$ integer, the Casimir operator $ Q^{(1)}_l$ acts on a rank-$l$ symmetric tensor filed ${\cal K}_{\alpha_1 \alpha_2 ...\alpha_l}(x)\equiv {\cal K}(x) $ as \cite{gaha,ta97,taazba}:
 \b \label{casimirl} Q_l^{(1)}{\cal K}(x)=Q_0^{(1)}{\cal K}(x)-2\Sigma_1 \partial x \cdot {\cal K}(x)+2\Sigma_1 x \partial\cdot
           {\cal K}(x)+2\Sigma_2 \eta {\cal K}'(x)-l(l+1){\cal K}(x),\e
where
\b Q_0^{(1)}=-\frac{1}{2}M_{\alpha \beta}M^{\alpha \beta}=-H^{-2}\partial^\top\cdot\partial^\top \equiv -H^{-2} \square_H\;.\e  $\square_H$ is the Laplace-Beltrami operator of dS space-time. ${\cal K}'$
is the trace of the rank-$l$ tensor field ${\cal K}(x)$ and $\Sigma_p$ is the non-normalized symmetrization operator:
         \b {\cal K}'_{\alpha_1...\alpha_{l-2}}=\eta^{\alpha_{l-1}\alpha_l}
          {\cal K}_{\alpha_1...\alpha_{l-2} \alpha_{l-1}\alpha_l},\e
         \b \label{sigmap} (\Sigma_p AB)_{\alpha_1...\alpha_l}=\sum_{i_1<i_2<...<i_p}
          A_{\alpha_{i_1}\alpha_{i_2}...\alpha_{i_p}}
          B_{\alpha_1...\not\alpha_{i_1}...\not\alpha_{i_2} ...\not\alpha_{i_p}...\alpha_l}\;.\e
The tensor field ${\cal K}(x)$ on the dS hyperboloid is a homogeneous function of the variables $x^{\alpha}$ with degree $\lambda$ \cite{dirds} $$ x\cdot\partial {\cal K}(x)=\lambda {\cal K}(x), \;\; \mbox{or}\;\; {\cal K}(lx)=l^\lambda{\cal K}(x).$$

For half-integer case $j=l+\frac{1}{2}$, the Casimir operator becomes
          $$ Q^{(1)}_j=-\frac{1}{2}\left(M_{\alpha \beta}+S_{\alpha \beta}^{(l)}+
                S_{\alpha\beta}^{(\frac{1}{2})}\right)
      \left(M^{\alpha \beta}+S^{\alpha \beta(l)}+S^{\alpha \beta(\frac{1}{2})}\right).$$ 
Using the identity
$${\cal S}^{(\frac{1}{2})}_{\alpha\beta}{\cal S}^{\alpha\beta(l)}\Psi(x)=l \Psi(x)-\Sigma_1\gamma(\gamma\cdot
\Psi(x)),$$  the action of the Casimir operator $ Q^{(1)}_j$ on a rank-$l$ symmetric tensor-spinor filed $\Psi_{\alpha_1 \alpha_2 ...\alpha_l}(x)\equiv \Psi(x) $ is \cite{gaha,ta97,taazba}:
      $$ Q^{(1)}_j\Psi(x)=\left(-\frac{1}{2}M_{\alpha \beta}M^{\alpha
      \beta}+\frac{i}{2} \gamma_{\alpha}\gamma_{\beta}
       M^{\alpha \beta}-l(l+2)-\frac{5}{2}\right)\Psi(x)$$ \b \label{casimirj} -2\Sigma_1
         \partial x \cdot \Psi(x)+2\Sigma_1 x \partial\cdot \Psi(x)+2\Sigma_2 \eta \Psi'(x)+\Sigma_1 \gamma \left(\gamma \cdot\Psi(x)\right).\e

\subsection{Unitary irreducible representation}

The Casimir operators commute with the generators of group and as a consequence, they are constant on each UIR of the dS group. The UIR of the dS group are classified by the eigenvalue of the Casimir operators \cite{tho,new,dix,tak}:
             \b \label{qe1} Q^{(1)}_{j,p}=\left(-j(j+1)-(p+1)(p-2)\right)I_d\equiv Q^{(1)}_{j} , \e
           \b Q^{(2)}_{j,p}=(-j(j+1)p(p-1))I_d, \e
$I_d$ is the identity operator. Three types of representations, corresponding to the different value of the parameters $j$ and $p$, exist:
\begin{itemize}
\item{Principal series representation},
 \b\label{ps}  j=0,\frac{1}{2},1,\frac{3}{2},2,....,\;\;\;p=\frac{1}{2}+i \nu,\; \nu \in \R \;
         \;\;
       \left\{ \begin{array}{clcr} \nu \geq 0 \;  \;\mbox{for}\; \; j=0,1,2,....\\
  \nu > 0\; \;\mbox{for} \;\; j=\frac{1}{2},\frac{3}{2},\frac{5}{2},.... \\ \end{array} \right. .\e
\item{Discrete series representation},
$$  j\geq p\geq1\;\mbox{ or}\; \frac{1}{2}, \;\;\; j-p=\mbox{integer number}, $$
     \b     j=1,2,3,...,\;\;\;\; \;\; p=0.\e
\item{Complementary series representation},
$$  j=0,\;\;\;\;-2 <  p-p^2<\frac{1}{4},$$
          \b  j=1,2,3,...,\;\;\;\;\;0<p-p^2 <\frac{1}{4}.\e
\end{itemize}

The principal series representation of dS group was constructed on the compact homogeneous three-sphere space, $S^3$ or ${\bf u}$-space \cite{tak}:
\b \label{principals} U^{(j,p)}(g)\left|{\bf u},m_j;j,p \right\rangle=|{\bf c}{\bf u}+{\bf d}|^{-2(1+p)}\sum_{m_j'}D^{(j)}_{m_jm_j'}\left(\frac{({\bf c}{\bf u}+{\bf d})^{-1}}{|{\bf c}{\bf u}+{\bf d}|} \right) \left|g^{-1}\cdot {\bf u},m_j';j, p \right\rangle,  \e
where $p=\frac{1}{2}+i\nu$ and $g^{-1}\cdot {\bf u}= ({\bf a}{\bf u}+{\bf b})({\bf c}{\bf u}+{\bf d})^{-1}$ with $g^{-1}=\left( \begin{array}{clcr} {\bf a} & {\bf b} \\ {\bf c} & {\bf d} \\    \end{array} \right) \in Sp(2,2)$. The action of the group element on quaternion ${\bf u}$ is defined by (\ref{qt}). $D_{m_jm_j'}^{(j)}$ furnish a certain representation of $SU(2)$ group in a $(2j+1)-$dimensional Hilbert space $V^j$ which is defined explicitly in \cite{ru70,ru73,vikl}:
$$ D^{(j)}_{mm'}({\bf u})=\left[\frac{(j+m)!(j-m)!}{(j+m')!(j-m')!}\right]^{\frac{1}{2}} \times $$
\b \label{su2} \sum_n \frac{(j+m')!}{n!(j+m'-n)!}\frac{(j-m')!}{(j+m-n)!(n-m-m')!}  u_{11}^{n}u_{12}^{j+m-n}u_{21}^{j+m'-n}u_{22}^{n-m-m'} .\e
The representation $D^{(j)}$ can be also defined in terms of the representation of little group $SO(3)$, which preserves \cite{wei} 
\b \label{littlmassiv}  \xi^\alpha_u(0)=\xi^0\left( 1, 0,0,0,1\right), \;\;\;\; {\bf u}(0)=(0,0,0,1),\e and then, one can use the "Lorentz boost" to obtain 
$\xi^\alpha_u \equiv \xi^0\left(1, \frac{\vec{k}}{k^0} , \; \frac{H\nu}{k^0}\right)$. The representation $U^{(j,p)}$ acts on an infinite dimensional Hilbert space ${\cal H}^{(j,p)}_{u}$:
$$ \left|{\bf u},m_j;j,p \right\rangle \in {\cal H}^{(j,p)}_{u}, \;\;\; \xi_u=(\xi^0, \xi^0 {\bf u}),\; \;\; \xi^0 >0,\;\;\;|{\bf u}|=1, \;\;\; -j\leq m_j\leq j.$$ 
The two representations with $\sigma_1=1+p=\frac{3}{2}+i\nu$ and  $\sigma_2=2-p=\frac{3}{2}-i\nu=\sigma_1^*$ are unitary equivalent (Th\'eor\'eme $2.1.$ page $389$ in \cite{tak}). They satisfy $\sigma_1+\sigma_2=3$ or $p_1+ p_2= 1$. It means that there exist a unitary operator $S$ ($S S^\dag=1$), so that: 
\b \label{uep} S \left|{\bf u},m_j;j,p \right\rangle=\left|{\bf u},m_j;j,1-p \right\rangle,\;\;\;\;\; SU^{(j,p)}(g) S^\dag=U^{(j,1-p)}(g),\e
where$$ \left\langle \tilde{{\bf u}}',m'_j;j,p^* |{\bf u},m_j;j,p \right\rangle=\left\langle \tilde{{\bf u}}',m'_j;j,1-p^* |{\bf u},m_j;j,1-p \right\rangle.$$
For the principal series, one has $1-p=p^*$. These two unitary equivalent representations are needed to construct the quantum field operator in terms of creation and annihilation operators.

The UIR of the dS group for discrete series is constructed on the unit ball homogeneous space $B$ or ${\bf q}$-space \cite{tak}:
$$ T^{(j_1,j_2,p)}(g)\left|{\bf q},m_{j_1},m_{j_2};j_1,j_2,p\right\rangle=|{\bf c}{\bf q}+{\bf d}|^{-2(1+p)} \times $$ \b \label{dseris2}\sum_{m_{j_1}'m_{j_2}'}D^{(j_1)}_{m_{j_1}'m_{j_1}}\left( \frac{({\bf a}+{\bf b}\bar {\bf q})^{-1}}{|{\bf c}{\bf q}+{\bf d}| }\right)
D^{(j_2)}_{m_{j_2}m_{j_2}'}\left(\frac{({\bf c}{\bf q}+{\bf d})^{-1}}{|{\bf c}{\bf q}+{\bf d}|} \right)
 \left|g^{-1}\cdot {\bf q},m_{j_1}',m_{j_2}';j_1,j_2,p\right\rangle.  \e
In this case, one has an infinite dimensional Hilbert space  ${\cal H}^{(j_1,j_2,p)}_{q}$,
$$ \left|{\bf q},m_{j_1},m_{j_2};j_1,j_2,p\right\rangle \in {\cal H}^{(j_1,j_2,p)}_{q},\;\;\; {\bf q}=r{\bf u} \in \R^4, |{\bf q}|=r<1, \;\; |{\bf u}|=1,\;\; -j\leq m_j\leq j.$$
The discrete series representations $T^{(j,0,p)}$ and $T^{(0,j,p)}$ are proportional to the two representations $\Pi^+_{j,p}$ and $\Pi^-_{j,p}$ in the Dixmier notation \cite{dix}:
\b \label{dis0jp} T^{(0,j;p)}(g)\left|{\bf q},m_j;j,p\right\rangle=|{\bf c}{\bf q}+{\bf d}|^{-2p-2}\sum_{m_j'}D^{(j)}_{m_jm_j'}\left(\frac{({\bf c}{\bf q}+{\bf d})^{-1}}{|{\bf c}{\bf q}+{\bf d}|} \right)\left|g^{-1}\cdot {\bf q},m_j';j,p\right\rangle, \e and 
\b \label{dis0jp2} T^{(j,0;p)}(g)\left|{\bf q},m_j;j,p\right\rangle=|{\bf c}{\bf q}+{\bf d}|^{-2p-2}\sum_{m_j'}D^{(j)}_{m_j'  m_j}\left(\frac{({\bf a}+{\bf b}\bar {\bf q})^{-1}}{|{\bf c}{\bf q}+{\bf d}|} \right)\left|g^{-1}\cdot 
{\bf q},m_j';j,p\right\rangle. \e
The representations $\Pi^\pm_{j,j}$ in the null curvature limit correspond to the Poincar\'e massless field with helicity $\pm j$ \cite{babo,micknied}. For a massless field, $D^{(j)}$ furnishes a certain representation of the little group $ISO(2)$ \cite{wei} so that $m_j=m'_j=j$ for $\Pi^+_{j,j}$ and 
$m_j=m'_j=-j$ for $\Pi^-_{j,j}$, the other cases of $m_j$ and $m'_j$ vanish. In this case the little group $ISO(2)$ preserves the five-vector $\xi^\alpha_B(0)$,  
\b \label{littlmasles} \xi^\alpha_B(0)=\xi^0\left( 1, 0,0,\frac{1}{\sqrt{2}},\frac{1}{\sqrt{2}}\right),\e
 and then, one can use the "Lorentz boost" to obtain
$\xi^\alpha_B \equiv \xi^0\left(1, \frac{\vec{k}}{k^0} , \; \frac{H}{k^0}\right)$. These representations were constructed previously \cite{ma77,fr59,wei}.

For the representations (\ref{dis0jp}) and (\ref{dis0jp2}), similar to the principal series, one can show the two representations with the values $p_1$ and  $p_2$ are unitary equivalent ($p_1+p_2=1$) \cite{tak}:
\b \label{ued} {\cal S} \left|{\bf q},m_j;j,p \right\rangle=\left|{\bf q},m_j;j,1-p \right\rangle,\;\;\;\; {\cal S}T^{(0,j;p)}(g) {\cal S}^\dag=T^{(0,j;1-p)}(g), \e
with ${\cal S}{\cal S}^\dag=1$. It means that the representation $T^{(0,j;p)}(g)$ for the values $p=\frac{1}{2},1,\frac{3}{2},2,...$ is unitary equivalent with $p=\frac{1}{2},0,-\frac{1}{2},-1,...$. The eigenvalues of the Casimir operators for these two set of value or under the transformation $p \longrightarrow 1-p$, do not change. The equations (\ref{uep}) and (\ref{ued}) are used to define the quantum field operators with various spin in sections IV and V.

The complementary series representations can be only associated with the tensor fields with $j=0,1,2,..$. These representation are constructed on three-sphere $S^3$ or ${\bf u}-$space \cite{dix,tak}. Among these representations the scalar representation with $j=p=0$ has only corresponding Poincar\'e group representation in the null curvature limit. This representation relates to the massless conformally coupled scalar field \cite{brmo}. The massless conformally coupled scalar field is the building block of the massless fields in dS space, since the other massless spin fields can be constructed by this field. The complementary series representations with $j=p=0$ is defined as \cite{tak}:
\b \label{comserp} U^{(0,0)}(g)\left|{\bf u},0;0,0 \right\rangle=|{\bf c}{\bf u}+{\bf d}|^{-2} \left|g^{-1}\cdot {\bf u},0;0, 0\right\rangle. \e
The scalar product in these Hilbert spaces are presented by a function ${\cal W}({\bf u},{\bf u}')$ which is defined on $S^3\times S^3$ \cite{tak}. The unitary equivalent representation (\ref{comserp}) is corresponding to $j=0,\; p=1$ and it is:
\b \label{comserpueq} U^{(0,1)}(g)\left|{\bf u},0;0,1 \right\rangle=|{\bf c}{\bf u}+{\bf d}|^{-4} \left|g^{-1}\cdot {\bf u},0;0, 1\right\rangle. \e 
Using an unitary operator $S$, the following relations between these two representations and corresponding states can be defined:
\b \label{ueqcomp} S \left|{\bf u},0;0,0 \right\rangle=\left|{\bf u},0;0,1 \right\rangle,\;\;\; SU^{(0,0)}(g) S^\dag=U^{(0,1)}(g). \e

One of the most important result of this formalism is that the total volume of the homogeneous spaces, which the UIR of the dS group constructed on, {\it i.e.} ${\bf q}$-space, ${\bf u}$-space or $\xi$-space, are finite:
$$ \int_{S^3}d\mu({\bf u})=1, \;\; \int_{B}d\mu({\bf q})=\frac{\pi^2}{2},$$
where $d\mu({\bf u})$ is the Haar measure or $SO(4)$-invariant normalized volume on three-sphere $S^3$ and  $d\mu({\bf q})=2\pi^2r^3drd\mu({\bf u})$ is the Euclidean measure on the unit ball $B$ \cite{tak}. We have used the definition  ${\bf q}=r{\bf u}\in B$ with ${\bf u}\in S^3, \;r<1$. In the following section, we discus that the two $x$- and $\xi$-spaces play a similar role as the space-time and the energy-momentum in Minkowskian space-time. Since a maximum length for an observable (or an even horizon in dS space or $x$-space) exists then a minimum size in the $\xi$-space (or the parameters in Hilbert space) can be defined by using the Heisenberg uncertainty principle. Each point in $\xi$-space represents a vector in Hilbert space and the number of points is infinite mathematically. Since the total volume of $\xi$-space is finite and a minimum length in $\xi$-space exists from uncertainty principle, the total number of points become finite physically. It means that the total number of quantum states in these Hilbert spaces is finite \cite{ta13}.  

As an example, we consider the compact space $S^1$, where the total volume of space is finite ($2\pi R$) but the total number of points is infinite. If a minimum length such as Planck length $l_p$ exists, the total number of points becomes finite ${\cal N}=2\pi R(l_p)^{-1}$. Therefore, the total number of quantum states depends on the scale of energy in the system or on the minimum length.

Although one can mathematically define an infinite dimensional Hilbert space, by accepting a minimum length in $\xi$-space, the total number of quantum states is physically finite. For principal series (${\cal H}^{(j,p)}_{u}$), we have \cite{ta13}
\b {\cal N}_{{\cal H}^{(j,p)}_{u}} =(2j+1) \int_{S^3} d\mu(\xi_u)=f(H,j,\nu,\xi^0)=\mbox{finite value},\e
and for discrete series ${\cal H}^{(j_1,j_2,p)}_{q}$, it is
\b {\cal N}_{{\cal H}^{(j_1,j_2,p)}_{q}}=(2j_1+1)(2j_2+1) \int_B d\mu(\xi_B)=\mbox{finite value}.\e
The total number of quantum states is a function of $H$, $p$, $j$ and $\xi^0$. This result is due to the existence of a minimum length and the compactness of the homogeneous spaces in which the Hilbert spaces (or the UIR) are constructed on. Then the total number of quantum states becomes finite \cite{ta13}.

\subsection{Physical conditions}
 
The tensor or tensor-spinor fields on dS hyperboloid in the ambient space formalism which are associated with the UIR's of dS group, must satisfy the following physical conditions:
\begin{itemize}
\item[i)] The field equations are: \b \label{fexs} \left( Q^{(1)}_l- \left<Q^{(1)}_{l,p}\right>\right){\cal K}(x)=0,\;\; \;\; \left( Q^{(1)}_j- \left<Q^{(1)}_{j,p}\right>\right)\Psi(x)=0.\e
\item[ii)] The tensor or tensor-spinor fields are homogeneous functions with degree $\lambda$: $$x\cdot\partial {\cal K}(x)=\lambda {\cal K}(x),\;\;\;\; x \cdot \partial \Psi(x)=\lambda \Psi(x).$$
\item[iii)] The transversality condition: $$ x\cdot{\cal K}(x)=0.\;\; \;\; x\cdot\Psi(x)=0.$$
\item[iv)] The divergencelessness condition: \b \label{divergence} \partial_l^\top \cdot {\cal K}(x)=0,\;\; \;\; \partial_l^\top  \cdot \Psi(x)=0. \e  $(\partial_l^\top \cdot)$ is defined as a generalized divergence on the dS hyperboloid, which acts on a tensor or a tensor-spinor field of rank-$l$ and results in a transverse tensor or a tensor-spinor field of rank-$l-1$.
\item[v)] The tracelessness condition: $${\cal K}'(x)=\eta\cdot\cdot
{\cal K}(x)=0, \;\;\;\; \Psi'(x)=\eta\cdot\cdot\Psi(x)=0.$$
\item[vi)] The index symmetrization condition: $${\cal K}_{\alpha\beta}={\cal K}_{\beta\alpha},\;\;\;\; \Psi_{\alpha\beta}=\Psi_{\beta\alpha}.$$
\item[vii)] For the tensor-spinor fields we have $$\gamma  \cdot \Psi(x)=0.$$
\end{itemize}
These conditions preserve the dS invariance and present the simplest solutions for the wave equations in the dS space time. By the principle (B) and the above conditions, the field operators must transform as the UIRs of the dS group and they are correspond to the elementary systems \cite{ta97}. In the next section, by using the second-order Casimir operator, we present the field equations and the homogeneous degrees of the tensor (-spinor) fields on dS hyperboloid.


\setcounter{equation}{0}
\section{Field equations}

In Minkowskian space-time, one can define the Fourier transformation between two variables $X^\mu$ and $k^\mu$, when the four-vector $X^\mu$ represents space-time coordinates in Minkowski space and the transformed variable $k^\mu$ represents energy-momentum four-vector. Generally, one cannot define the Fourier transformation on a curved manifold, but in the dS space-time, due to the maximally symmetric properties of dS hyperboloid, one can define a similar transformation, namely, the Fourier-Helgason-type transformation \cite{hel62,hel94} or the Bros-Fourier transformation \cite{brmo03}. Then, corresponding to any space-time variable $x^\alpha$ exists another variable or the transform variable, $\xi^\alpha$, which is defined in the positive cone $C^+$. In what follows, these spaces are called $x$-space (dS hyperboloid $M_H$) and $\xi$-space (positive cone $C^+$). The $\xi$-space can be written in terms of two subspaces: the three-sphere $S^3$ ($\xi_u$) and the closed unit ball $B$ ($\xi_B$). The necessary relations concern to these three different homogeneous spaces are presented in \cite{hel62,hel94}.

The dS invariant volume element in the $x$-space on the dS hyperboloid is \cite{brmo}:
\b \label{mainds} d\mu(x)=\left.\frac{dx^{(0)}\wedge dx^{(1)}\wedge dx^{(2)}\wedge dx^{(3)} \wedge dx^{(4)}}{d(x.x+H^{-2})}\right|_{M_H}=d\mu (x').\e
The dS invariant volume element on three-sphere is \cite{tak}:
\b \label{mains3} \frac{d\mu({\bf u})}{|1+{\bf u}|^6}= \frac{d\mu({\bf u}')}{|1+{\bf u}'|^6},  \;\;\;\;\;\; d\mu({\bf u}')=\frac{d\mu({\bf u})}{|{\bf c}{\bf u}+{\bf d}|^6} ,\e
where ${\bf u}'=g^{-1}\cdot {\bf u}$, $g^{-1}=\left( \begin{array}{clcr} {\bf a} & {\bf b} \\ {\bf c} & {\bf d} \\    \end{array} \right)$and $|1+{\bf u}'|=\frac{|1+{\bf u}|}{|{\bf c}{\bf u}+{\bf d}|}.$
Also the dS invariant volume element on the unit ball $B$ is \cite{tak}: 
\b \label{mainb} \frac{d\mu({\bf q})}{(1-|{\bf q}|^2)^4}= \frac{d\mu({\bf q}')}{(1-|{\bf q}'|^2)^4}, \;\;\;\;\;\; d\mu({\bf q}')=\frac{d\mu({\bf q})}{|{\bf c}{\bf q}+{\bf d}|^8}, \e
where ${\bf q}'=g^{-1}\cdot {\bf q}$ and $1-|{\bf q}'|^2=\frac{1-|{\bf q}|^2}{|{\bf c}{\bf q}+{\bf d}|^2 }$. These invariant volume elements are used for defining the transformation of the quantum field operators in dS ambient space formalism. The total volumes of these three spaces are:
$$ \int_{dS}d\mu(x)=\infty,\;\;\;\; \int_{S^3}d\mu({\bf u})=1, \;\;\;\; \int_{B}d\mu({\bf q})=\frac{\pi^2}{2}.$$  
The following integral is used for defining the scalar product in the Hilbert space of the discrete series:
$$ \int_{B} \left( 1-|{\bf q}|^2\right)^{2p-2}d\mu({\bf q}) =\frac{\pi^2}{2p(2p-1)}.$$

By using the volume element on theses spaces, the Dirac delta function  can be defined as \cite{gesh}:
$$ \int_{dS} \delta_{dS}(x-x') d\mu(x)=1,$$
$$ \int_{S^3} \delta_{S^3}({\bf u}-{\bf u}') d\mu({\bf u})=1,\;\;\;\; \int_{B} \delta_{B}({\bf q}-{\bf q}') d\mu({\bf q})=1.$$
Using the above Dirac delta function, one can define the orthogonal basis for corresponding Hilbert spaces \cite{god,vikl3}:
\b \label{ketbra} \left< x|x'\right>\equiv N(x)\delta_{dS}(x-x'),\;\; \left< \tilde{\bf u}|{\bf u}'\right>\equiv N({\bf u})\delta_{S^3}({\bf u}-{\bf u}'),\;\; \left< \tilde{\bf q}|{\bf q}'\right>\equiv N({\bf q})\delta_{B}({\bf q}-{\bf q}'),\e
where $N(x)$, $N({\bf q})$ and $N({\bf u})$ are the normalization constants which can be fixed by imposing the physical conditions such as the local Hadamard condition. In the following two subsections, by using the definition of the second-order Casimir operator in $x$-space and ${\bf q}$-space, the field equations for various spin fields are presented.

\subsection{Field equation in $x$-space}

The tensor field ${\cal K}_{\alpha_1 ...\alpha_l}(x)$ or tensor-spinor fields $\Psi_{\alpha_1 ...\alpha_l}(x)$ can be associated with the UIR of the dS group if they satisfy the conditions (i-vii) in section II-D. The homogeneity condition allows us to write the tensor or tensor-spinor fields in the following forms \cite{brgamo,gata,gagata,taazba,berotata,derotata}:
$$ {\cal K}_{\alpha_1 ...\alpha_l}(x)={\cal D}_{\alpha_1 ...\alpha_l}(x,\partial;\lambda)\left(x\cdot\xi\right)^\lambda={\cal U}_{\alpha_1 ...\alpha_l}(x,\xi;\lambda)\left(x\cdot\xi\right)^\lambda , $$
$$ \Psi_{\alpha_1 ...\alpha_l}(x)= D_{\alpha_1 ...\alpha_l}(x,\partial;\lambda)\left(x\cdot\xi\right)^\lambda=U_{\alpha_1 ...\alpha_l}(x,\xi;\lambda)\left(x\cdot\xi\right)^\lambda,$$ where
$\xi \in C^+$ and the homogeneous degree of ${\cal U}$ and $U$ is zero: $$ x\cdot \partial\; {\cal U}=0=x\cdot \partial\; U, \;\;\;\; x \cdot\partial \left(x\cdot\xi\right)^\lambda=\lambda \left(x\cdot\xi\right)^\lambda.$$ 

Utilizing the equations (\ref{casimirl}), (\ref{qe1}), (\ref{fexs}) and the conditions (i-vii) in section II-D, the field equation for the tensor field ${\cal K}_{\alpha_1 ...\alpha_l}(x)$ ($j=l$) reads as:
\b \label{fexl} \left[Q^{(1)}_0+(p+1)(p-2)\right]{\cal K}_{\alpha_1 ...\alpha_l}(x)=0.\e
This equation can be written in the following form:
\b \label{fexlmass} \left[\square_H-H^2(p+1)(p-2)\right]{\cal K}_{\alpha_1 ...\alpha_l}(x)=0,\e
and the associated mass parameter to the tensor field or boson field is 
\b \label{mbp}  m^2_{b,p}=-H^2(p+1)(p-2).\e
For principal series $(p=\frac{1}{2}+i\nu)$, we have $m^2_{b,\nu}=H^2(\frac{9}{4}+\nu^2)$. The mass parameter in the cases of the complementary and discrete series $(p=0,1,2,3,...)$ are $m^2_{b,0}=2H^2$, $m^2_{b,1}=2H^2$, $m^2_{b,2}=0,  ...$. One can see for the values $p>2$ the mass square is negative or mass parameter is imaginary. These fields may be called the tackyon fields. 

Since the tensor field ${\cal K}$ is a homogeneous function with degree of $\lambda$ we obtain \cite{brgamo,ta97}: \b \label{hodeg} Q^{(1)}_0\left(x\cdot \xi\right)^\lambda=-\lambda(\lambda+3)\left(x\cdot\xi\right)^\lambda, \e then from the equation (\ref{fexl}), the homogeneous degree $\lambda$ must satisfy the following equation:
$$ -\lambda(\lambda+3)+(p+1)(p-2)=0 .$$
It has two solutions as follows:
\b \label{hdll} \lambda_{1,2}= -\frac{3}{2}\pm \sqrt{\frac{9}{4}+(p+1)(p-2)}.\e
One can relate these solutions to the two unitary equivalent representations of dS group (\ref{uep}), (\ref{ued}) and (\ref{ueqcomp}). 
In terms of mass parameter, it would be $$ \lambda_{1,2}= -\frac{3}{2}\pm \sqrt{\frac{9}{4}-\frac{m^2_{b,p}}{H^2}}. $$
For principal series, we have:
\b \label{hdmlp} p=\frac{1}{2}+i\nu: \;\;\; \lambda_1= -\frac{3}{2}+i\nu, \;\;\; \lambda_2=-\frac{3}{2}-i\nu, \;\; m^2_{b,\nu}=H^2\left(\frac{9}{4}+\nu^2\right).\e 
In the null curvature limit, one obtains the mass in the Minkowsakian space-time \cite{brgamo,brmo,ta97}:
$$ \lim_{H\rightarrow 0,\nu\rightarrow \infty} m^2_{b,\nu}=\lim_{H\rightarrow 0,\nu\rightarrow \infty} \left(\frac{9}{4}H^2+H^2\nu^2\right)=\lim_{H\rightarrow 0,\nu\rightarrow \infty} (H\nu)^2=m^2 .$$
For complementary ($p=0$) and discrete series ($p\leq l=j=1,2,3,..$), we have:
$$ p=0:\;\;\;\; \lambda_1= -2,\;\;\lambda_2=-1,\;\; m^2_{b,1}=2H^2, $$
 $$ p=1:\;\;\;\; \lambda_1= -2,\;\;\lambda_2=-1,\;\; m^2_{b,1}=2H^2, $$
$$p=2:\;\;\;\; \;\;\lambda_1= -3,\;\;\;\lambda_2=0,\;\;\;\; m^2_{b,2}=0,$$
\b \label{hdllp} p=3:\;\;\;\; \lambda_1= -4,\;\;\lambda_2=1,\;\; m^2_{b,3}=-4H^2,....  .\e

The minimally coupled scalar field corresponds to the value $j=0$ and $p=2$ or $p=-1$, therefore, $\lambda_1= -3,\;\;\lambda_2=0$ and $m^2_{b,-1}=0$, but these are not acceptable values for $p$ in the UIR of the dS group. The constant solution ($\lambda=0$) leads to the zero mode problem. To construct a covariant quantum field, one must use the Krein space quantization, so that the field operators transform according to the indecomposable representation of the dS group \cite{gareta}. This problem is also appeared for the rank-$2$ symmetric tensor field with $p=j=2$ (linear gravity) \cite{derotata}. In section VII-B, we present another method for solving this problem.

For $p>2$, one of the homogeneous degrees of tensor field becomes positive and the plan wave is $(x.\xi)^n, n>0$. This plane wave is infinite for large $x$:
\b \label{pws2i} \lim_{x\rightarrow \infty} (x\cdot\xi)^n\longrightarrow \infty.\e
Consequently, one cannot define a massless boson particle with spin $j=p\geq 3$. The class of functions of the type $(x\cdot\xi)^n, n>0$ is infinite in a large $x$. One cannot define the field operators for these cases in a distributions sense \cite{stwi,brmo}.

Using the equations (\ref{casimirj}), (\ref{qe1}), (\ref{fexs}) and the conditions (i-vii) in section II-D, the field equation for tensor-spinor field $\Psi_{\alpha_1 ...\alpha_l}(x)$ ($j=l+\frac{1}{2}$) reads as:
\b \label{fexj} \left[Q^{(1)}_0+\frac{i}{2}\gamma_{\alpha}\gamma_{\beta}
       M^{\alpha \beta}+(p+1)(p-2)-\frac{7}{4}\right]\Psi_{\alpha_1 ...\alpha_l}(x)=0.\e
The following identity:
\b \label{identity1} \left( \frac{i}{2}\gamma_{\alpha} \gamma_{\beta}M^{\alpha\beta}-2\right)^2=-Q^{(1)}_0-\frac{i}{2}\gamma_{\alpha} \gamma_{\beta}M^{\alpha\beta}+4,\e 
allows us to obtain the first-order field equations in the following form:
\b \label{fieqjlin} \left[\frac{i}{2}\gamma_{\alpha} \gamma_{\beta}M^{\alpha\beta} -2 \pm\sqrt{\frac{9}{4}+(p+1)(p-2)}\right]\Psi_{\alpha_1 ...\alpha_l}(x)=0.\e
By replacing above equation in the equation (\ref{fexj}), the second-order field equation can be rewritten as:
\b \label{fexj2} \left[Q^{(1)}_0+\frac{1}{4}+(p+1)(p-2)\mp  \sqrt{\frac{9}{4}+(p+1)(p-2)}\right]\Psi_{\alpha_1 ...\alpha_l}(x)=0.\e
The meaning of the $\pm$ signs, in the first-order spinor field equations (\ref{fieqjlin}) may be interpreted by the dS Dirac equation and its dual equation \cite{ta97,bagamota}, but the interpretation of these signs in the second-order equation (\ref{fexj2}) is completely different. We obtain two second-order field equations with two different mass parameters associated with the two spinor fields. This problem will be discussed in section VII-E.

The equation (\ref{fexj2}) can be written in the following form:
\b \label{fexj2mass} \left[\square_H-H^2\left(\frac{1}{4}+(p+1)(p-2)\mp  \sqrt{\frac{9}{4}+(p+1)(p-2)}\right)\right]\Psi_{\alpha_1 ...\alpha_l}(x)=0,\e
then the mass parameter associated with the tensor-spinor or fermion fields, is: 
\b \label{mfp} m^2_{f,p}=-H^2\left(\frac{1}{4}+(p+1)(p-2)\mp  \sqrt{\frac{9}{4}+(p+1)(p-2)}\right).\e
 
For principal series, we have $m^2_{f,\nu}=H^2(2+\nu^2\mp\sqrt{- \nu^2})=H^2(2+\nu^2\pm i\nu)$. The mass square has two parts: a real part $(2H^2+H^2\nu^2)$ and an imaginary part $(\pm H^2\nu)$. In the null curvature limit, we obtain the mass in the Minkowsakian space-time \cite{brgamo,brmo,ta97}:
$$ \lim_{H\rightarrow 0,\nu\rightarrow \infty} m^2_{f,\nu}=\lim_{H\rightarrow 0,\nu\rightarrow \infty}(2H^2+H^2\nu^2\pm iH H\nu)=m^2.$$ 
It is known that the mass is a well-defined quantity in the Minkowskian space-time contrary to the dS space, where it is only a parameter, which turns to be real mass in the null curvature limit. Although the mass parameter is imaginary in the dS space for spinor fields, the probability amplitude can be defined properly for these fields.

There is no complementary series representation for the spinor fields, whereas, for the discrete series $(p=\frac{1}{2}, \frac{3}{2}, \frac{5}{2},...)$, we have:
\b  m^2_{f,\frac{1}{2}}=2H^2, \;\;\; m^2_{f,\frac{3}{2}}=
\left\{ \begin{array}{clcr} 2H^2\\
 0 \\ \end{array} \right.
 ,\;\;\;\;m^2_{f,\frac{5}{2}}=\left\{ \begin{array}{clcr} 0\\
 -4H^2 \\ \end{array} \right.. \e
There are two different tensor-spinor fields with two different mass parameters and different field equations for each value of $p\geq \frac{3}{2}$. Beginning with the spinor field equation (\ref{fexj}), one can obtain two different spinor field equations (\ref{fexj2}). Relations between these spinor fields and the UIR's of the dS group will be discussed in the section VII-E.

$\Psi$ is a homogeneous function with degree $\lambda$ and from the equations (\ref{fexj2}) and (\ref{hodeg}), we obtain:
$$ -\lambda(\lambda+3)+\frac{1}{4}+(p+1)(p-2)\mp\sqrt{\frac{9}{4}+(p+1)(p-2)}=0,$$
which contains two separate equations with two solutions for each ones:
$$ \lambda^{(+)}_{1,2}=-\frac{3}{2} \pm \sqrt{\frac{9}{4}+\frac{1}{4}+(p+1)(p-2)+\sqrt{\frac{9}{4}+(p+1)(p-2)}},$$
\b \label{hdf} \lambda^{(-)}_{1,2}=-\frac{3}{2} \pm \sqrt{\frac{9}{4}+\frac{1}{4}+(p+1)(p-2)-\sqrt{\frac{9}{4}+(p+1)(p-2)}}.\e
For principal series, one obtains:
$$  \lambda^{(-)}_1=-2+ i\nu, \;\; \lambda^{(-)}_2=-1 -i\nu,\;\;\; m^2_{f,\nu}=H^2(2+\nu^2\pm i\nu), $$
\b \label{hdmjp} \lambda^{(+ )}_1= -2- i\nu, \;\; \lambda^{(+ )}_2=-1 +i\nu ,\;\;\; m^2_{f,\nu}=H^2(2+\nu^2\pm i\nu), \e
with the same mass parameter for both cases and $\left[\lambda^{(- )}_1\right]^*=\lambda^{(+ )}_1 $ and $\left[\lambda^{(- )}_2\right]^*=\lambda^{(+ )}_2 $. For discrete series, we have: 
$$   p=\frac{1}{2}, \;\;\; \lambda^{(+)}=\lambda^{(-)}= -2, \;\; -1, \;\;\;  m^2_{f,\frac{1}{2}}=2H^2,$$ $$
p=\frac{3}{2},\;\;\;  \;\;
\left\{ \begin{array}{clcr} \lambda^{(-)}=-2,\;\; -1,\;\; m^2_{f,\frac{3}{2}}=2H^2\\ \lambda^{(+)}=\;\; 0,\;\; -3,\;\;\;\;\;
 m^2_{f,\frac{3}{2}}=0\;\;\;\; \\ \end{array} \right.
 ,$$
\b \label{hdmjd}
p=\frac{5}{2},\;\;\;  \;\;
\left\{ \begin{array}{clcr} \lambda^{(-)}=0,\;\; -3,\;\;\;\;\; m^2_{f,\frac{3}{2}}=0\;\;\;\;\;\;\\ \lambda^{(+)}=\;\; 1,\;\; -4,\;\;\;\;\;
 m^2_{f,\frac{3}{2}}=-4H^2 \\ \end{array} \right.
 .\e
Note that $\lambda_1+\lambda_2=-3$. The only possible values of the homogeneous degree for discrete series are $\lambda=-1,\;-2$ and $\lambda=0,\;-3$. The first two values correspond to the massless conformally coupled scalar fields and the second ones to the minimally coupled scalar fields. The minimally coupled scalar field can be written in terms of the conformally coupled scalar field (VII-B). For the other values of the homogeneous degree, an infra-red divergence appears (\ref{pws2i}) which can not be eliminated by the Krein space quantization \cite{gareta} or the other method. Using the identity 
\b \label{spinide} \frac{i}{2}\gamma_{\alpha} \gamma_{\beta}M^{\alpha\beta}=\not x \not \partial^\top, \e the spinor field equation can be rewritten as follows:
\b  \left[\not x \not \partial^\top -2 \pm \sqrt{\frac{9}{4}+(p+1)(p-2)}\right]\Psi_{\alpha_1 ...\alpha_l}(x)=0,\e
where for principal series, we obtain:
\b \label{fieqjlin2} \left[\not x \not \partial^\top -2 \pm i\nu\right]\Psi_{\alpha_1 ...\alpha_l}(x)=0.\e 
The field equation for $\Psi^\dag$ is:
$$x_\alpha \partial^\top_\beta \Psi^\dag_{\alpha_1 ...\alpha_l}(x)\gamma^0\gamma^\beta\gamma^\alpha+(-2\mp i\nu)\Psi^\dag_{\alpha_1 ...\alpha_l}(x)\gamma^0=0. $$ 
This equation can be rewritten in the following compact form:
$$ \Psi^\dag_{\alpha_1 ...\alpha_l}(x)\gamma^0 \left[\overleftarrow{ \not \partial^\top}\not x-2\mp i\nu \right]=0.$$

One can obtain proper solutions for these field equations, using the conditions (i-vii) in section II-D. Over the years in a series of papers, the solutions of various tensor or tensor-spinor fields have been studied \cite{brgamo,gareta,bagamota,gagarota,taazba,derotata}. The dS plane-wave formalism for discrete series representations with $p\geq 3$ is not a suitable solution, since it becomes infinite for large values of $x$ (\ref{pws2i}). 

The field equations can be derived through a variational procedure from the action which is defined in terms of the Lagrangian density ${\cal L}$:
$$S[\Phi]=\int d\mu(x) {\cal L}(\Psi,\nabla^\top_\alpha \Psi),\;\;\; \frac{\delta {\cal L}}{\delta \Psi}-\partial^\top_\alpha \frac{\delta {\cal L}}{\delta \partial_\alpha^\top \Psi}=0,$$
where
$d\mu(x)$ is the dS-invariant volume element (\ref{mainds}). $\nabla_\alpha^\top$ is the covariant derivative on dS hyperboloid, which is defined in the following sections. The Lagrangian density for free boson fields ($j=l$ integer) in principal series and discrete series with $j =l \neq p$ can be defined as: 
\b \label{lfbf} S[{\cal K},j,p]=\int d\mu(x) {\cal L}({\cal K},j,p)=\int  d\mu(x)  {\cal K}\cdot\cdot\left(Q_{0}^{(1)}+(p+1)(p-2)\right){\cal K}, \e 
where $\cdot \cdot$ is a shortened notation for total contraction. For free fermion fields ($j$ half-integer) in principal series and discrete series with $j \neq p $ and $j=p=\frac{1}{2}$, there are two possibility for defining the action or the Lagrangian density:
\b \label{lfsf} S_1[\Psi,\tilde{\Psi};j,p]=\int d\mu(x) {\cal L}_1(\Psi, \tilde{ \Psi};j,p)=\int  d\mu(x) \tilde{ \Psi} \cdot\cdot \left[\not x \not \partial^\top -2 \pm \sqrt{\frac{9}{4}+(p+1)(p-2)}\right]\Psi, \e
\b \label{lfsf2} S_2[\Psi,\tilde{\Psi};j,p]=\int  d\mu(x) \tilde{ \Psi} \cdot\cdot \left[Q^{(1)}_0+\not x \not \partial^\top+(p+1)(p-2)-\frac{7}{4}\right]\Psi, \e
where $\tilde{\Psi}=\Psi^\dag \gamma^0$ and $\tilde{\Psi}\cdot\cdot\Psi$ is a scalar field \cite{ta97,bagamota}. For $j=p \geq 1$, one cannot solve the field equations, since a gauge invariance appears in the field equations. For these fields, the gauge invariant field equations and then the Lagrangian must be defined, after which, the gauge fixing parameter will be introduced. This procedure will be considered in section VI.

\subsection{Field equation in $\xi$-space}

There are three different Hilbert spaces, namely, ${\cal H}_{u}^{(j,p)}$, ${\cal H}_{q}^{(j,0,p)}$ and ${\cal H}_{q}^{(0,j,p)}$ correspond to the three equations (\ref{principals}), (\ref{dis0jp}) and (\ref{dis0jp2}) respectively. Here, we only consider one in a sense that the procedure for others is utterly the same. The action of the Casimir operator $Q_j^{(1)}$ on the Hilbert space ${\cal H}_{q}^{(0,j,p)}$, equation (\ref{dis0jp}), is:
\b \label{casimirinxi}  Q_j^{(1)} \left|{\bf q};j,p\right>= \left[-j(j+1)-(p+1)(p-2)\right]\left |{\bf q};j,p\right>,\e
where $\left |{\bf q};j,p\right>\in {\cal H}_{q}^{(0,j,p)}$ is a $(2j+1)$-component vector with the component $\left |{\bf q},m_j;j,p\right>$. Using the definition  ${\bf q}=r{\bf u}$ with $|{\bf u}|=1$ and $|{\bf q}|=r$, one can simply obtain the ${\bf u}$-space equation. Here, only the ${\bf q}$-space equation is discussed. The differential form of the Casimir operator $Q_j^{(1)}$  on the Hilbert space ${\cal H}_{q}^{(0,j,p)}$ is \cite{tak}: $$ - Q_j^{(1)} \left|{\bf q};j,p\right>=\left[\left(\frac{1}{2}(1-|{\bf q}|^2)\right)^2\triangle -p(1-|{\bf q}|^2){\cal D} \right.-\frac{1}{2}(1-|{\bf q}|^2)(A_j{\cal D}_1+B_j{\cal D}_2+C_j{\cal D}_3)$$
\b  \label{casimidi} \left.+(p(p+1)-j(j+1))|{\bf q}|^2+2(j(j+1)-(p+1))\right]\left.|{\bf q};j,p\right>,\e
where the differential operators $\triangle, {\cal D}, {\cal D}_1,{\cal D}_2$ and ${\cal D}_3$ are defined explicitly in ${\bf q}$-space by Takahashi \citep{tak}: 
$$ \triangle=\frac{\partial^2}{\partial q_1^2}+\frac{\partial^2}{\partial q_2^2}+\frac{\partial^2}{\partial q_3^2}+\frac{\partial^2}{\partial q_4^2}=\frac{\partial^2}{\partial r^2}+\frac{3}{r}\frac{\partial}{\partial r}+\frac{1}{r^2}\triangle_{S^3},$$
$${\cal D}=q_1\frac{\partial}{\partial q_1}+q_2\frac{\partial}{\partial q_2}+q_3\frac{\partial}{\partial q_3}+q_4\frac{\partial}{\partial q_4}=r\frac{\partial}{\partial r}+{\bf u}\cdot \frac{\partial}{\partial {\bf u}},$$
$${\cal D}_1=q_1\frac{\partial}{\partial q_2}-q_2\frac{\partial}{\partial q_1}+q_4\frac{\partial}{\partial q_3}-q_3\frac{\partial}{\partial q_4}=u_1\frac{\partial}{\partial u_2}-u_2\frac{\partial}{\partial u_1}+u_4\frac{\partial}{\partial u_3}-u_3\frac{\partial}{\partial u_4},$$
$${\cal D}_2=q_1\frac{\partial}{\partial q_3}-q_3\frac{\partial}{\partial q_1}+q_2\frac{\partial}{\partial q_4}-q_4\frac{\partial}{\partial q_2}=u_1\frac{\partial}{\partial u_3}-u_3\frac{\partial}{\partial u_1}+u_2\frac{\partial}{\partial u_4}-u_4\frac{\partial}{\partial u_2},$$
\b\label{relations} {\cal D}_3=q_1\frac{\partial}{\partial q_4}-q_4\frac{\partial}{\partial q_1}+q_3\frac{\partial}{\partial q_2}-q_2\frac{\partial}{\partial q_3}=u_1\frac{\partial}{\partial u_4}-u_4\frac{\partial}{\partial u_1}+u_3\frac{\partial}{\partial u_2}-u_2\frac{\partial}{\partial u_3}.\e
 $\triangle_{S^3} $ is the Laplace operator on the hypersphere $S^3$. The $A_j,\; B_j$ and $C_j$ are $(2j+1)\times (2j+1)$-matrix representations of the generators of the $SU(2)$ group \cite{tak}. By using the equations (\ref{casimirinxi}) and (\ref{casimidi}), the following partial differential equation for the quantum state $\left.|{\bf q};j,p\right>$  is obtain: 
\b\label{difequ} \left[\frac{1}{4}(1-|{\bf q}|^2)\triangle -p{\cal D}-\frac{1}{2}(A_j{\cal D}_1+B_j{\cal D}_2+C_j{\cal D}_3)+(j(j+1)-p(p+1))\right]\left|{\bf q};j,p\right>=0.\e 
By solving this equation, the state $\left|{\bf q};j,p\right>$ is determined. This partial differential equation is a $(2j+1)\times (2j+1)$-matrix equation which has complicated solutions \cite{tak}. Takahashi has proved that this equation has a non-trivial solution in the following form, (page 399, Proposition 3.1. in \cite{tak}):
\b \label{qinru2} \left|{\bf q},m;j,p\right>=F(p-j,j+p+1;2;r^2)\sum_{m'} C_{mm'} \left|{\bf u},m';j,p\right>.\e
 $F$ is a hyper-geometric function, which is a polynomial function of degree $j-p$. $ \left|{\bf u},m';j,p\right>$ is a base-vector of a $(2j+1)$-dimensional Hilbert space $V^j$. 

Multiplying above equation with bra $\left<x\right|$, and stating the result in terms of $\left<x|{\bf q};j,p\right>\equiv f(x;{\bf q};j,p)$, one can show that this function has a specific relation with the tensor field ${\cal K}$ or tensor-spinor field $\Psi$ which are presented in the previous subsection. This process is similar to the first quantization in the usual quantum mechanics, which is not relevant here. In the next section, the second quantization or the quantization of these fields is discussed in the ambient space formalism.
 
It is interesting to note that the tensor fields on dS hyperboloid in terms of the five-variable $x$ or $\xi$ ($f(x)$ or $g(\xi)$) are homogeneous functions:
$$ f(lx)=l^\lambda f(x), \;\;\;\; g(l\xi)=l^\sigma g(\xi),$$
where $ \lambda$ and $\sigma$ are the homogeneous degrees. The homogeneous degree $\lambda$ was obtained by using the field equation in $x$-space in the previous subsection and the homogeneous degree $\sigma$ for each tensor (or spinor) field may be fixed by using the UIR of dS group which will be discussed seperatly for various spin fields in the following. For tensor fields, one obtains $\lambda=\sigma$ and for tensor-spinor fields $\lambda=\sigma \pm\frac{1}{2}$.


\setcounter{equation}{0} 
\section{Massive quantum free field operators}

\subsection{General case}

The massive fields in the dS space-time correspond to the principal series representation of the dS group $(p=\frac{1}{2}+i\nu)$. For constructing the quantum field operator, one must first define the creation and annihilation operators on the Hilbert space ${\cal H}^{(j,p)}_{u}$. A creation operator $a^\dag({\bf u},m_j;j,p)$ is defined as an operator that simply adds a state with quantum numbers $({\bf u},m_j;j,p)$ to the vacuum state $|\Omega >$
\b \label{cractionpe} a^\dag ({\bf u},m_j;j,p) \left| \Omega \right> \equiv\left|{\bf u},m_j;j,p \right\rangle,\e
where the vacuum state $\left| \Omega \right>$ is invariant under the action of the UIR of the dS group:
\b \label{vacumtrp} U^{(j,p)}(g) \left| \Omega \right>=\left| \Omega \right>.\e
Here, the norm of the vacuum state can be fixed as $\left< \Omega \right.\left| \Omega \right>=1 $, contrary to the norms of the other states which are fixed by the local Hadamard behavior \cite{brmo}. In what follows, it has been shown that, the vacuum state $\left| \Omega \right>$ can be identified with the Bunch-Davies or Hawking-Ellis vacuum state. 

The adjoint operator $a^\dag({\bf u},m_j;j,p)$ is $a({\bf u},m_j;j,p)$ and it can be defined from the equation (\ref{cractionpe}):
\b   \left< \Omega\right|\left[a^\dag({\bf u},m_j;j,p)\right]^\dag \equiv\left<\tilde{\bf u},m_j;j,p^* \right|,  \e
where 
\b \label{adjp} \left[a^\dag({\bf u},m_j;j,p)\right]^\dag\equiv a({\bf \tilde{u}},m_j;j,p^*)\equiv a({\bf \tilde{u}},m_j;j,1-p). \e 
The orthogonality condition on the Hilbert space ${\cal H}^{(j,p)}_{u}$ is defined explicitly by Takahashi \cite{tak};
$$ \left<\tilde{\bf u},m_j;j,p^*|{\bf u}',m'_j;j,p \right\rangle=N({\bf u},m_j) \delta_{S^3}({\bf u}-{\bf u}')\delta_{m_j m'_j},$$
where $N({\bf u}, m_j)$ is the states normalization, one can show that the operator $a({\bf u},m_j;j,p)$ removes a state from any state in which it acts on: 
\b \label{anist} \left< \Omega|a({\bf \tilde{u}},m_j;j,1-p)|{\bf u}',m'_j;j,p \right\rangle=N({\bf u},m_j) \delta_{S^3}({\bf u}-{\bf u}')\delta_{m_j m'_j}\left< \Omega \right.\left| \Omega \right>.\e
$a({\bf u},m_j;j,p)$ is called annihilation operator so that:
\b \label{anihilip} a({\bf u},m_j;j,p) \left| \Omega \right>=0. \e
Using the equations (\ref{principals}), (\ref{cractionpe}) and (\ref{vacumtrp}), the creation and annihilation operators under the action of the dS group transform as:
$$ U^{(j,p)}(g) a^\dag ({\bf u},m_j;j,p)\left[ U^{(j,p)}(g)\right]^\dag= |{\bf c}{\bf u}+{\bf d}|^{-2(1+p)}\;\;\;\;\;\;\;\;\;\;\;\;\;\;\;\;\;\;\;\;\;\;\;\;$$
\b \label{creationtra} \;\;\;\;\;\;\;\;\;\;\;\;\;\;\;\;\;\;\;\;\;\;\;\times \sum_{m_j'}D^{(j)}_{m_jm_j'}\left(\frac{({\bf c}{\bf u}+{\bf d})^{-1}}{|{\bf c}{\bf u}+{\bf d}|} \right) a^\dag(g^{-1}.{\bf u},m_j';j, p), \e
$$ U^{(j,p)}(g) a({\bf \tilde{u}},m_j;j,1-p)\left[ U^{(j,p)}(g)\right]^\dag= |{\bf c}{\bf u}+{\bf d}|^{-2(2-p)}\;\;\;\;\;\;\;\;\;\;\;\;\;\;\;\;\;\;\;\;\;\;\;$$
\b \label{anihltiontra} \;\;\;\;\;\;\;\;\;\;\;\;\;\;\;\;\;\;\;\;\;\;\;\times \sum_{m_j'}\left[D^{(j)}_{m_jm_j'}\left(\frac{({\bf c}{\bf u}+{\bf d})^{-1}}{|{\bf c}{\bf u}+{\bf d}|} \right) \right]^* a(g^{-1}.{\bf \tilde{u}},m_j';j, 1-p). \e
Also, with regard to (\ref{uep}), the annihilation operator $a({\bf u},m_j;j,p)$ transforms by the the unitary equivalent representation of principal series:
$$ U^{(j,1-p)}(g) a({\bf u},m_j;j,p)\left[ U^{(j,1-p)}(g)\right]^\dag= |{\bf c}{\bf \tilde{u}}+{\bf d}|^{-2(2-p)}\;\;\;\;\;\;\;\;\;\;\;\;\;\;\;\;\;\;\;\;\;\;\;$$
\b \label{anihltiontra2} \;\;\;\;\;\;\;\;\;\;\;\;\;\;\;\;\;\;\;\;\;\;\;\times \sum_{m_j'}\left[D^{(j)}_{m_jm_j'}\left(\frac{({\bf c}{\bf \tilde{u}}+{\bf d})^{-1}}{|{\bf c}{\bf \tilde{u}}+{\bf d}|} \right) \right]^* a(g^{-1}.{\bf u},m_j';j,p). \e

The quantum state $\left| {\bf u},m_j;j,p\right>$ is called ''one-particle'' state. The N-particle state can be obtained by acting on the vacuum with N creation operators. For identical particles, under the action of a permutation operator, N-particle state is either symmetric or anti-symmetric (bosonic or fermionic respectively). Therefore, similar to the Minkowsky space, one can prove that the following relation is hold:
\b \label{cumtre} a({\bf \tilde{u}}',m_j';j,1-p)a^\dag ({\bf u},m_j;j,p)\pm a^\dag ({\bf u},m_j;j,p)a ({\bf \tilde{u}}',m_j';j,1-p)=N({\bf u}, m_j)\delta_{S^3}({\bf u}'-{\bf u})\delta_{m_jm_j'},\e
with the $+$ and $-$ signs, corresponding to fermionic or bosonic states, respectively.

The quantum field operators for the integer cases $j=l=0,1,2,...$ can be defined in terms of creation and annihilation operators as follows:
\b \label{qfopb} {\cal K}_{\alpha_1..\alpha_l}(x)\equiv \sum_{m_j}\int_{S^3}d\mu({\bf u})\left[a({\bf \tilde{u}},m_j;j,1-p){\cal U}_{\alpha_1..\alpha_l}(x;{\bf u},m_j;j,\nu )+a^\dag({\bf u},m_j;j,p){\cal V}_{\alpha_1..\alpha_l}(x;{\bf u},m_j;j,\nu )\right],\e
and for the half-integer cases $j=l+\frac{1}{2}=\frac{1}{2}, \frac{3}{2},...$ one has:
\b \label{qfops} \Psi_{\alpha_1...\alpha_l}(x)\equiv \sum_{m_j}\int_{S^3}d\mu({\bf u})\left[a({\bf \tilde{u}},m_j;j,1-p)U_{\alpha_1..\alpha_l}(x;{\bf u},m_j;j,\nu )+a^\dag({\bf u},m_j;j,p)V_{\alpha_1..\alpha_l}(x;{\bf u},m_j;j,\nu )\right],\e
where $ \Psi_{\alpha_1...\alpha_l}, \; U_{\alpha_1..\alpha_l} $ and  $ V_{\alpha_1..\alpha_l}$ are four-component spinors. The particles that are annihilated and created by these fields may be carried non-zero values of one or more conserved quantum numbers like the electric charge. For these cases, the quantum field operator can be written in terms of creation and annihilation operators as:
\b \label{qfopss2} \Psi_{\alpha_1...\alpha_l}(x)=\sum_{m_j}\int_{S^3}d\mu({\bf u})\left[a({\bf \tilde{u}},m_j;j,1-p)U_{\alpha_1..\alpha_l}(x;{\bf u},m_j;j,\nu )+a^{c\dag}({\bf u},m_j;j,p)V_{\alpha_1..\alpha_l}(x;{\bf u},m_j;j,\nu )\right],\e 
where the label 'c' denotes the ''charge conjugate''. $a^\dag$ creates a particle, whereas $a^{c\dag}$ creates an anti-particle with the opposite electric charge.

The coefficients ${\cal U}_{\alpha_1..\alpha_l},\; {\cal V}_{\alpha_1..\alpha_l},\; U_{\alpha_1..\alpha_l}$ and $V_{\alpha_1..\alpha_l}$ are chosen so that under the dS transformations the field operators transform by the UIR of dS group (principle B) \cite{brgamo,gata,bagamota}:
\b \label{fioptrb}  U^{(j,p)}(g)  {\cal K}_{\alpha_1..\alpha_l}(x) \left[ U^{(j,p)}(g)\right]^\dag= \Lambda_{\alpha_1}^{\;\;\alpha'_1}..\Lambda_{\alpha_l}^{\;\;\alpha'_l}  {\cal K}_{\alpha'_1..\alpha'_l}(\Lambda x), \e
\b \label{fioptrs}  U^{(j,p)}(g) \Psi_{\alpha_1..\alpha_l}(x) \left[ U^{(j,p)}(g)\right]^\dag= \Lambda_{\alpha_1}^{\;\;\alpha'_1}..\Lambda_{\alpha_l}^{\;\;\alpha'_l}g^{-1}  \Psi_{\alpha'_1..\alpha'_l}(\Lambda x),\e
where $\Lambda \in SO(1,4)$ and $g \in Sp(2,2)$. To obtain the Minkowskian spinor field in the null curvature limit $(H=0)$, in the ambient space formalism, the adjoint spinor should be defined as follows  \cite{ta97,ta96,bagamota}:
$$\bar\Psi(x)\equiv \Psi^\dag(x) \gamma^0\gamma^4.$$ Therefore it transforms by the following relation \cite{ta97,bagamota,taazba}:
\b \label{fioptrbc} U^{(j,p)}(g)\bar \Psi_{\alpha_1..\alpha_l}(x) \left[ U^{(j,p)}(g)\right]^\dag=\Lambda_{\alpha_1}^{\;\;\alpha'_1}..\Lambda_{\alpha_l}^{\;\;\alpha'_l} \bar \Psi_{\alpha'_1..\alpha'_l}(\Lambda x)\left(-\gamma^4 g\gamma^4\right).\e

Using the equations (\ref{creationtra}), (\ref{anihltiontra}), (\ref{qfopb}), (\ref{fioptrb}) and the dS invariant volume elements on three-sphere (\ref{mains3}) the coefficients ${\cal U}_{\alpha_1..\alpha_l}(x;{\bf u},m_j;j,\nu )$ and ${\cal V}_{\alpha_1..\alpha_l}(x;{\bf u},m_j;j,\nu )$ satisfy the following relations:
 $$ |{\bf c}{\bf u}+{\bf d}|^{-2(1+p)+6} \sum_{m_j}D^{(j)}_{m_jm_j'}\left(\frac{({\bf c}{\bf u}+{\bf d})^{-1}}{|{\bf c}{\bf u}+{\bf d}|} \right){\cal V}_{\alpha_1..\alpha_l}(x;{\bf u},m_j;j,\nu )\;\;\;\;\;\;\;\;\;\;\;\;\;\;\;\;\;\;\;\;\;
$$
\b \label{vtbp} \;\;\;\;\;\;\;\;\;\;\;\;\;\;\;\;\;\;\;\;\;\;\;\;\;\;\;\;\;\;\;\;\;\;\;\;\;\;\;\;
=\Lambda_{\alpha_1}^{\;\;\alpha'_1}..\Lambda_{\alpha_l}^{\;\;\alpha'_l}{\cal V}_{\alpha'_1..\alpha'_l}(\Lambda x; g^{-1}.{\bf u},m'_j;j,\nu ),
\e
$$ |{\bf c}{\bf u}+{\bf d}|^{-2(2-p)+6} \sum_{m_j}\left[D^{(j)}_{m_jm_j'}\left(\frac{({\bf c}{\bf u}+{\bf d})^{-1}}{|{\bf c}{\bf u}+{\bf d}|} \right)\right]^*{\cal U}_{\alpha_1..\alpha_l}(x;{\bf u},m_j;j,\nu )\;\;\;\;\;\;\;\;\;\;\;\;\;\;\;\;\;\;\;\;\;\;
$$
\b \label{utbp} \;\;\;\;\;\;\;\;\;\;\;\;\;\;\;\;\;\;\;\;\;\;\;\;\;\;\;\;\;\;\;\;\;\;\;\;\;\;\;\;
=\Lambda_{\alpha_1}^{\;\;\alpha'_1}..\Lambda_{\alpha_l}^{\;\;\alpha'_l}{\cal U}_{\alpha'_1..\alpha'_l}(\Lambda x; g^{-1}.{\bf u},m'_j;j,\nu ).
\e
These equations are important for the calculation of the coefficients ${\cal U}_{\alpha_1..\alpha_l}(x;{\bf u},m_j;j,\nu )$ and ${\cal V}_{\alpha_1..\alpha_l}(x;{\bf u},m_j;j,\nu )$.
After making use of the equations (\ref{creationtra}), (\ref{anihltiontra}), (\ref{qfopss2}), (\ref{fioptrs}) and the dS invariant volume elements on three-sphere (\ref{mains3}) it turns out that the coefficients $U_{\alpha_1..\alpha_l}(x;{\bf u},m_j;j,\nu )$ and $V_{\alpha_1..\alpha_l}(x;{\bf u},m_j;j,\nu )$ will satisfy the following relations:
 $$ |{\bf c}{\bf u}+{\bf d}|^{-2(1+p)+6} \sum_{m_j}D^{(j)}_{m_jm_j'}\left(\frac{({\bf c}{\bf u}+{\bf d})^{-1}}{|{\bf c}{\bf u}+{\bf d}|} \right)V_{\alpha_1..\alpha_l}(x;{\bf u},m_j;j,\nu )\;\;\;\;\;\;\;\;\;\;\;\;\;\;\;\;\;\;\;\;\;\;\;
$$
\b \label{vtfp0}\;\;\;\;\;\;\;\;\;\;\;\;\; \;\;\;\;\;\;\;\;\;\;\;\;\;\;\;\;\;\;\;\;\;\;\;\;\;\;
=\Lambda_{\alpha_1}^{\;\;\alpha'_1}..\Lambda_{\alpha_l}^{\;\;\alpha'_l}g^{-1}V_{\alpha'_1..\alpha'_l}(\Lambda x; g^{-1}.{\bf u},m'_j;j,\nu ),
\e
$$ |{\bf c}{\bf u}+{\bf d}|^{-2(2-p)+6} \sum_{m_j}\left[D^{(j)}_{m_jm_j'}\left(\frac{({\bf c}{\bf u}+{\bf d})^{-1}}{|{\bf c}{\bf u}+{\bf d}|} \right)\right]^*U_{\alpha_1..\alpha_l}(x;{\bf u},m_j;j,\nu )\;\;\;\;\;\;\;\;\;\;\;\;\;\;\;\;\;\;\;\;\;\;\;
$$
\b \label{utfp0} \;\;\;\;\;\;\;\;\;\;\;\;\;\;\;\;\;\;\;\;\;\;\;\;\;\;\;\;\;\;\;\;\;\;\;\;\;\;\;\;\;
=\Lambda_{\alpha_1}^{\;\;\alpha'_1}..\Lambda_{\alpha_l}^{\;\;\alpha'_l}g^{-1}U_{\alpha'_1..\alpha'_l}(\Lambda x; g^{-1}.{\bf u},m'_j;j,\nu ).
\e

The equations (\ref{vtbp}), (\ref{utbp}), (\ref{vtfp0}) and (\ref{utfp0}) and the conditions (i-vii) in section II-D allow us to calculate the functions ${\cal U}$, ${\cal V},\; V$ and $U$ explicitly, up to a normalization constant. By using the homogeneity condition, one can write these coefficients in the following form:
\b \label{vht} {\cal U}_{\alpha_1..\alpha_l}(x;{\bf u},m_j;j,\nu )= u_{\alpha_1..\alpha_l}(x,{\bf u},m_j;j,\nu )(x.\xi_u)^\lambda ,\e 
where $\xi_u^\alpha=\xi^0(1,{\bf u})$ and $\lambda$ is given by equation (\ref{hdmlp}) for tensor fields and equation (\ref{hdmjp}) for tensor-spinor fields. Utilizing the equations (\ref{xi0t}), (\ref{x.xit}), (\ref{utbp}) and (\ref{vht}), one can obtain the transformation law of $u_{\alpha_1..\alpha_l}(x,{\bf u},m_j;j,\nu )$. 

Similar to the Minkowskian space-time, by defining the functions ${\cal U}$, ${\cal V},\; V$ and $U$ for the values $\xi^\alpha_{u}(0)=(1,0,0,0,1)$ and $x_0^\alpha=(0,0,0,0,H^{-1})$, one can calculate these functions for the arbitrary values of  $\xi_u^\alpha$ and $x^\alpha$, in terms of $\xi_{u}^\alpha(0)$ and $x_0^\alpha$, using equations (\ref{vtbp}), (\ref{utbp}), (\ref{vtfp0}) and (\ref{utfp0}). To do so, first one should determine the transformations $\Lambda^R$ and $g^R$ which make $\xi_{u}^\alpha(0)$ invariant (little group $SO(3)$):
$$\left(\Lambda^R\right)_{\;\;\alpha}^{\beta}\xi^\alpha_{u}(0)=\xi^\beta_{u}(0),\;\; \left(\Lambda^R\right)^{\;\;\alpha}_{\beta}\gamma^\beta=g^R \gamma^\alpha \left(g^R\right)^{-1},$$ or
$$ {\bf u}'(0)=g^R.{\bf u}(0)= ({\bf a}^R{\bf u}(0)+{\bf b}^R)({\bf c}^R{\bf u}(0)+{\bf d}^R)^{-1}=I,$$
where  ${\bf u}(0)=\left( \begin{array}{clcr} 1 & 0\\ 0 & 1 \\    \end{array} \right)$. Afterwards, by replacing $\Lambda^R$ and $g^R$ in the equations (\ref{vtbp}), (\ref{utbp}), (\ref{vtfp0}) and (\ref{utfp0}), one obtains:
 $$ |{\bf c}^R+{\bf d}^R|^{-2(1+p)+6} \sum_{m_j}D^{(j)}_{m_jm_j'}\left(\frac{({\bf c}^R+{\bf d}^R)^{-1}}{|{\bf c}^R+{\bf d}^R|} \right){\cal V}_{\alpha_1..\alpha_l}(x_0;{\bf u}(0),m_j;j,\nu )\;\;\;\;\;\;\;\;\;\;\;\;\;\;\;\;\;\;\;\;\;
$$
\b \label{vtbpex} \;\;\;\;\;\;\;\;\;\;\;\;\;\;\;\;\;\;\;\;\;\;\;\;\;\;\;\;\;\;\;\;\;\;\;\;\;\;\;
=\left(\Lambda^R\right)_{\alpha_1}^{\;\;\alpha'_1}..\left(\Lambda^R\right)_{\alpha_l}^{\;\;\alpha'_l}{\cal V}_{\alpha'_1..\alpha'_l}( x_0; {\bf u}(0),m'_j;j,\nu ).
\e
In this matrix equation the only unknown function is ${\cal V}_{\alpha_1..\alpha_l}(x_0;{\bf u}(0),m_j;j,\nu )$. Solving this equation and respecting the conditions (i-vii) of section II-D, ${\cal V}_{\alpha_1..\alpha_l}(x_0;{\bf u}(0),m_j;j,\nu )$  can be obtained up to a normalization constant. Similar to the Poincar\'e group in the Minkowskian space-time \cite{wei}, one can obtain these polarization tensors (-spinors) using the above equations and conditions and their transformations under the discrete symmetries in the dS space-time \cite{morrota}. These calculations will be considered in a forthcoming paper.

Utilizing the representation theory, we define the quantum states and the quantum field operators in terms of the dS plane-waves $(x\cdot\xi)^\lambda$ and the polarization tensors (or spinors) $u_{\alpha_1..\alpha_l}(x,{\bf u},m_j;j,\nu )$. But these solutions are not globally defined on the dS hyperboloid due to the ambiguity of the
phase factor \cite{vikl}, and also for $\Re \lambda <0$, which is singular on $x\cdot\xi_{u}=0$. Nevertheless, one can use
the complexified dS manifold to obtain a globally well-defined solution \cite{brmo}. In order to obtain a well-defined function, one should consider a proper definition for the $\xi^\alpha_u$ and the complex dS space-time $z^\alpha,$ it is called $i \epsilon$ prescription \cite{brgamo,brmo}. To fulfill such purpose, one considers the
solution in a complex dS space-time $M_H^{(c)}$:
$$ M_H^{(c)}=\left\{ z=x+iy\in  \C^5;\;\;\eta_{\alpha \beta}z^\alpha z^\beta=(z^0)^2-\vec z\cdot\vec z-(z^4)^2=-H^{-2}\right\}$$
\b =\left\{ (x,y)\in  \R^5\times  \R^5;\;\; x^2-y^2=-H^{-2},\; x\cdot y=0\right\}.\e
Let $T^\pm= \R^5+iV^\pm$ to be the forward and backward tubes in $ \C^5$. The domain $V^+$(resp. $V^-)$ stems from the causal structure on $M_H$:
\b \label{v+-} V^\pm=\left\{ x\in \R^5;\;\; x^0\stackrel{>}{<} \sqrt {\parallel \vec x\parallel^2+(x^4)^2}\right\}.\e
Then we introduce their respective intersections with $M_H^{(c)}$,
   \b {\cal T}^\pm=T^\pm\cap M_H^{(c)},\e
which are called forward and backward tubes of the complex dS space-time. Finally, the ``tuboid'' above $M_H^{(c)}\times M_H^{(c)}$ is defined by
\b \label{tuboid} {\cal T}_{12}=\left\{ (z,z');\;\; z\in {\cal T}^+,z' \in {\cal T}^- \right\}, \e
for more details, see \cite{brmo}. If $z$ varies in ${\cal T}^+$ (or ${\cal T}^-$) and $\xi$ lies in the positive cone ${\cal C}^+$
$$\xi \in {\cal C}^+=\left\{ \xi \in {\cal C}; \; \xi^0>0 \right\},$$
the plane wave solutions are globally defined since the imaginary part of $(z.\xi)$ has  a fixed sign. The phase is chosen such as
 \b \mbox{boundary value of} \; (z.\xi)^\lambda \mid_{x.\xi>0}>0.\e

In the following subsections, we briefly recall the quantum field operators and the Wightman two-point functions for the massive spin $j\leq 2$ fields which have been calculated previously in the $x$-space \cite{ta97,brgamo,gata,gagarota,taazba}.

\subsection{Massive scalar field}

A massive scalar field is a simple case and associates with the principal series representation ($j=0$ and $p=\frac{1}{2}+i\nu,\;\;\nu \geq 0$). The eigenvalue of the Casimir operator for this field is 
$<Q_{0,\nu}^{(1)}>=\nu^2+\frac{9}{4}$ with the corresponding
mass parameter as $m_{b,\nu}^2=H^2(\nu^2+\frac{9}{4})$. The field equation is:
$$  \left[ Q_0^{(1)}-\left(\nu^2+\frac{9}{4}\right)\right]\phi(x)=0,\;\; \mbox{or}\; \;\left[ \square_H
+H^2\left(\nu^2+\frac{9}{4}\right)\right]\phi(x)=0. $$ 
Lagrangian for this field equation is defined by:
$$S[\phi]=\int d\mu(x){\cal L}(\phi, \nabla^\top_\alpha\phi)=\int d\mu(x)\left[\frac{1}{2}\nabla_\alpha^\top \phi \nabla^{\top\alpha}\phi-\frac{1}{2}<Q_{0,\nu}^{(1)}>\phi^2\right],$$
where the covariant derivative for scalar field is $\nabla^\top_\alpha \phi=\partial^\top_\alpha \phi$. The field operator is defined by:
\b \label{scaqfop} \phi(x)=\int_{S^3}d\mu({\bf u})\left[a({\bf \tilde{u}},0;0,1-p){\cal U}(x;{\bf u},0;0,\nu )+a^\dag({\bf u},0;0,p){\cal V}(x;{\bf u},0;0,\nu )\right],\e
and so the equations (\ref{vtbp}) and (\ref{utbp}) become:
$$ {\cal U}(x;{\bf u}; \nu)=|{\bf c}{\bf u}+{\bf d}|^{2(2-p)-6} {\cal U}(\Lambda x;g^{-1}\cdot {\bf u}; \nu),$$
$$ {\cal V}(x;{\bf u}; \nu)=|{\bf c}{\bf u}+{\bf d}|^{2(1+p)-6} {\cal V}(\Lambda x;g^{-1}\cdot {\bf u}; \nu).$$
Using the transformation of $\xi_u^\alpha$ in our notation (\ref{xi0t}),
$$ \xi^\alpha_u\equiv(1,{\bf u}) \Longrightarrow \xi^\alpha_{u'}\equiv|{\bf c}{\bf u}+{\bf d}|^{2}(1,{\bf u}'),$$
and the equation (\ref{vht}), the coefficient ${\cal U}$ and ${\cal V}$ can be rewritten in terms of $\xi_u^\alpha$ as follows:
$${\cal U}(x;{\bf u};\nu )=c_1 (x\cdot\xi_u)^{-\frac{3}{2}-i\nu}, $$ 
$${\cal V}(x;{\bf u};\nu )=c_2 (x.\cdot\xi_u)^{-\frac{3}{2}+i\nu}.$$
$c_1$ and $c_2$ are arbitrary constants.
One can see that the homogeneous degree of massive scalar field is $\lambda=-\frac{3}{2}\pm i\nu$ (\ref{hdmlp}). One can simply show that this field operator satisfies the following transformation rule, under the dS group,
$$ U^{(0,p)}(g) \phi(x) \left[ U^{(0,p)}(g)\right]^\dag=\phi(\Lambda x).$$ 

The well-defined field operator is obtained by taking the boundary value of the field operator in the complex dS space-time \cite{brmo}:
$$ \phi(x)=\lim_{y\rightarrow 0} \Phi(z)=\lim_{y\rightarrow 0} \Phi(x+iy),$$
where
$$\Phi(z)=\int_{S^3}d\mu({\bf u})\left[a({\bf \tilde{u}},0;0,1-p){\cal U}(z;{\bf u},0;0,\nu )+a^\dag({\bf u},0;0,p){\cal V}(z;{\bf u},0;0,\nu )\right].$$
Therefore the quantum field operator in this notation can be written as \cite{brgamo,brmo}:
$$ \phi(x)=\int_{S^3}   d\mu({\bf u}) \left\lbrace\; c_1 a({\bf \tilde{u}},0;0,1-p)\left[(x\cdot\xi_u)_+^{-\frac{3}{2}-i\nu}
        +e^{-i\pi(-\frac{3}{2}-i\nu)}(x\cdot\xi_u)_-^{-\frac{3}{2}-i\nu}\right]\right.\;\;\;\;\;\;\;\;\;\;\;\;\;\;\;\;\;\;\;\;$$
     \b\label{foinam} \;\;\;\;\;\;\;\;\;\;\;\;\;\;\;\;\;\;\;\;\;\;\ \left. +c_2 a^\dag({\bf u},0;0,p)\left[(x\cdot\xi_u)_+^{-\frac{3}{2}+i\nu}
        +e^{i\pi(-\frac{3}{2}+i\nu)}(x\cdot\xi_u)_-^{-\frac{3}{2}+i\nu}\right]\; \right\rbrace ,\e
where \cite{gesh}: $$ (x\cdot \xi)_+=\left\{\begin{array}{clcr} 0 & \;\;\;\mbox{for} \;
x\cdot
\xi\leq 0\\ (x\cdot \xi) & \;\;\;\mbox{for} \;x\cdot \xi>0 \\ \end{array}
\right. .$$  For a real scalar field, we have $c_1=c_2\equiv \sqrt{ c_{0,\nu}}$.

A two-point function ${\cal W}(x,x')$ is boundary value (in the distributional sense) of an analytic function $W(z,z')$.
$W(z,z')$ is maximally analytic, i.e., it can be analytically continued to the cut domain \cite{brmo}
$$\Delta =\{ (z,z') \in M_H^{(c)} \times M_H^{(c)} \;\; :\;\; (z-z')^2\leq 0 \}. $$
The two-point Wightman function ${\cal W}(x,x')$ is boundary value of $W(z,z')$
from ${\cal T}_{12}$ and the ``permuted Wightman function'' ${\cal W}(x',x)$
is boundary value of $W(z,z')$ from the domain
$$ {\cal T}_{21}=\{ (z,z');\;\; z'\in {\cal T}^+,z \in {\cal T}^- \}. $$
Therefore the analytic function $W$ is \cite{brgamo,brmo}:
$$ W(z,z')=\left<\Omega|\Phi(z)\Phi(z')|\Omega\right>=\int_{S^3}d\mu({\bf u}_1)\int_{S^3}d\mu({\bf u}_2){\cal U}(z,{\bf u}_1,0;0,\nu){\cal V}(z',{\bf u}_2,0;0,\nu)<{\bf u}_1;\nu|{\bf u}_2;\nu>
$$
\b \label{tpfsp} = \int_{S^3}d\mu({\bf u}){\cal U}(z,{\bf u},0;0,\nu){\cal V}(z',{\bf u},0;0,\nu)=c_{0,\nu}\int_{S^3}d\mu({\bf u}) (z\cdot \xi_u)^{-\frac{3}{2}-i\nu}(z'\cdot \xi_u)^{-\frac{3}{2}+i\nu}.\e
This integral can be calculated in terms of a generalized Legendre function of the first kind \cite{brmo}
\b \label{stpfp} W(z,z') = 2\pi^2e^{\pi \nu} H^3 c_{0,\nu} P^{(5)}_{-\frac{3}{2}+i\nu}(H^2z.z') ,\e
where $c_{0,\nu}$ is fixed by imposing the local Hadamard condition \cite{brmo}:
\b \label{ncsv}
c_{0,\nu}=\frac{e^{-\pi
\nu}\Gamma(\frac{3}{2}+i\nu)\Gamma(\frac{3}{2}-i\nu)}{2^5\pi^4H}.\e

The phenomenon of non-uniqueness of vacuum state in a general curved space-time appears here in the normalization constant $c_{0,\nu}$. In the case of the dS space-time, it has been discovered that the Hadamard condition selects a unique vacuum state for the quantum field operators. The Hadamard condition postulates that the short distance behaviour of two-point function of the field operator on a curved space-time should be the same as of the Minkowskian counterparts. In the dS case the preferred vacuum state coincides with the so called Euclidean or Bunch-Davies vacuum state, and singles out one vacuum in the two-parameter family of
quantizations constructed by Allen \cite{al}.

The generalized Legendre function of the first kind $P^{(d+1)}_{\lambda}$ is defined by
the two following integral representations \cite{bagamota}
\begin{equation}
P^{(d+1)}_{\lambda}(\cos \theta)
=\frac{2\omega_{d-1}}{\omega_d}(\sin\theta)^{2-d}
\int_0^\theta\cos\left[ \left(\lambda +\frac{d-1}{2}\right)\tau\right] \Bigl[
2(\cos\tau-\cos\theta)\Bigr]^{\frac{d-3}{2}}d\tau,
\label{Legendre1}
\end{equation}
and
\begin{equation}
P^{(d+1)}_{\lambda}(z)=\frac{\Gamma(d/2)}{{\sqrt
\pi}\Gamma({\frac{d-1}{2}})} \int_{0}^{\pi}\left[z+(z^{2}-1)^{1/2}\cos
t\right]^{\lambda}(\sin t)^{d-2} dt, \label{Legendre2}
\end{equation}
which are valid on the cut complex plane ${\mathbb C} \setminus (-\infty,
-1]$ with $\omega_{d}=\frac{2\pi^{d/2}}{\Gamma({d/2})}$. These
functions can be written as follows: 
\b
P^{(d+1)}_{\lambda}(z)=
\frac{\Gamma(d-1)\Gamma(\lambda+1)}{\Gamma(\lambda+d-1)}\,C_{\lambda}^{\frac{d-1}{2}}(z)=F\left(\lambda+d-1,-\lambda;\,{d\over 2};\,{{1-z}\over
2}\right), \label{F}
\e
$F$ is the hyper-geometric function and $C_{\lambda}^{\kappa}(z)$ is Gegenbauer function of the first
kind:
\begin{equation}
C_{\lambda}^{\kappa}(z)=\frac{\Gamma(\lambda+2\kappa)}{\Gamma(\lambda+1)\Gamma(2\kappa)}
F\left(\lambda+2\kappa,-\lambda;\kappa+{1\over 2};{{1-z}\over
2}\right).
\end{equation}

\subsection{Massive spinor field}

A massive spinor field associates with the principal series representation with $j=\frac{1}{2}$ and $p=\frac{1}{2}+i\nu$, $\nu>0$. The eigenvalue of the Casimir operator and the corresponding mass parameter are  $<Q_{\frac{1}{2},\nu}^{(1)}>=\nu^2+\frac{3}{2}$ and $m_{f,\nu}^2=H^2(\nu^2+2\pm i\nu)$, respectively. 
The second-order field equation is:
$$  \left[ Q_{\frac{1}{2}}^{(1)}-<Q_{\frac{1}{2},\nu}^{(1)}>\right]\psi(x)=0,\;\; \mbox{or}\; \;\left[ Q_{0}^{(1)}+\not x \not \partial^\top-\frac{5}{2}-<Q_{\frac{1}{2},\nu}^{(1)}>\right]\psi(x)=0. $$ 
The action for this field equation can be defined as:
$$S[\psi, \tilde{\psi}]=\int d\mu(x){\cal L}(\psi,\nabla^\top_{\alpha}\psi;\tilde{\psi},\tilde{\nabla}_{\alpha}^\top\tilde{\psi})=\int d\mu(x)\left[\tilde{\nabla}_{\alpha}^\top\tilde{ \psi} \nabla^{\top\alpha}\psi-\left(<Q_{\frac{1}{2},\nu}^{(1)}>+\frac{5}{2}\right)\tilde{\psi}\psi\right],$$
where $\tilde{\psi}=\psi^\dag \gamma^0$ and the covariant derivatives for spinor fields are defined by:
 \b\label{codes12anbar} \nabla_{\alpha}^\top \psi=\left(\partial^\top_\alpha+\gamma^\top_\alpha \not x\right)\psi,\;\; \tilde{ \nabla}_{\alpha}^\top \tilde{\psi}=\partial^\top_\alpha \tilde{\psi}.\e The field equation is obtained from  the usual Euler-Lagrange equations: 
$$\left[\square_H+H^2\left(2+\nu^2\pm i\nu\right)\right]\psi(x)=0.$$
This field satisfies also the following dS-Dirac first-order field equation:
\b \label{dsdiracfe} \left(\not x \not \partial^\top-2 \pm i\nu\right)\psi(x)=0.\e
Using the following identity in dS ambient space formalism:
$$ \left(\not x \not \partial^\top-2+i\nu\right)\left(\not x \not \partial^\top-1-i\nu\right)= \square_H+2+\nu^2+i\nu,$$
one can write the spinor field in terms of a scalar field
\b \label{sminscm} \psi(x)= \left(\not x \not \partial^\top-1-i\nu\right){\cal U} \phi(x),\e
where ${\cal U}$ is a four-component constant spinor, which can be fixed in the null curvature limit, and $\phi$ satisfies: 
$$ \left[\square_H+H^2\left(2+\nu^2+ i\nu\right)\right]\phi(x)=0.$$
One can simply replace $\nu$ with $-\nu$ and obtain the other equation for spinor field. The charged spinor field operator is:
\b \label{spinqfop} 
\psi(x)=\int_{S^3}d\mu({\bf u})\sum_{m}\left[a({\bf \tilde{u}},m;\frac{1}{2},1-p)U(x;{\bf u},m;\frac{1}{2},\nu )+a^{c\dag}({\bf u},m;\frac{1}{2},p)V(x;{\bf u},m;\frac{1}{2},\nu )\right],\e
where $ m_{\frac{1}{2}}\equiv m=-\frac{1}{2},\frac{1}{2}$ and $\psi, \; U$ and $V$ are four-components spinors. Afterwards, the equations (\ref{vtfp0}) and (\ref{utfp0}) become:
$$ |{\bf c}{\bf u}+{\bf d}|^{3-2i\nu} \sum_{m}D^{(\frac{1}{2})}_{mm'}\left(\frac{({\bf c}{\bf u}+{\bf d})^{-1}}{|{\bf c}{\bf u}+{\bf d}|} \right)V(x;{\bf u},m;\frac{1}{2},\nu )\;\;\;\;\;\;\;\;\;\;\;\;\;\;\;\;\;\;\;\;\;\;
$$
\b \label{vtfp12} 
\;\;\;\;\;\;\;\;\;\;\;\;\;\;\;\;\;\;\;\;\;\;\;\;\;\;\;\;\;\;\;\;\;\;\;\;\;\;=g^{-1}V(\Lambda x; g^{-1}\cdot{\bf u},m';\frac{1}{2},\nu ),
\e
$$ |{\bf c}{\bf u}+{\bf d}|^{3+2i\nu} \sum_{m}\left[D^{(\frac{1}{2})}_{mm'}\left(\frac{({\bf c}{\bf u}+{\bf d})^{-1}}{|{\bf c}{\bf u}+{\bf d}|} \right)\right]^*U(x;{\bf u},m;\frac{1}{2},\nu )\;\;\;\;\;\;\;\;\;\;\;\;\;\;\;\;\;\;\;\;\;\;\;
$$
\b \label{utfp12} 
\;\;\;\;\;\;\;\;\;\;\;\;\;\;\;\;\;\;\;\;\;\;\;\;\;\;\;\;\;\;\;\;\;\;\;\;\;\;=g^{-1}U(\Lambda x; g^{-1}\cdot{\bf u},m';\frac{1}{2},\nu ).
\e
There are two possibilities for the homogeneous degrees of spinor field: $\lambda^+= -2-i\nu, \; -1+i\nu$ and $\lambda^-= -2+i\nu, \; -1-i\nu$ (\ref{hdmjp}). Similar to the case of the scalar fields, the above equations, (\ref{vtfp12}) and (\ref{utfp12}), can completely determine the homogeneity degree of the spinor fields and yield to the following relations for the spinors  $U$ and $V$:
$$ U(x;{\bf u},m;\frac{1}{2},\nu )=(x.\xi_u)^{-2-i\nu}u(x,{\bf u},m;\frac{1}{2},\nu),$$ $$  V(x;{\bf u},m;\frac{1}{2},\nu )=(x.\xi_u)^{-1+i\nu}v(x,{\bf u},m;\frac{1}{2},\nu) . $$ 
Therefore, only one of these two possibilities ($\lambda^+$ and $\lambda^-$), can transform according to the UIR $U^{(\frac{1}{2},p)}(g)$ of the dS group. The acceptable homogeneous degree is $\lambda^+$ and consequently, one of the second order field equations and mass parameters are also suitable in this case. For detailed information about the explicit forms of $U$ and $V$  spinors, the reader may refer to the previously published paper \cite{bagamota}. Following our survey, the equation (\ref{fioptrs}) becomes:
$$ U^{(\frac{1}{2},p)}(g) \psi(x) \left[ U^{(\frac{1}{2},p)}(g)\right]^\dag=g^{-1}\psi(\Lambda x).$$ 
In this formalism, the charge conjugation spinor is defined as \cite{morrota}:
$$ \psi^c=\eta_c C\left(\gamma^4\right)^t\left(\bar \psi\right)^t,$$ where $\eta_c$ is an arbitrary unobservable phase value, generally chosen to be a unity, and $C=\gamma^2\gamma^4$ in the $\gamma$ representation (\ref{gammam}). 

The two-point function for the spinor field has been calculated in maximally symmetric space in \cite{allu}, in addition, the analytic two-point function in the dS ambient space formalism has been also calculated in \cite{bagamota}:
$$ S_{i_1i_2}(z_1,z_2)=\left<\Omega\right|\psi_{i_1}(z_1)\bar \psi_{i_2}(z_2)\left|\Omega\right> $$ \b \label{tpffp} =\int_{S^3} d\mu({\bf u}) (z_1.\xi_u)^{-2-i\nu}(z_2.\xi_u)^{-2+i\nu}\sum_{m} u_{i_1}(z_1,{\bf u},m;\frac{1}{2},\nu)\bar u_{i_2}(z_2,{\bf u},m;\frac{1}{2},\nu). \e
Therefore, the "permuted" two-point function $S'$ is given by: 
$$ S'_{i_1i_2}(z_1,z_2)=-\left<\Omega\right|\bar \psi_{i_2}(z_2)\psi_{i_1}(z_1)\left|\Omega\right> $$ \b \label{tpffp2} =\int_{S^3} d\mu({\bf u}) (z_1.\xi_u)^{-1-i\nu}(z_2\cdot\xi_u)^{-1+i\nu}\sum_{m}\bar v_{i_2}(z_2,{\bf u},m;\frac{1}{2},\nu) v_{i_1}(z_1,{\bf u},m;\frac{1}{2},\nu), \e
where \citep{ta96,ta97,bagamota}
$$ \sum_{m} u_i(z_1,{\bf u},m;\frac{1}{2},\nu)\bar u_j(z_2,{\bf u},m;\frac{1}{2},\nu)=\frac{c_{\frac{1}{2},\nu}}{2}\left(\not \xi_u \gamma^4\right)_{ij},$$
$$\sum_{m}\bar v_{j}(z_2,{\bf u},m;\frac{1}{2},\nu) v_{i}(z_1,{\bf u},m;\frac{1}{2},\nu)= -\frac{c_{\frac{1}{2},\nu}}{2} \left(\frac{\not z_1\not \xi_u }{z_1.\xi_u}\frac{\not z_2\gamma^4}{x_2\cdot \xi_u}\right)_{ij}.$$ $c_{\frac{1}{2},\nu}$ is the normalization constant. In terms of the generalized Legendre function of the first kind, the analytic spinor two-point function can be written as \cite{bagamota}:
\b \label{tpfsm} S(z_1,z_2)=A_\nu\left[ (2-i\nu) \nu P^{(7)}_{-2-i\nu}(H^2z_1\cdot z_2)\not z_1-(2+i\nu) P^{(7)}_{-2+i\nu}(H^2z_1\cdot z_2)\not z_2\right]\gamma^4, \e $A_\nu=2i\pi^2e^{\pi \nu} c_{\frac{1}{2},\nu}$. The values of the constants $c_{\frac{1}{2},\nu}$ and $A_\nu$ are determined by imposing the local Hadamard condition: the short-distance behaviour of $S(z_1,z_2)$ should coincide with the leading behaviour of the corresponding Minkowskian two-point function \cite{bagamota}:
$$c_{\frac{1}{2},\nu} =\frac{\nu(\nu^2+1)}{(2\pi)^3(e^{2\pi \nu}-1)}    , \;\;\;\; A_\nu=\frac{i}{64\pi}\frac{\nu(i+\nu^2)}{\sinh(\pi\nu)}.$$
The 'permuted' two-point function $S'$ is \cite{bagamota}:
$$S'(z_1,z_2)=-\not z_1 S(z_1,z_2)\gamma^4 \not z_2 \gamma^4.$$

The mass square associated with the spinor field is a complex number. Bearing in mind that the mass is not a well-defined physical quantity in the dS space-time and it is only a parameter, one finds out that this parameter actually has a relation with the homogeneous degree of the spinor field. In contrary, one obtains a real mass in the null curvature limit ($H=0$). A massive spinor field, which is a causal field, propagates in the dS light cone. A spinor field in the dS space in the null curvature limit becomes exactly equivalent to its Minkowskian counterpart \cite{bagamota}. The homogeneous degree of tensor or spinor field play the role of the mass for causal propagation in the dS space and it must to satisfy $\Re \lambda < 0$.

\subsection{Massive vector field}

A massive vector field associates with the principal series representation of the dS group with $j=1$ and $p=\frac{1}{2}+i\nu$, $\nu \geq 0$. The eigenvalue of the Casimir operator and the corresponding mass parameter are $<Q_{1,\nu}^{(1)}>=\nu^2+\frac{1}{4},$ and $m_{b,\nu}^2=H^2(\nu^2+\frac{9}{4})$, respectively. The field equation is:
$$ \left[ Q_1^{(1)}-\left(\nu^2+\frac{1}{4}\right)\right]K_\alpha(x)=0, \;\; \mbox{or}\;\;\left[ \square_H
+H^2\left(\nu^2+\frac{9}{4}\right)\right]K_\alpha(x)=0. $$
Lagrangian for this field equation can be defined as:
$$ S[K]=\int d\mu(x) \left[ \frac{1}{4}F_{\alpha\beta}F^{\alpha\beta}-\frac{1}{2}<Q_{1,\nu}^{(1)}>K_\alpha K^\alpha  \right],$$ where $ F_{\alpha\beta}=\nabla^\top_\alpha K_\beta-\nabla^\top_\beta K_\alpha$ and $\partial \cdot K=0$. Transverse-covariant derivative which acts on tensor field is \cite{tata}:
\b \label{dscdrt}
\nabla^\top_\beta T_{\alpha_1....\alpha_l}\equiv 
\partial^\top_\beta
T_{\alpha_1....\alpha_l}-H^2\sum_{n=1}^{l}x_{\alpha_n}T_{\alpha_1..\alpha_{n-1}\beta\alpha_{n+1}..\alpha_l}.\e

The massive vector field can be written in terms of the massive scalar field (\ref{scaqfop}) \cite{gata}:
 $$ K_{\alpha}(x)=\left[ Z^\top_{\alpha}+\frac{1}{\nu^2+\frac{1}{4}}
D_{1\alpha}\left(Z\cdot \partial^\top+2H^2x\cdot Z\right)\right]\phi(x).$$
$Z$ is a constant five-vector, which defines the polarization states and $D_1=\partial^\top$. The massive vector field operator can be written in the following form:
\b \label{vectqfop} K_\alpha(x)=\int_{S^3}d\mu({\bf u})\sum_{m} \left[a({\bf \tilde{u}},m;1,1-p){\cal U}_\alpha(x;{\bf u},m;1,p )+a^{\dag}({\bf u},m;1,\nu){\cal V}_\alpha(x;{\bf u},m;1,\nu )\right],\e
with $ m_1=m=-1,0,1$. The equations (\ref{vtbp}) and (\ref{utbp}) can indeed fix the homogeneous degrees of the vector field and one obtains:
$$ {\cal U}_\alpha(x;{\bf u},m;1,\nu )=(x\cdot\xi_u)^{-\frac{3}{2}-i\nu}u_\alpha(x,{\bf u},m;1,\nu),$$ $$  {\cal V}_\alpha(x;{\bf u},m;1,\nu )=(x\cdot\xi_u)^{-\frac{3}{2} +i\nu}v_\alpha(x,{\bf u},m;1,\nu) . $$ 
In the previous paper \cite{gata}, we have calculated these two functions by solving the field equation and considering the conditions (i-vii) of section II-D. Their explicit forms also can be calculated, using the equations (\ref{vtbp}) and (\ref{utbp}). The massive vector field operator satisfies the following relation:
$$ U^{(1,p)}(g) K_\alpha(x) \left[ U^{(1,p)}(g)\right]^\dag=\Lambda^{\;\;\alpha'}_{\alpha}K_{\alpha'}(\Lambda x).$$ 

The analytic two-point function for the massive vector field is defined by:
$$ W_{\alpha\alpha'}(z,z')=<\Omega|K_{\alpha}(z)K_{\alpha'}(z')|\Omega>$$ \b \label{tpfvm} =\int_{S^3}(z\cdot\xi_u)^{-\frac{3}{2}-i\nu}(z'\cdot\xi_u)^{-\frac{3}{2}+i\nu}\sum_m u_{\alpha}(z,{\bf u},m;1,\nu)v_{\alpha'}(z',{\bf u},m;1,\nu), \e
where we have \cite{gagata}:
$$ \sum_m u_{\alpha}(z,{\bf u},m;1,\nu)v_{\alpha'}(z',{\bf u},m;1,\nu)=$$
\b \label{eqprop1}
c_{1,\nu}\left[- \theta_\alpha\cdot \theta'_{\alpha'}+\frac{(\theta_{\alpha}\cdot
z')\bar\xi'_{\alpha'}}{z'\cdot\xi}+\frac{(\theta'_{\alpha'}\cdot
z)\bar\xi_{\alpha}}{z\cdot\xi}-\frac{(z\cdot
z') \bar\xi_{\alpha} \bar\xi'_{\alpha'}}{H^{2}(z\cdot\xi)
(z'\cdot\xi)}\right].
\e
$c_{1,\nu}$ is the normalization constant and $\theta'_{\alpha\beta}=\eta_{\alpha\beta}+H^2z'_\alpha z'_\beta$. In this formalism, the analytic two-point function can be easily written in terms of the scalar analytic two-point function \cite{gata}: \b \label{tpmsve}  W_{\alpha\alpha'}(z,z')=\left[\theta_{\alpha}\cdot\theta'_{\alpha'}+\frac{1}{H^2(\nu^2+\frac{1}{4})}
\partial_{\alpha}^\top\left(\theta'_{\alpha'}\cdot\partial^\top+2H^2\theta'_{\alpha'}\cdot
z\right) \right]W(z,z'),\e where $W(z,z')$ is given by (\ref{stpfp}).

For the complementary series
($<Q_{1,p}^{(1)}>=p-p^2,\;\;0<p-p^2 <\frac{1}{4}$) and the discrete series
($<Q_{1,1}^{(1)}>=0,\;\; j=p=1$), one can replace $\nu$ respectively
with $\pm \sqrt{p-p^2-\frac{1}{4}}$ and $\pm \frac{i}{2}$. In the case of the
complementary series, the associated mass is 
$m_{b,p}^2=H^2(p-p^2)$, but there is no physically meaningful representation belonging to the Poincar\'e group that can be interpreted as a contraction limit $H \longrightarrow 0$ for these representations. Therefore, the physical meaning of a complementary vector field is not clear. 

The associated mass for the vector field ($j=1$) which corresponds to the discrete series representations $\Pi^\pm_{1,1}$, is $m_{b,1}^2=2H^2 $. These representations ($\Pi^\pm_{1,1}$) have the physically meaningful Poincar\'e limit. It is a massless vector field so the parameter $\nu$ should be replaced by $\pm \frac{i}{2}$ in the equations (\ref{vectqfop}) and (\ref{tpmsve}). Note that the generalized
polarization vector $u_\alpha(x,{\bf u},m;1,\nu)$ or the related two-point function (\ref{tpmsve}) diverges at this limit. This type of singularity is actually due to the divergencelessness condition. This condition is imposed on the massless vector field in order to associate this type of field with the UIR's $\Pi^\pm_{1,1}$. To overcome this problem, the divergencelessness condition should be dropped out. Then, the massless vector field associates with an indecomposable
representation of the dS group and hence a gauge invariance appears. This field will be discussed in sections VII. 

\subsection{Massive spin-$\frac{3}{2}$ field}

A massive vector-spinor field is associated with the principal series representation of the dS group with $j=\frac{3}{2}$ and $p=\frac{1}{2}+i\nu$, $\nu>0$. The corresponding mass parameter and the eigenvalue of the Casimnir operator are $m_{f,\nu}^2=H^2(\nu^2+2\pm i\nu)$ and 
$<Q_{\frac{3}{2},\nu}^{(1)}>=\nu^2-\frac{3}{2}$, respectively. The second-order field equation is:
$$  \left[ Q_{\frac{3}{2}}^{(1)}-<Q_{\frac{3}{2},\nu}^{(1)}>\right]\Psi_\alpha(x)=0,\;\; \mbox{or}\; \;\left[ Q_{0}^{(1)}+\not x \not \partial^\top-\frac{5}{2}-<Q_{\frac{3}{2},\nu}^{(1)}>\right]\Psi_\alpha(x)=0. $$ 
The action for this field equation can be defined as:
$$S[\Psi, \tilde{\Psi}]=\int d\mu(x)\left[\frac{1}{2}\left(\tilde{\nabla}^\top_\alpha\tilde{ \Psi}_\beta-\tilde{\nabla}^\top_\beta\tilde{ \Psi}_\alpha\right)\left( \nabla^{\top\alpha} \Psi^\beta-\nabla^{\top \beta} \Psi^\alpha\right)-\left(<Q_{\frac{3}{2},\nu}^{(1)}>+\frac{5}{2}\right)\tilde{\Psi}_\alpha\Psi^\alpha\right],$$
where $\tilde{\Psi}_\alpha=\Psi^\dag_\alpha \gamma^0$ and $\partial\cdot \Psi=0=\gamma \cdot \Psi$. The transverse-covariant derivative which acts on tensor-spinor field is defined by the following relation
\b \label{cdsa} \nabla^\top_\beta \Psi_{\alpha_1....\alpha_l}\equiv 
\left(\partial_\beta^\top+\gamma^\top_\beta\not x\right)
\Psi_{\alpha_1....\alpha_l}-H^2\sum_{n=1}^{l}x_{\alpha_n}\Psi_{\alpha_1..\alpha_{n-1}\beta\alpha_{n+1}..\alpha_l}, \e 
\b \label{cdsa2} \tilde{\nabla}^\top_\beta\tilde{\Psi}_{\alpha_1....\alpha_l}\equiv 
\partial_\beta^\top
\tilde{\Psi}_{\alpha_1....\alpha_l}-H^2\sum_{n=1}^{l}x_{\alpha_n}\tilde{\Psi}_{\alpha_1..\alpha_{n-1}\beta\alpha_{n+1}..\alpha_l}.\e 

The first-order and second-order field equations are:
$$ \left(\not x \not \partial^\top-2 \pm i\nu\right)\Psi_\alpha(x)=0,\;\;\;\left[\square_H+H^2\left(2+\nu^2\pm\nu\right)\right]\Psi_\alpha(x)=0.$$
The spin-$\frac{3}{2}$ field can be written in terms of the spinor field (\ref{spinqfop}) \cite{taazba}:
$$ \Psi_\alpha(x)=\left[Z^\top_\alpha-\frac{1}{4}\gamma^\top_\alpha\not Z^\top-\frac{1}{\nu^2+1}\left( -\frac{1}{4}(1-i\nu)\gamma^\top_\alpha\not Z^\top+2\partial^\top _\alpha x.Z \right.\right.$$ $$\left. +\frac{2}{3}\partial^\top _\alpha Z\cdot \partial^\top +\frac{1}{3}(i\nu+1)\partial^\top _\alpha \not Z^\top \not x-\frac{1}{3}(i\nu+1)\gamma^\top_\alpha Z\cdot x \not x-\frac{2}{3}\gamma^\top_\alpha Z\cdot x \not \partial^\top  \right.  $$ $$ \left.\left.   -\frac{1}{6}\gamma_\alpha^\top Z\cdot \partial^\top  \not \partial^\top -\frac{1}{6}i\nu\gamma^\top_\alpha \not x Z\cdot \partial^\top-\frac{1}{12}(i\nu+1)\gamma^\top_\alpha\not Z^\top \not x \not \partial^\top\right)\right]\psi(x),$$ 
where $Z$ is a constant five-vector. Vector-spinor field may also be given in terms of a scalar field by using the equation (\ref{sminscm}). The charged vector-spinor field operator can be defined as:
\b \label{spinvqfop1} \Psi_\alpha(x)=\int_{S^3}d\mu({\bf u})\sum_{m}\left[a({\bf \tilde{u}},m;\frac{3}{2},1-p)U_\alpha(x;{\bf u},m;\frac{3}{2},\nu )+a^{c\dag}({\bf u},m;\frac{3}{2},p)V_\alpha(x;{\bf u},m;\frac{3}{2},\nu )\right],\e
with $m=-\frac{3}{2},-\frac{1}{2},\frac{1}{2}, \frac{3}{2}$ and $\Psi_\alpha, U_\alpha$ and $V_\alpha$ are four-components spinors. Similar to the case of the spinor field, only the homogeneity degree $\lambda^+=-2-i\nu, \;-1+i\nu$ is allowed:
$$ U_\alpha(x;{\bf u},m;\frac{3}{2},\nu )=(x\cdot\xi_u)^{-2-i\nu}u_\alpha(x, {\bf u},m;\frac{3}{2},\nu),$$ $$  V_\alpha(x;{\bf u},m;\frac{3}{2},\nu )=(x\cdot\xi_u)^{-1+i\nu}v_\alpha(x,{\bf u},m;\frac{3}{2},\nu) . $$
Following our previous discussions, we are not interested to the explicit forms of $u_\alpha$ and $v_\alpha$, because they are already established in \cite{gata}. It is easy to show that the vector-spinor field operator (\ref{spinvqfop1}) satisfies the following transformation law: 
$$ U^{(\frac{3}{2},p)}(g) \Psi_\alpha(x) \left[ U^{(\frac{3}{2},p)}(g)\right]^\dag=\Lambda^{\;\;\alpha'}_{\alpha}g^{-1}\Psi_{\alpha'}(\Lambda x).$$ 
The adjoint vector-spinor field is defined by:
$$\bar\Psi_\alpha(x)\equiv \Psi_\alpha^\dag(x) \gamma^0\gamma^4,$$ which transforms as:
$$ U^{(\frac{3}{2},p)}(g)\bar \Psi_\alpha(x) \left[ U^{(\frac{3}{2},p)}(g)\right]^\dag=\Lambda^{\;\;\alpha'}_{\alpha} \bar \Psi_{\alpha'}(\Lambda x)\left(-\gamma^4 g\gamma^4\right).$$

The analytic two-point function for the massive vector-spinor field is:
$$ S^{ii'}_{\alpha\alpha'}(z,z')=\left<\Omega\right|\Psi_{\alpha}^{i}(z)\bar \Psi_{\alpha'}^{i'}(z')\left|\Omega\right> $$ \b \label{tpfvfp} =\int_{S^3} d\mu({\bf u}) (z\cdot\xi_u)^{-2-i\nu}(z'\cdot\xi_u)^{-2+i\nu}\sum_{m} u_{\alpha}^{i}(z,{\bf u},m;\frac{3}{2},\nu)\bar u_{\alpha'}^{i'}(z',{\bf u},m;\frac{3}{2},\nu). \e 
Note that the equation (\ref{tpfvfp}) can be written in terms of the analytic spinor two-point function as \cite{taazba}:
\begin{equation} \label{5.6} S_{\alpha \alpha'}(z,z')=
D_{\alpha \alpha'}(z, \partial^\top ;z', {\partial'}^\top)
S(z,z'),\end{equation} where
$$ D_{\alpha \alpha'}=\theta_\alpha \cdot \theta'_{\alpha'}-\frac{1}{4}\gamma^\top_\alpha\gamma^\top\cdot \theta'_{\alpha'}-\frac{1}{\nu^2+1}\left(\frac{(i\nu-1)}{4}\gamma^\top_\alpha\gamma^\top\cdot \theta'_{\alpha'}+2\partial^\top_\alpha z\cdot\theta'_{\alpha'}\right.$$ $$\left. +\frac{2}{3}\partial^\top_\alpha\theta'_{\alpha'}\cdot\partial^\top -\frac{2}{3}\gamma^\top_\alpha \theta'_{\alpha'} \cdot z \not \partial^\top+ \frac{(i\nu+1)}{3}\partial^\top_\alpha\gamma^\top\cdot\theta'_{\alpha'}\not z-\frac{(i\nu+1)}{3}\gamma^\top_\alpha \theta'_{\alpha'}\cdot z \not z \right.  $$ \b\left. -\frac{1}{6}\gamma^\top_\alpha\theta'_{\alpha'} \cdot\partial^\top \not\partial^\top-\frac{1}{6}i\nu\gamma^\top_\alpha \not z \theta'_{\alpha'} \cdot\partial^\top -\frac{(i\nu+1)}{12}\gamma^\top_\alpha\gamma^\top\cdot\theta'_{\alpha'} \not z \not \partial^\top\right) ,\e
with $S(z,z')$ as the analytic spinor two-point function (\ref{tpfsm}).

There are two vector-spinor field ($j=\frac{3}{2}$) in the discrete series representations. They are correspond to $p=\frac{1}{2}$ and $ p=\frac{3}{2}$ and their corresponding eigenvalues of the Casimir operators are:
$$<Q^{(1)}_{\frac{3}{2},\frac{1}{2}}>=-\frac{3}{2} ,\;\;\;\; <Q^{(1)}_{\frac{3}{2},\frac{3}{2}}>=-\frac{5}{2}.$$ 
The case $p=\frac{1}{2}$ which corresponds to the representations
$\Pi^{\pm}_{\frac{3}{2},\frac{1}{2}}$, does not have a corresponding representation in Poincar\'e group at the null curvature limit. Therefore, the corresponding field is called an auxiliary vector-spinor field. Then one can replace $\nu$ by $\nu=0$ in the equation (\ref{spinvqfop1}) and (\ref{5.6}) for the massive vector-spinor field in this case. This field will be considered in section V-A. 

The second case ($p=\frac{3}{2}$) which corresponds to the representations
$\Pi^{\pm}_{\frac{3}{2},\frac{3}{2}}$, has a physically meaningful
Poincar\'e limit. This is precisely the massless vector-spinor
field and $\nu$ should be replaced by $\pm i$ in the equation (\ref{spinvqfop1}) and (\ref{5.6}) for the massive vector-spinor field. Nevertheless, the field operator and also the two-point function become divergent at this limit ($\nu=\pm i$). This type of singularity is actually due to the presence of the imposed auxiliary conditions $\partial \cdot \Psi=0$ and $ \gamma \cdot\Psi=0$, in order to associate the massless vector-spinor field with the UIR's $\Pi^{\pm}_{\frac{3}{2},\frac{3}{2}}$. To solve this problem, the auxiliary conditions should be dropped out. Then, the massless vector-spinor field associates with an indecomposable
representation of the dS group and hence a gauge invariance appears. This field will be considered in section VII.

\subsection{Massive spin-$2$ rank-2 symmetric tensor field}

A massive spin-$2$ rank-$2$ symmetric tensor field associates with the principal series representation of the dS group with $j=2$ and $p=\frac{1}{2}+i\nu$, $\nu \geq 0$. The eigenvalue of the Casimir operator and its corresponding mass parameter are $<Q_{2,\nu}^{(1)}>=\nu^2-\frac{15}{4}$ and
$m_{b,\nu}^2=H^2(\nu^2+\frac{9}{4})$, respectively. The field equation is:
$$ \left[ Q_2^{(1)}-\left(\nu^2-\frac{15}{4}\right)\right]{\cal K}_{\alpha\beta}(x)=0, \;\; \mbox{or}\;\;\left[ \square_H
+H^2\left(\nu^2+\frac{9}{4}\right)\right]{\cal K}_{\alpha\beta}(x)=0. $$
With the conditions $x \cdot {\cal K}=0=\partial \cdot {\cal K},\;{\cal K}_\alpha^{\;\;\alpha}=0$ and ${\cal K}_{\alpha\beta}={\cal K}_{\beta\alpha}$, the action of this field equation can be defined as:
\b \label{acsp2ran}  S[{\cal K}]=\int d\mu(x)\left[ \frac{1}{8}\left(\nabla^\top_\alpha {\cal K}_{\beta\gamma}-\nabla^\top_{(\beta} {\cal K}_{\gamma)\alpha}\right) \left(\nabla^{\top\alpha} {\cal K}^{\beta\gamma} -\nabla^{\top(\beta} {\cal K}^{\gamma)\alpha}\right)-\frac{1}{2}\left(<Q_{2,\nu}^{(1)}>-4\right) {\cal K}_{\alpha\gamma}{\cal K}^{\alpha\gamma}\right],\e
where $A_{(\alpha\beta)}=A_{\alpha\beta}+A_{\beta\alpha}$. The massive spin-$2$ field can be written in terms of a polarization tensor and a massive scalar field (\ref{sminscm}):
$$ {\cal K}_{\alpha\beta}(x)={\cal D}_{\alpha\beta}(x,\partial;Z_1,Z_2,\nu)\phi(x).$$
$Z_1$ and $Z_2$ are two constant five-vectors and the explicit form of ${\cal D}$ was given in \cite{gagarota}. The field operator is define by:
\b \label{ms2r2s} {\cal K}_{\alpha\beta}(x)=\sum_{m}\int_{S^3}d\mu({\bf u})\left[a({\bf \tilde{u}},m;2,1-p){\cal U}_{\alpha\beta}(x;{\bf u},m;2,\nu )+a^{\dag}({\bf u},m;2,p){\cal V}_{\alpha\beta}(x;{\bf u},m;2,\nu )\right],\e
with $ m=-2,-1,0,1,2$. The homogeneous degrees of ${\cal U}$ and ${\cal V}$ are 
$$ {\cal U}_{\alpha\beta}(x;{\bf u},m;2,\nu )=(x\cdot\xi_u)^{-\frac{3}{2}-i\nu} u_{\alpha\beta}(x,{\bf u},m;2,\nu),$$ 
$$ {\cal V}_{\alpha\beta}(x;{\bf u},m;2,\nu )=(x\cdot\xi_u)^{-\frac{3}{2}+i\nu}v_{\alpha\beta}(x,{\bf u},m;2,\nu).$$ 
The coefficients $u$ and $v$ satisfy the conditions: 
$$ v_{\alpha\beta}(x,{\bf u},m;2,\nu)= v_{\beta \alpha}(x,{\bf u},m;2,\nu),\; \;\; v_{\alpha}^{\;\;\alpha}(x,{\bf u},m;2,\nu)=0.$$
Again, the explicit forms of $u_{\alpha\beta}$ and $v_{\alpha\beta}$ are given in \cite{gagata}. They can also be calculated by using the equations (\ref{vtbp}) and (\ref{utbp}). The field operator satisfies the following transformation rule:
$$ U^{(2,p)}(g) {\cal K}_{\alpha\beta}(x) \left[ U^{(2,p)}(g)\right]^\dag=\Lambda^{\;\;\alpha'}_{\alpha}\Lambda^{\;\;\beta'}_{\beta}{\cal K}_{\alpha'\beta'}(\Lambda x).$$ 

The analytic two-point function for this field is
$$ W_{\alpha\beta\alpha'\beta'}(z,z')=<\Omega|{\cal K}_{\alpha\beta}(z){\cal K}_{\alpha'\beta'}(z')|\Omega>$$ \b \label{tpftm1} =\int_{S^3}(z\cdot\xi_u)^{-\frac{3}{2}-i\nu}(z'\cdot\xi_u)^{-\frac{3}{2}+i\nu}\sum_m u_{\alpha\beta}(z,{\bf u},m;2,\nu)v_{\alpha'\beta'}(z',{\bf u},m;2,\nu), \e
which for the current formalism can be written in terms of the analytic scalar two-point function \cite{gagata}: \b \label{tpftm2} W_{\alpha\beta\alpha'\beta'}(z,z')=D_{\alpha\beta\alpha'\beta'}(z,z',\partial,\partial')  W(z,z').\e
$W(z,z')$ being the analytic scalar two-point function (\ref{stpfp}). The polarization tensor is:
$$ D_{\alpha\beta\alpha'\beta'}(z,z',\partial,\partial')=\frac{4}{3}\frac{ \nu^2+\frac{9}{4} }{
\nu^2+\frac{1}{4}}\left[
\theta-\frac{H^{2}D_{2}^\top D_{1}^\top}{2\lambda^{2}}\right]\left[
\theta'-\frac{H^{2}D'^\top_{2}D'^\top_{1}}{2\lambda^{*2}}\right]
$$
$$   + \frac{\nu^2+\frac{25}{4}}{\nu^2+\frac{9}{4} 
}\left[-\,\Sigma_1\Sigma_1' \theta\cdot\theta'+\frac{H^{2}\Sigma_1(\theta\cdot z')D'^\top_{2}}{\lambda^{*}-1}+\frac{H^{2}\Sigma_1'
(\theta'\cdot z)D^\top_{2}}{\lambda-1}+\frac{z H^{2}D^\top_{2}D'^\top_{2}}{
(\lambda-1)(\lambda^{*}-1)}\right]\times
$$
 $$ \left[\frac{ \nu^2+\frac{9}{4}}{
\nu^2+\frac{1}{4}}\left(-\theta_{\alpha}\cdot\theta'_{\alpha'}+
\frac{H^{2}\lambda(\theta\cdot z' )D'^\top_{1}}{ \nu^2+\frac{9}{4} }+
\frac{H^{2}\lambda^{*}(\theta'\cdot z) D^\top_{1}}{ \nu^2+\frac{9}{4}
}+ \frac{ H^2 z D^\top_{1}D'^\top_{1}}{ \nu^2+\frac{9}{4}}\right)\right],$$
where $\lambda=-\frac{3}{2}+i\nu$, $D^\top_1=\partial^\top$, $D^\top_{2}=\Sigma_1(\partial^\top-x)$ and $\Sigma_1$ is the index symmetrizer (\ref{sigmap}).

For a spin-$2$ rank-$2$ symmetric tensor
field, which associate with the complementary
series representation with $j=2$ and $0<p-p^2<\frac{1}{4}$, the eigenvalue of the Casimir operator and its corresponded mass parameter are $<Q_{2,p}^{(1)}>=p-p^2-4$ and $m_{b,p}^2=H^2(p-p^2+2)$, respectively. Although the
associated mass is strictly positive, the physical meaning of its associated field remains unclear. These representations in the complementary series at the $H=0$ limit, do not correspond to any physical representations of the Poincar\'e group. The field operator and the corresponded two-point function are obtainable simply by replacing the parameter $\nu$ with $\nu=i(p-\frac{1}{2})$.

The spin-2 fields ($j=2$) are corresponded to discrete series representation with $p=1$ ( $\Pi^\pm_{2,1}$) or $p= 2$ ( $\Pi^\pm_{2,2}$). The corresponding eigenvalues of their Casimir operators are respectively:
$$<Q^{(1)}_{2,1}>=-4 ,\;\;\;\; <Q^{(1)}_{2,2}>=-6.$$ 
The representations  $\Pi^\pm_{2,1}$  at the null curvature limit $H=0$, do not correspond to any physical representations of the Poincar\'e group.
The corresponded mass is $m_{b,1}^2=2H^2$. We call them the auxiliary spin-2 fields and they will be considered in section V-A. 

The second case ($p=2$) which corresponds to the representation
$\Pi^{\pm}_{2,2}$, has a physical meaning in the
Poincar\'e limit. Its associated mass being $m_{b,2}^2=0$. Hence $\Pi^\pm_{2,2}$ corresponds to the massless spin-$2$ rank-$2$ symmetric tensor fields (linear quantum gravity in the dS space). In this case, $\nu $ should
be replaced by $\pm \frac{3i}{2}$ in the field operators and then in their corresponded two-point functions. The projection operator $D_{\alpha\beta\alpha'\beta'}$ on the classical level (\ref{tpftm2}) and the normalization
constant $c_{0,\nu}$ on the quantum level (\ref{ncsv}), become singular in this limit. The first singularity is actually due to the divergencelessness condition in which necessarily associates the rank-$2$ symmetric tensor field with the discrete series representation $\Pi^{\pm}_{2,2}$. In order to remove this singularity, the
divergencelessness condition should be dropped out. Then the field
equation becomes gauge invariant, and the field operator transforms under  an indecomposable representation
of the dS group. By fixing the gauge, the field can eventually be
quantized where unfortunately, a second type of singularity appears, which is in fact due to the so-called zero mode
problem of the Laplace-Beltrami operator of dS space-time inherited
from the minimally coupled scalar field \cite{gareta}.
Nevertheless, using the Krein space quantization \cite{gareta}, this singularity can be successfully removed and one can defined the quantum  
massless spin-$2$ rank-$2$ symmetric tensor fields in the dS space-time
\cite{ta09,taro12} but the theory is not analytic. In section VII, another method for solving this problem (analiticity) is presented. In this method the tensor field can be written in terms of a polarization tensor and a conformally coupled scalar field. Then we present a linear quantum gravity based on a rank-$2$ symmetric tensor field which are analytic and free of any infra-red divergence. But this tensor field also breaks the conformal invariance and then it is not a real massless spin-$2$ field in the dS space-time. For solving the problem of conformal invariance, one is obligated to use a rank-$3$ mixed-symmetric tensor field for massless spin-$2$ field which will be considered in section VII.


\setcounter{equation}{0}
\section{Quantum field operators for discrete series}

The construction of quantum field operators in terms of creation and annihilation operators for principal series can be directly generalized to the discrete series representation with $j \neq p$. For the values $j=p\geq 1$, the quantum state $|{\bf q}; j,j>$ cannot be defined uniquely by differential equation (\ref{difequ}), since a constant vector in $V^j$ is an arbitrary solution in this case:
$$ \left|{\bf q};j,j\right> \Longrightarrow \left|{\bf q};j,j\right>_g= \left|{\bf q};j,j\right>+\left|{\bf q}_0;j,j\right>,$$
where ${\bf q}_0$ is a constant. Because of this arbitrariness, one cannot construct the quantum field operators on creation and annihilation operators on this state. Neglecting this non-uniqueness, at least one of the functions ${\cal U}_l$ or ${\cal V}_l$ blows up for this values, in addition the two-point function becomes singular in the limit of $j=p\geq 1$. 
A massless field on the dS space associates with the value $j=p$  and in the null curvature limit corresponds to a massless field of the Poincar\'e group.
In fact, creation and annihilation operators for discrete series with $j=p \geq 1$ cannot be used to construct all of the irreducible quantum fields.
Therefore, one may remove some of the physical conditions (iv-vii) of the section II-D and consequently, the quantum field transforms with an indecomposable representation of the dS group. Applying this peculiar limitation on different types of the fields naturally leads to the introduction of gauge invariance or gauge principle. 

In the present section, the cases with $j\neq p$, which called auxiliary fields, are considered and afterwards, two auxiliary quantum fields characterized with the sets ($j=\frac{3}{2},\;\; p=\frac{1}{2}$) and ($j=2,\;\;p=1$) are constructed. They appear in the indecomposable representations of the massless fields \cite{fatata,azam,petata}. Next, the important case of  $j=p \geq 1$ and the appearance of the gauge invariance will be discussed.

\subsection{The case $j\neq p$}

In order to construct the quantum field operators, one first define creation and annihilation operators on the proposed Hilbert spaces ${\cal H}^{(0,j,p)}_{q}$ and ${\cal H}^{(j,0,p)}_{q}$. For discrete series, similar to the principal series, creation operator $a^\dag({\bf q},m_j;j,p)$ is defined as the operator that simply adds a state with quantum numbers $({\bf q},m_j;j,p)$ 
\b \label{cractionp2} a^\dag ({\bf q},m_j;j,p) \left| \Omega \right> \equiv\left|{\bf q},m_j;j,p \right\rangle.\e
$\left| \Omega \right>$ is vacuum state which is invariant under the action of the discrete UIR of dS group:
\b \label{vacumtr2} T^{(0,j;p)}(g) \left| \Omega \right>=\left| \Omega \right>.\e
In what follows, it has been shown that, this vacuum state $|\Omega>$ can be identified with the vacuum state in the principal series representation case.
Due to the similarity of the cases UIR $T^{(0,j;p)}$ and $T^{(j,0;p)}$, only the former would be considered here. 

$a({\bf q},m_j;j,p)$ is the adjoint operator of the creation operator $a^\dag({\bf q},m_j;j,p)$ which can be defined using the equation (\ref{cractionp2}):
$$<\Omega|\left[a^\dag({\bf q},m_j;j,p)\right]^\dag=<{\bf \tilde{q}},m_j;j,p|.$$
For the principal series $p$ is imaginary ($\frac{1}{2}+i\nu$) and one has $p^*=1-p$ but for discrete series $p$ is real. Since the two representations $T^{(0.j;p)}$ and $T^{(0,j;1-p)}$ are unitary equivalent (\ref{ued}), then the annihilation operator may be defined similar to the principal series (\ref{adjp}) as:
\b \label{adjd}\left[a^\dag({\bf q},m_j;j,p)\right]^\dag\equiv
a({\bf \tilde{q}},m_j;j,1-p).\e

By the help of the orthogonality condition on the Hilbert space ${\cal H}^{(0,j,p)}_{q}$ (was defined explicitly by Takahashi \cite{tak}), one shows that the operator $a({\bf q},m_j;j,p)$ removes a state from any state in which it acts on, similar to the massive case (\ref{anist}). It is called the annihilation operator and it annihilates the vacuum state:
\b \label{anihilip2} a({\bf q},m_j;j,p) \left| \Omega \right>=0. \e
From (\ref{dis0jp}), (\ref{cractionp2}) and (\ref{vacumtr2}), one obtains:
$$ T^{(0,j;p)}(g) a^\dag ({\bf q},m_j;j,p)\left[ T^{(0,j;p)}(g)\right]^\dag= |{\bf c}{\bf q}+{\bf d}|^{-2(p+1)}\;\;\;\;\;\;\;\;\;\;\;\;\;\;\;\;\;$$
\b \label{creationtra2} \;\;\;\;\;\;\;\;\;\;\;\;\;\;\;\;\;\;\;\;\; \times \sum_{m_j'}D^{(j)}_{m_j'm_j}\left(\frac{({\bf c}{\bf q}+{\bf d})^{-1}}{|{\bf c}{\bf q}+{\bf d}|} \right) a^\dag\left(g^{-1}\cdot {\bf q},m_j';j, p\right). \e
Similar to the principal series, for the anihiliation operator, one obtains:
$$ T^{(0,j;p)}(g) a({\bf \tilde{q}},m_j;j,1-p)\left[ T^{(0,j;p)}(g)\right]^\dag= |{\bf c}{\bf q}+{\bf d}|^{-2(2-p)}\;\;\;\;\;\;\;\;\;\;\;\;\;\;\;\;\;$$
\b \label{anihltiontra2}\;\;\;\;\;\;\;\;\;\;\;\;\;\;\;\;\;\;\;\;\;  \times \sum_{m_j'}\left[D^{(j)}_{m_j m'_j}\left(\frac{({\bf c}{\bf q}+{\bf d})^{-1}}{|{\bf c}{\bf q}+{\bf d}|} \right)\right]^* a\left(g^{-1}\cdot{\bf \tilde{q}},m_j';j, 1-p\right).\e
Using the equation (\ref{ued}), one can see the annihilation operator $a({\bf q},m_j;j,p)$ transforms by the unitary equivalent representation of discrete series:
$$ T^{(0,j;1-p)}(g) a({\bf q},m_j;j,p)\left[ T^{(0,j;1-p)}(g)\right]^\dag= |{\bf c}{\bf q}+{\bf d}|^{-2(2-p)}\;\;\;\;\;\;\;\;\;\;\;\;\;\;\;\;\;$$
\b \label{anihltiontra23}\;\;\;\;\;\;\;\;\;\;\;\;\;\;\;\;\;\;\;\;\;  \times \sum_{m_j'}\left[D^{(j)}_{m_j m'_j}\left(\frac{({\bf c}{\bf q}+{\bf d})^{-1}}{|{\bf c}{\bf q}+{\bf d}|} \right)\right]^* a\left(g^{-1}\cdot{\bf q},m_j';j, p\right).\e

One can prove that the following relation (similar to the principal series) will hold:
\b  a({\bf \tilde{q}}',m_j';j,1-p)a^\dag ({\bf q},m_j;j,p)\pm a^\dag ({\bf q},m_j;j,p)a ({\bf \tilde{q}}',m_j';j,1-p)=N({\bf q}, m_j)\delta_{S^3}({\bf q}'-{\bf q})\delta_{m_jm_j'},\e
with  $+$ or $-$ sign for fermionic or bosonic states, respectively.

The quantum field operator in terms of creation and annihilation operators can be written as the following forms, for tensor field $j=l$,
\b \label{qfop2} {\cal K}_{\alpha_1..\alpha_l}(x)=\int_{B}d\mu({\bf q})\sum_{m_j}\left[a({\bf \tilde{q}},m_j;j,1-p){\cal U}_{\alpha_1..\alpha_l}(x;{\bf q},m_j;j,p )+a^\dag({\bf q},m_j;j,p){\cal V}_{\alpha_1..\alpha_l}(x;{\bf q},m_j;j,p )\right],\e
and for tensor-spinor field $j=l+\frac{1}{2}$,
\b \label{qfops2} \Psi_{\alpha_1..\alpha_l}(x)=\int_{B}d\mu({\bf q})\sum_{m_j}\left[a({\bf \tilde{q}},m_j;j,1-p)U_{\alpha_1..\alpha_l}(x;{\bf q},m_j;j,p )+a^\dag({\bf q},m_j;j,p)V_{\alpha_1..\alpha_l}(x;{\bf q},m_j;j,p )\right].\e

The coefficients ${\cal U}$, ${\cal V}$, $U$ and $V$ are chosen so that under the dS transformations the field operators transform by the UIR of the dS group (principle B). For tensor fields ($j=l$), one writes:
\b \label{fioptrbd}  T^{(0,j;p)}(g)  {\cal K}_{\alpha_1..\alpha_l}(x) \left[ T^{(0,j;p)}(g)\right]^\dag= \Lambda_{\alpha_1}^{\;\;\alpha'_1}..\Lambda_{\alpha_l}^{\;\;\alpha'_l}  {\cal K}_{\alpha'_1..\alpha'_l}(\Lambda x), \e
which has an equivalent for the tensor-spinor fields ($j=l+\frac{1}{2}$), by the form \cite{brgamo,gata,bagamota}:
\b \label{fioptrsd}   T^{(0,j;p)}(g) \Psi_{\alpha_1..\alpha_l}(x) \left[ T^{(0,j;p)}(g)\right]^\dag= \Lambda_{\alpha_1}^{\;\;\alpha'_1}..\Lambda_{\alpha_l}^{\;\;\alpha'_l}g^{-1}  \Psi_{\alpha'_1..\alpha'_l}(\Lambda x),\e
with $\Lambda \in SO(1,4)$ and $g\in Sp(2,2)$. After making use of (\ref{creationtra2}), (\ref{anihltiontra2}), (\ref{qfop2}), (\ref{fioptrbd}) and the dS invariant volume element on the unit ball B (\ref{mainb}), it is easy to show that the coefficients ${\cal U}_{\alpha_1..\alpha_l}(x;{\bf q},m_j;j,p )$ and ${\cal V}_{\alpha_1..\alpha_l}(x;{\bf q},m_j;j,p )$ satisfy the following equations:
 $$ |{\bf c}{\bf q}+{\bf d}|^{-2p+6} \sum_{m_j}D^{(j)}_{m_j'm_j}\left(\frac{({\bf c}{\bf q}+{\bf d})^{-1}}{|{\bf c}{\bf q}+{\bf d}|} \right){\cal V}_{\alpha_1..\alpha_l}(x;{\bf q},m_j;j,p )\;\;\;\;\;\;\;\;\;\;\;\;\;\;\;\;\;\;\;\;
$$
\b \label{vt2} \;\;\;\;\;\;\;\;\;\;\;\;\;\;\;\;\;\;\;\;\;\;\;\;\;\;\;\;\;\;\;\;\;\;\;\;\;\;\;\;\;\;\;\;\;
=\Lambda_{\alpha_1}^{\;\;\alpha'_1}..\Lambda_{\alpha_l}^{\;\;\alpha'_l}{\cal V}_{\alpha'_1..\alpha'_l}(\Lambda x; g^{-1}\cdot{\bf q},m'_j;j,p ),
\e
$$|{\bf c}{\bf q}+{\bf d}|^{2p+4} \sum_{m_j}\left[D^{(j)}_{m_j'm_j}\left(\frac{({\bf c}{\bf q}+{\bf d})^{-1}}{|{\bf c}{\bf q}+{\bf d}|} \right)\right]^*{\cal U}_{\alpha_1..\alpha_l}(x;{\bf q},m_j;j,p )\;\;\;\;\;\;\;\;\;\;\;\;\;\;\;\;\;\;\;\;\;\;
$$
\b \label{ut}\;\;\;\;\;\;\;\;\;\;\;\;\;\; \;\;\;\;\;\;\;\;\;\;\;\;\;\;\;\;\;\;\;\;\;\;\;\;\;\;\;\;\;\;\;\;\;\;
=\Lambda_{\alpha_1}^{\;\;\alpha'_1}..\Lambda_{\alpha_l}^{\;\;\alpha'_l}{\cal U}_{\alpha'_1..\alpha'_l}(\Lambda x; g^{-1}\cdot{\bf q},m'_j;j,p ).
\e
These equations and the conditions (i-vii) in section II-D allow us to calculate the functions ${\cal U}$ and ${\cal V}$ explicitly. One can write similar equations for a tensor-spinor field. The homogeneity condition results to:
$${\cal V}_{\alpha_1..\alpha_l}(x;{\bf q},m_j;j,p )=(x\cdot\xi_B)^\lambda v_{\alpha_1..\alpha_l}(x,{\bf q},m_j;j,p),$$ where $\xi^\alpha_B =\xi^0(1, \coth \kappa \; {\bf q})$.  For these fields, we can choose $\xi^\alpha_{B}(0)=(1,0,0,0,1)$ similar to the massive case where the little group is $SO(3)$. The null curvature limit of these field operators does not exist {\it i.e.} these fields do not have any counterparts in the null curvature limit.

Similar to the principal series, one can explicitly calculate the quantum field operator and its corresponded two-point function for these fields only for $p<3$. As it has been already known, one of the homogeneous degrees, $\lambda$, is positive for $p\geq 3$ (\ref{hdll}) and (\ref{hdf}) and also, the plane wave becomes infinite for the large values of $x$ (\ref{pws2i}). Consequently, one cannot define the field operator in a sense of a distribution function for these cases ($p\geq 3$). The mass associated with these fields is also imaginary or $m^2<0$ in (\ref{mbp}) and (\ref{mfp})

In these cases ($j\neq p$), two auxiliary fields are important to construct the massless conformal quantum fields in the dS space, one of which being the auxiliary spin-$\frac{3}{2}$ field, is defined by $j=\frac{3}{2}$ and $p=\frac{1}{2}$ and the other one, by $j=2$ and $p=1$. In the following subsection, these two auxiliary field have been described.

\subsection{Auxiliary spin-$\frac{3}{2}$ field}

An auxiliary vector-spinor field associates with the discrete series representation with $j=\frac{3}{2}$,  $p=\frac{1}{2}$ and its corresponding mass is $m^2_{f,\frac{1}{2}}=2H^2$. It satisfies the conditions (i-vii) of the section II-D and its relevant eigenvalue of the Casimir operator is $<Q^{(1)}_{\frac{3}{2},\frac{1}{2}}>=-\frac{3}{2}$.
This field satisfies the following field equation
$$ \left(\not x \not \partial^\top-2 \right)\Psi_\alpha(x)=0,\;\;\;\left(\square_H+2H^2\right)\Psi_\alpha(x)=0.$$
The field operator of the auxiliary vector-spinor field is defined by:
\b \label{spinvqfop} \Psi_\alpha(x)=\int_{B}d\mu({\bf q})\sum_{m}\left[a({\bf \tilde{q}},m;\frac{3}{2},\frac{1}{2})U_\alpha(x;{\bf q},m;\frac{3}{2},\frac{1}{2} )+a^{c\dag}({\bf q},m;\frac{3}{2},\nu)V_\alpha(x;{\bf q},m;\frac{3}{2},\frac{1}{2} )\right],\e
where $m=\frac{3}{2}, \frac{1}{2},-\frac{1}{2},-\frac{3}{2}$ and $\Psi_\alpha, U_\alpha$ and $V_\alpha$ are four-component spinors. From the equation (\ref{hdmjd}), it is well-known that the homogeneous degrees of spinor-vector field with $ p=\frac{1}{2}$ are $\lambda=-1$ and $-2$:
$$ U_\alpha(x;{\bf q},m;\frac{3}{2},\frac{1}{2} )=(x\cdot\xi_B)^{-2}u_\alpha(x, {\bf q},m;\frac{3}{2},\frac{1}{2} ),$$ $$  V_\alpha(x;{\bf q},m;\frac{3}{2},\frac{1}{2} )=(x\cdot\xi_B)^{-1}v_\alpha(x,{\bf q},m;\frac{3}{2},\frac{1}{2} ) . $$ 
Through defining ${\bf q}=r{\bf u}$, one obtains $\xi_B=(1,\coth \kappa {\bf q})=(1,\coth \kappa r{\bf u})=(1,\pm {\bf u})$. Then the field operator, the four-component spinors $U$ and $V$ and also the two-point function can be obtained simply from the principal series counterpart by replacing $\nu=0$ in their corresponding relations.
 
The analytic two-point function for auxiliary vector-spinor field is:
$$ S^{ii'}_{\alpha\alpha'}(z,z')=\left<\Omega\right|\Psi_{\alpha}^{i}(z)\bar \Psi_{\alpha'}^{i'}(z')\left|\Omega\right> $$ \b \label{tpfvfpau} =\int_{B} d\mu({\bf q}) (z\cdot\xi_B)^{-2}(z'\cdot\xi_B)^{-2}\sum_{m} u_{\alpha}^{i}(z,{\bf q},m;\frac{3}{2},\frac{1}{2})\bar u_{\alpha'}^{i'}(z',{\bf q },m;\frac{3}{2},\frac{1}{2}). \e  
This function can be obtained from the principal counterpart (\ref{5.6}) by replacing $\nu=0$:
\begin{equation} S_{\alpha \alpha'}(z,z')=
D_{\alpha \alpha'}(z, \partial^\top ;z', {\partial'}^\top)
S_c(z,z'),\label{tps32au}\end{equation} where
$$ D_{\alpha \alpha'}(z, \partial^\top ;z', {\partial'}^\top)=\theta_\alpha \cdot \theta'_{\alpha'}-\frac{1}{4}\gamma^\top_\alpha\gamma^\top\cdot \theta'_{\alpha'}-\left(-\frac{1}{4}\gamma^\top_\alpha\gamma^\top\cdot \theta'_{\alpha'}+2\partial^\top_\alpha z\cdot\theta'_{\alpha'}\right.$$ $$\left. +\frac{2}{3}\partial^\top_\alpha\theta'_{\alpha'}\cdot\partial^\top -\frac{2}{3}\gamma^\top_\alpha \theta'_{\alpha'} \cdot z \not \partial^\top+ \frac{1}{3}\partial^\top_\alpha\gamma^\top\cdot\theta'_{\alpha'}\not z-\frac{1}{3}\gamma^\top_\alpha \theta'_{\alpha'}\cdot z \not z \right.  $$ \b\left. -\frac{1}{6}\gamma^\top_\alpha\theta'_{\alpha'} \cdot\partial^\top \not\partial^\top-\frac{1}{12}\gamma^\top_\alpha\gamma^\top\cdot\theta'_{\alpha'} \not z \not \partial^\top\right).\e
$S_c(z,z')$ is the analytic two-point function of conformally massless spinor field \cite{bagamota} which will be considered in the section VII-C:
\b \label{tscon}  S_c(z,z')=\frac{i}{2\pi^2}\frac{(\not z-\not z')\gamma^4}{[(z-z')^2]^2}. \e

\subsection{Auxiliary spin-$2$ rank-2 symmetric tensor field}

An auxiliary spin-$2$ rank-$2$ symmetric traceless tensor field corresponds to the discrete series representation with $j=2$ and $p=1$. The eigenvalue of the Casimir operator and its corresponding mass are $<Q^{(1)}_{2,1}>=-4$ and $m^2_{b,1}=2H^2$. The field equation is:
$$ \left( Q_2^{(1)}+4\right){\cal K}_{\alpha\beta}(x)=0, \;\; \mbox{or}\;\;\left( \square_H+2H^2\right){\cal K}_{\alpha\beta}(x)=0. $$
The tensor field $\K_{\alpha\beta}$ can be written in terms of a massless conformally coupled scalar field \cite{petata}:
\b \K_{\alpha \beta}(x)=\left(-\frac{2}{3}\theta Z_1.+\Sigma_1
Z^\top_1+\frac{1}{3}D_2 \left[ \frac{1}{9} D_1 Z_1.+x.Z_1
\right]\right)\left( Z^\top_{2}+\ D_{1}\
(x.Z_2)\right)\phi_c,\e
where $Z_1$ and $Z_2$ are constant five-vectors and $\Sigma_1$ is defined in equation (\ref{sigmap}). $\phi_c$ will be considered in section VII.A. The field operator can be written in the following form:
\b \label{as2r2s} K_{\alpha\beta}(x)=\sum_{m}\int_{B}d\mu({\bf 	q})\left[a({\bf \tilde{q}},m;2,0){\cal U}_{\alpha\beta}(x;{\bf q},m;2,1)+a^{\dag}({\bf q},m;2,1){\cal V}_{\alpha\beta}(x;{\bf q},m;2,1 )\right],\e
with $m=2,1,0,-1,-2$. The homogeneous degrees of such field are $\lambda=-1$ and $-2$ (\ref{hdllp}):
$$ {\cal U}_{\alpha\beta}(x;{\bf q},m;2,1 )=(x\cdot\xi_B)^{-2} u_{\alpha\beta}(x,{\bf q},m;2,1),$$ 
$$ {\cal V}_{\alpha\beta}(x;{\bf q},m;2,1 )=(x\cdot\xi_B)^{-1}v_{\alpha\beta}(x,{\bf q},m;2,1).$$ 
The coefficients $u$ and $v$ satisfy the conditions (v-vi) of section II-D: 
$$ v_{\alpha\beta}(x,{\bf q},m;2,1)= v_{\beta \alpha}(x,{\bf q},m;2,1),\; \;\; v_{\alpha}^{\;\;\alpha}(x,{\bf q},m;2,1)=0,$$
and, the field operator satisfies the following transformation role:
$$   T^{(0,2;1)}(g) K_{\alpha\beta}(x) \left[ T^{(0,2;1)}(g)\right]^\dag= \Lambda_{\alpha}^{\;\;\alpha'}\Lambda_{\beta}^{\;\;\beta'}  K_{\alpha'\beta'}(\Lambda x).$$

The analytic two-point function is defined by:
$$ W_{\alpha\beta\alpha'\beta'}(z,z')=<\Omega|K_{\alpha\beta}(z)K_{\alpha'\beta'}(z')|\Omega>$$ \b \label{tpfvm} =\int_{B} d\mu({\bf q})(z\cdot\xi_B)^{-2}(z'\cdot\xi_B)^{-1}\sum_m u_{\alpha\beta}(z,{\bf q},m;2,1)v_{\alpha'\beta'}(z',{\bf q},m;2,1). \e
This analytic two-point function has been calculated in the previous paper \cite{petata}: $$W_{\alpha\beta\alpha'\beta'}(z,z')=D_{\alpha\beta\alpha'\beta'}(z,\partial;z',\partial';1)  W_c(z,z'),$$
where $$ D_{\alpha\beta\alpha'\beta'}(z,\partial;z',\partial';1)=-\frac{2}{3}\Sigma_1'\theta
\theta'.\left(\theta.\theta'+\ D_1^\top(\theta'.z)\right)$$
$$ +\Sigma_1\Sigma_1'\theta.\theta'\left(\theta.\theta'
    +\ D_1^\top(\theta'.z)\right) +\frac{1}{3} D_2^\top\Sigma_1'\left(\frac{1}{9}D_1^\top\theta'.+z.\theta'\right)\left(\theta.\theta'
    +\ D_1^\top(\theta'.z)\right),$$
and $W_c(z,z')$ is the analytic two-point function of massless conformally coupled scalar field \cite{brmo,ta96} which will be considered in the section VII-A:
\b \label{stpci} W_c(z,z')=\frac{-iH^2}{2^4\pi^2} P_{-1}^{(5)}(H^2 z.z')=
\frac{H^2}{8\pi^2}\frac{-1}{1-{\cal Z}(z,z')}. \e
${\cal Z}(z,z')=-H^2 z\cdot
z'$ is the geodesic distance between two points on the complex dS hyperboloid.

\subsection{The case $j=p\geq 1$ }

The case $j=p\geq 1$ is of particular interest, since it corresponds to the massless fields in the dS space with existing equivalent massless fields in the Minkowski space-time, in the null curvature limit. Here, we only consider three sets of values, namely $j=p= 1, \frac{3}{2}$ and $ 2$. Even for these cases, one has trouble with the construction of the quantum field operators, which transform as an UIR of the dS group.

The field equation in $\xi$-space (\ref{difequ}) for these cases become:
\b\label{difequj=p} \left[\frac{1}{4}(1-|{\bf q}|^2)\triangle -j{\cal D}-\frac{1}{2}(A_j{\cal D}_1+B_j{\cal D}_2+C_j{\cal D}_3)\right]\left|{\bf q};j,j\right>=0.\e 
Since $\triangle, {\cal D}, {\cal D}_1,{\cal D}_2$ and ${\cal D}_3$ are differential operators, the quantum state $|{\bf q};j,j>$ cannot be uniquely defined by this equation, and also, it is invariant under the following transformation:
\b \label{gatrq} \left|{\bf q};j,j\right> \Longrightarrow \left|{\bf q};j,j\right>^g= \left|{\bf q};j,j\right>+\left|{\bf q}_0;j,j\right>,\e
where $\left|{\bf q}_0;j,j\right> \in V^j$ and ${\bf q}_0$ is constant.

If we assume  ${\bf q}=r{\bf u}$ with $|{\bf u}|=1$ and $|{\bf q}|=r$, then the  $\left |{\bf q};j,p\right>$ can be written as (\ref{qinru2}): 
$$ \label{qinru} \left|{\bf q},m;j,p\right>=F(p-j,j+p+1;2;r^2)\sum_{m'} C_{mm'} \left|{\bf u},m';j,p\right>,$$
with $\left |{\bf u};j,p\right> \in V^j$ and $F(p-j,j+p+1;2;r^2)$ being the hyper-geometric function. $F(p-j,j+p+1;2;r^2)$
satisfies the following differential equation \cite{tak} which can be simply obtained, using equation (\ref{difequj=p}) and relations (\ref{relations}):  
\b \label{xiesegau} \left[ \frac{1-r^2}{4}\left(\frac{d^2}{dr^2}+\frac{3}{r}\frac{d}{dr}\right)-pr\frac{d}{dr}+(j-p)(j+p+1)\right]F(p-j,j+p+1;2;r^2)=0.\e
One of the solutions of the above equation for $j=p$ is constant and therefore the hyper-geometric function becomes constant:
$$ F(0,2j+1;2;r^2)=\mbox{constant}.$$
This is precisely alike to a problem, previously occurred in the establishment of the minimally coupled scaler fields in the dS space, where one needed to introduce an indecomposable representation of the dS group to define field operators in an indefinite inner product space (Krein space quantization) \cite{ta97,dere,gareta}. It means that the creation operator $a^\dag({\bf q},m_j;j,j)$ cannot be defined properly whenever it is being transformed by an UIR $T^{(0,j;j)}(g)$ of the dS group. 

Similar to the previous section, by ignoring this difficulty, one can define a field operator which suffers the lack of the capability of being transformed under an UIR of the dS group \cite{gareta,gagarota,ta09}. For such a case, one has:
\b   T^{(0,j;j)}(g) {\cal K}_{\alpha_1..\alpha_l}(x) \left[ T^{(0,j;j)}(g)\right]^\dag= \Lambda_{\alpha_1}^{\;\;\alpha'_1}..\Lambda_{\alpha_l}^{\;\;\alpha'_l}{\cal K}_{{\alpha'_1..\alpha'_l}}(\Lambda x)+\left[{\cal D}^\top_l \Omega\right]_{\alpha_1..\alpha_l}, \e
where ${\cal D}^\top_l$ is a generalized-gradient which preserves the trasversality and symmetric type of tensor field ${\cal K}$ on the dS hyperboloid. $\Omega(x)$ is a linear combination of annihilation and creation operators whose precise form is not our concern here. It is sufficient to know that $\Omega(x)$ is a rank-$(l-1)$ tensor field with some constituents. There are two types of tensor field $\Omega$: the first type satisfies the following field equation: 
$$\left[\partial^\top_l\cdot {\cal D}^\top_l \Omega_3(x)\right]_{\alpha_1..\alpha_{l-1}}=0,$$ 
where $(\partial^\top_l\cdot) $ is the generalized divergence in the ambient space formalism on the dS hyperboloid. It is called the pure gauge state. The second type is 
$$\left[\partial^\top_l\cdot {\cal D}^\top_l \Omega_1(x)\right]_{\alpha_1..\alpha_{l-1}}\neq 0,$$ 
which satisfies some conditions (ii-vii) of section II-D. It is the gauge-dependent state. The explicit form of $\Omega_3$ and $\Omega_1$ can be obtained from the gauge invariant transformation and gauge fixing field equation which are being discussed in the next section. These two type of un-physical states with the physical state construct the Gupta-Bluler triplet states of the massless fields in the dS space-time.

The field operators act on a vector space which is constructed on an indecomposable representation of the dS group. For these fields a massless quantum state can be divided into three parts:
$$ {\cal M}=V_1 \oplus V_2 \oplus V_3, $$ 
where $V_1 $ is the space of the gauge dependent states and $V_3$ is the pure gauge states. The physical states appear in $V_2/V_3\equiv {\cal H}$, which are being constructed on the UIR of discrete series with $j=p \geq 1$, {\it i.e.} ${\cal H}_{q}^{(j,0,j)}$ and ${\cal H}_{q}^{(0,j,j)}$:
$$ |\;\mbox{physical states}> \in {\cal H}_{q}^{(j,0,j)} \oplus {\cal H}_{q}^{(0,j,j)}.$$
The precise forms of $V_3$ and $V_1$ spaces can be also defined by using the gauge invariant transformation and the gauge fixing field equation in $x$-space.

The gauge invariance in the $x$-space appears in a different form in comparison with the ${\bf q}$-space. When $j=p$, the coefficient $v_{\alpha_1\alpha_2...}(x,{\bf q},m_j;j,j )$ cannot be defined uniquely and it becomes singular.
From the viewpoint of the quantum field theory, the coefficient $v_{\alpha_1\alpha_2...}(x,{\bf q},m_j;j,j )$ has infinite normalization \cite{gagarota,taro12} and with the conditions (i-vii) of section II-D one cannot obtain it \cite{gagata,gagarota,berotata,derotata}. This problem appears due to the divergencelessness condition which is necessary for associating each field with an UIR of dS group. 

By using the second order Casimir operator and dS algebra, one can derive the gauge invariant field equations. Up to this point, based on the gauge theory approach, the three cases of vector fields, vector-spinor fields and spin-2 fields have been discussed.


\setcounter{equation}{0}
\section{Gauge invariant field equations}

Establishment of the gauge invariant field equation (\ref{difequj=p}) and the gauge invariant transformation (\ref{gatrq}) of the massless fields with $j=p\geq 1$ in the $\xi$-space were the subject of the previous section. In this section the gauge invariant field equations and corresponding gauge transformation in $x$-space are studied. By using the second-order Casimir operator in $x$-space and the dS algebra the gauge invariant field equations of the three important cases, namely the vector fields, the vector-spinor fields and the spin-$2$ rank-$2$ symmetric tensor fields have been presented \cite{ta97,gagarota,fatata,ta09,taro12}. The gauge invariant field equation for spin-$2$ rank-$3$ tensor field ${\cal K}_{\alpha\beta\gamma}=-{\cal K}_{\alpha\gamma\beta}$ is also obtained. The Lagrangian for massless fields with spin $1,\frac{3}{2}$, and $2$ are defined. Then the gauge invariant Lagrangian and the gauge covariant derivative are presented. 

We assume that the interactions between the elementary systems in the universe are governed by the gauge principle and formulated through the gauge-covariant derivative which is defined as a quantity that preserves the gauge invariant transformation of the Lagrangian. For an instructive review of gauge transformation and gauge potentials the reader can see \cite{ilio13}. The following includes only the three gauge fields, namely the vector fields ($j=p=1$), vector-spinor field ($j=p=\frac{3}{2}$) and spin-$2$ field ($j=p=2$). These gauge fields, in the language of the gauge theory, are the sources or potentials of the various forces. Such fact can be considered as a connection in the gauge group manifold. One can associate to these gauge transformations with the local symmetrical groups, noting that these gauge fields transform according to these symmetrical groups.

The gauge vector fields $(j=p=1)$, $ K_\alpha^{\;\;a}$ with $a=1,2,..., n^2-1$, may be associated with the gauge group $SU(n)$. The same association may be engaged between the gauge spin-$2$ fields $(j=p=2)$ in the gauge gravity framework and the gauge group $SO(1,4)$, or equivalently, the gauge group $SO(2,4)$ \cite{fr84}. Imposing some physical conditions on the gauge field or on the connection, such case can be described as a spin-$2$ rank-$3$ mix-symmetric tensor field ${\cal K}^M_{\alpha\beta\gamma}$. The vector-spinor gauge fields $(j=p=\frac{3}{2})$, $ \Psi_\alpha^{\;\;l}$ with $l=1,2,..., N$, are the spinor fields and consequently, their corresponding gauge group must have spinorial generators to justify a set of well defined gauge-covariant derivative. Therefore, a set of anti-commutative generators satisfy a super-algebra. Nevertheless, such algebra would not be closed, since its constituent generators are Grasmanian functions which will have usual functions as their multiplication products {\it i.e.} the anti-commutation of two spinor generators become a tensor generator. In this case, for obtaining a closed super-algebra, the Grasmanian generators must be coupled with the generators of the dS group. Additionally, in the language of the gauge theory, one may describe a vector-spinor field as a gauge field which must be coupled with a spin-$ 2$ gauge potential and the gauge group in this case is a super-group.

\subsection{Vector gauge theory}

This subsection is a direct generalization of the abelian gauge theory in the dS ambient space formalism \cite{rota05}. Since such formalism in the ambient space notation is utterly similar to its Minkowskian counterpart, only the differences and necessary considerations are presented here. In terms of Casimir operator (\ref{casimirl}), the vector-gauge invariant field equation can be rewritten in the following form \cite{gagarota,rota05}:
$$ Q^{(1)}_1 K_\alpha^{\;\;a}+ \nabla^\top_\alpha \partial\cdot K^{\;a}=0.$$
This field equation is invariant under the following gauge transformation:
$$ K_\alpha \longrightarrow (K^g)_\alpha^{\;\;a}=K_\alpha^{\;\;a} +\nabla^\top_\alpha \phi^a,$$
where $\phi^a$ is the arbitrary scalar fields. The following identities can be used to prove the gauge invariance:
$$ Q_1^{(1)}\nabla^\top_\alpha \phi(x)=\nabla^\top_\alpha Q_0^{(1)}\phi(x),\;\;\;\; \partial^\top \cdot \nabla^\top \phi(x)=-Q_0^{(1)}\phi(x).$$
The field equation can be obtained from the action integral:
\b \label{guvela2} S_1[K]=\int d\mu(x)\left( \nabla^\top_\alpha K_\beta^{\;\;a}-\nabla^\top_\beta K_\alpha^{\;\;a}\right)\left(\nabla^{\top\alpha} K^{\beta b}-\nabla^{\top\beta} K^{\alpha b}\right)\delta_{ab}.\e

Now we attempt to reformulate this action by using the gauge theory. The vector gauge potentials $K_\alpha^{\;\;a}$ with $a=1,2,..., n^2-1$, can be associated with the gauge group $SU(n)$. One can assume that $t^a$ is the generator of $SU(n)$ group, satisfying the following commutation relation
\b  [t_a,t_b]=C^c_{\;\;ab}t_c.\e
The notation of local gauge symmetry with its space-time-dependent transformation can be used to generate the gauge interaction. 
For obtaining a local gauge invariant Lagrangian, it is
necessary to replace the transverse-covariant derivative
$ \nabla^\top_\beta$ (\ref{dscdrt}) with the gauge-covariant derivative
$D_\beta^{K}$ which is defined by 
\b D_\beta^K \equiv\nabla^\top_\beta +iK_{\beta}^{\;\;a}t_a, \e
where the gauge potential or connection $K_{\beta}^{\;\;a}$ is a massless vector field. Considering a local infinitesimal gauge transformation generated by $\epsilon^a(x)t_a$, one has
$$ \delta_{\epsilon}K_\alpha^{\;\;a}=D^K_\alpha\epsilon^a=\partial^\top_\alpha \epsilon^a+ C_{cb}^{\;\;\;\;a}K_\alpha^{\;\;c}\epsilon^b.$$
The curvature ${\cal F}$, for the Lie algebra of the gauge group $SU(N)$, is defined by:
$${\cal F}(D^K_\alpha,D^K_\beta)=-[D^K_\alpha,D^K_\beta]=F_{\alpha\beta}^{\;\;\;\;a}t_a,$$
where
$$ F_{\alpha\beta}^{\;\;\;\;a}=\nabla^\top_\alpha K_\beta^{\;\;a}-\nabla^\top_\beta K_\alpha^{\;\;a}+K_\beta^{\;\;b}K_\alpha^{\;\;c}C_{bc}^{\;\;\;\;a},\;\;\; \;\; x^\alpha F_{\alpha\beta}^{\;\;\;\;a}=0= x^\beta F_{\alpha\beta}^{\;\;\;\;a}.$$

The $SU(n)$ gauge invariant action or Lagrangian in the dS background for the gauge field $K_\alpha^{\;\;a}$ is:
$$ S_1[K]=-\frac{1}{2}\int d\mu(x) \tr \left({\bf F}_{\alpha\beta}{\bf F}^{\alpha\beta} \right)=-\frac{1}{2}\int d\mu(x) F_{\alpha\beta}^{\;\;\;\;a}F^{\alpha\beta b} \tr\left(t_at_b\right) , $$ 
with ${\bf F}_{\alpha\beta}= F_{\alpha\beta}^{\;\;\;\;a}t_a$ and summing over the repeated indices. The normalization of the structure constant is usually
fixed by requiring that, in the fundamental representation, the corresponding
matrices of the generators $t_a$ are normalised such as \cite{ilio13}
$$  \tr\left(t_at_b\right) =\frac{1}{2} \delta_{ab}.$$
Then the action becomes
$$ S_1[K]=-\frac{1}{4}\int d\mu(x)\left[ \left( \nabla^\top_\alpha K_\beta^{\;\;a}-\nabla^\top_\beta K_\alpha^{\;\;a}\right)\left(\nabla^{\top\alpha} K^{\beta b}-\nabla^{\top\beta} K^{\alpha b}\right)\delta_{ab}+O({\cal K}^3)\right],$$ 
where $O[(K)^3]$ results in the second order of $K$ in the field equation and this term is defining the interaction between the gauge potentials $K_\alpha^{\;\;a}$. The first part is the action integral (\ref{guvela2}).

Similar to the Minkowskian space-time, the gauge fixing field equation can be obtained from the new Lagrangian with the additional gauge fixing terms:
$$ S_1[K]=\int d\mu(x) \left[ -\frac{1}{4}F_{\alpha\beta}^{\;\;\;\;a}F^{\alpha\beta a}-\frac{c'}{2} \partial\cdot K^{\;a}\partial\cdot K^{\;a} \right], $$
with summing over index $a$ and $c'$ being the gauge fixing parameter. The linear gauge fixing field equation is
\b \label{qifev} Q^{(1)}_1 K_\alpha^{\;\;a}+c \nabla^\top_\alpha \partial\cdot K^a=0,\e
where $c$ is a new gauge fixing parameter. The Faddeev-Popov ghost fields can also be added in the quantization procedure in an exact similar way as in the Minkowski space-time.

\subsection{Spin-$2$ gauge theory}

The spin $2$-gauge field can be described by a rank-$2$ symmetric tensor field ${\cal H}_{\alpha\beta}$ or a rank-$3$ mix-symmetric tensor field ${\cal K}_{\alpha\beta\gamma}$ \citep{fr84}. In this subsection we present the gauge invariant field equations for these two cases. The rank-$3$ mixed symmetric tensor field may be reformulated in the gauge-gravity model.

Utiyama was the first who proposed that the general relativity can be seen as a gauge theory based on the local Lorentz group \cite{ut}. Kibble comprehensively extended the Utiyama's gauge theory of gravitation by showing that the local Poincar\'e symmetry can generate a space with torsion as well as curvature \cite{ki}. It is well-known that the gauge gravity model, which is already built for the Poincar\'e group, becomes the Einstein general relativity by imposing the metric compatibility and torsion free conditions \cite{mama,hel}. For obtaining a similar construction with the Yang-Mills theory (or previous sub-section) one uses the Fronsdal paper for spin-two, conformal gauge theory on the Dirac $6$-cone formalism \cite{fr84} which can simply be mapped on the dS ambient space formalism.

In what follows, a brief introduction of the formalism of the gauge gravity model in the dS ambient space for gauge groups $SO(1, 4)$ will be presented.  The conformal consideration or $SO(2,4)$ gauge gravity will be discussed in section VIII-C. In this formalism, imposing some physical conditions on the gauge potentials or the gauge vector fields ($10$ vector fields for $SO(1,4)$ and $15$ vector fields for $SO(2,4)$), one may describe these $10$ gauge vector fields by a rank-$3$ tensor field ${\cal K}_{\alpha\beta\gamma}=-{\cal K}_{\alpha\gamma\beta}$, on the dS hyperboloid. The mix-symmetric part of ${\cal K}_{\alpha\beta\gamma}$ may be described as a spin-$2$ field which propagates on the dS light cone, and in the quantum level, the field operator transforms by an indecomposable representation of dS group.

\subsubsection{Rank-$2$ symmetric tensor field}

The gauge invariant field equation for rank-$2$ symmetric tensor field ${\cal H}_{\alpha\beta}={\cal H}_{\beta\alpha}$ is \cite{taro12}:
\b \label{festgi} \left( Q_2^{(1)}+6\right){\cal H}(x)+D^\top_2\partial_2^\top\cdot {\cal H}(x)=0,\e
which is invariant under the following gauge transformation:
\b \label{gfst3} {\cal H}_{\alpha\beta}\Longrightarrow {\cal H}_{\alpha\beta}^g={\cal H}_{\alpha\beta}+\left(D_2^\top A\right)_{\alpha\beta}. \e   
$A_\beta(x)$ is an arbitrary vector field, the generalized gradient $D_2^\top$ is defined by: $$\left(D_2^\top A\right)_{\alpha\beta}=\nabla^\top_\alpha A_\beta+\nabla^\top_\beta A_\alpha.$$ $\partial_2^\top\cdot {\cal H}$ is the generalized divergence: $$ \partial_2\cdot
 {\cal H}=\partial^\top\cdot {\cal H}-\frac{1}{2}H^2D_1 {\cal H}'=\partial \cdot {\cal H}- H^2 x
 {\cal H}'-\frac{1}{2} \bar  \partial {\cal H}'.$$  For simplicity we impose the traslessness condition ${\cal H}'=0$. This condition results in $\partial ^\top \cdot A=0$. The following identities are used to provide the gauge invariance:
$$ Q_2^{(1)}D_2^\top A(x)=D_2^\top Q_1^{(1)}A(x),\;\;\;\; \partial_2^\top \cdot D_2^\top A(x)=-\left(Q_1^{(1)}+6\right)A(x).$$
The action for this field equation can be defined from the massive case (\ref{acsp2ran}) by replacing $<Q_2^{(1)}>=-6$:
\b  S[{\cal K}]=\int d\mu(x)\left[ \frac{1}{8}\left(\nabla^\top_\alpha {\cal H}_{\beta\gamma}-\nabla^\top_{(\beta} {\cal H}_{\gamma)\alpha}\right) \left(\nabla^{\top\alpha} {\cal H}^{\beta\gamma} -\nabla^{\top(\beta} {\cal H}^{\gamma)\alpha}\right)+5 {\cal H}_{\alpha\gamma}{\cal H}^{\alpha\gamma}\right]. \e
This gauge invariant field equation or its Lagrangian cannot be reformulated in the conformal invariant way then the rank-$3$ tensor field is considered in the next subsection. 

\subsubsection{Rank-$3$ tensor field}

We begin first from the group theoretical point of view. Using the equations (\ref{casi1}), (\ref{genm}) and (\ref{gens}) the effect of the second order Casimir operator on the rank-$3$ tensor field with the subsidiary condition ${\cal K}_{\alpha_1\alpha_2\alpha_3}=-{\cal K}_{\alpha_1\alpha_3\alpha_2}$ on the dS hyperboloid can be calculated:
$$Q^{(1)}_3{\cal K}_{\alpha_1\alpha_2\alpha_3}= Q_0 {\cal K}_{\alpha_1\alpha_2\alpha_3}-4 {\cal K}_{\alpha_1\alpha_2\alpha_3}-2 {\cal K}_{\alpha_2\alpha_1\alpha_3}-2{\cal K}_{\alpha_3\alpha_2\alpha_1}$$
\b \label{casira3}  +2 \delta_{\alpha_1\alpha_2}{\cal K}_{..\alpha_3}+2 \delta_{\alpha_1\alpha_3}{\cal K}_{.\alpha_2.}+2x_{\alpha_1} \partial^\top   \cdot{\cal K}_{.\alpha_2\alpha_3}+2x_{\alpha_2} \partial^\top\cdot {\cal K}_{\alpha_1 .\alpha_3}+2x_{\alpha_3} \partial^\top\cdot {\cal K}_{\alpha_1\alpha_2.}, \e
where ${\cal K}_{..\alpha_3}={\cal K}_{\beta\;\;\;\;\alpha_3}^{\;\;\;\; \beta}$ is traced over the first and the second indices. For simplicity, we impose the tracelessness condition ${\cal K}_{..\alpha_3}=0$. Using the dS algebra, one can obtain the following gauge invariant field equation:
\b \label{ran3mieq} \left(Q^{(1)}_3+6\right) {\cal K}_{\alpha\beta\gamma}+  \nabla^\top_\alpha \left( \partial_3^\top\cdot {\cal K}\right)_{\beta\gamma}=0,\e
which is invariant under the gauge transformation:
\b \label{gts2r3} {\cal K}_{\alpha \beta\gamma} \longrightarrow {\cal K}_{\alpha \beta\gamma}^g={\cal K}_{\alpha \beta\gamma}+ \nabla^\top_\alpha \Lambda_{\beta\gamma},\e
where $\Lambda_{\alpha\beta}$ is an arbitrary rank-$2$ anti-symmetric tensor field ($x\cdot \Lambda_{.\alpha}=0, \; \Lambda_{\alpha\beta}=-\Lambda_{\beta\alpha}$). The tracelessness condition on ${\cal K}$ results in the divergencelessness condition on the pure gauge field $\Lambda$, $\partial^\top \cdot \Lambda_{.\alpha}=0$. The generalized divergence $\partial_3^\top\cdot$ is defined by:
$$  \left( \partial_3^\top\cdot {\cal K}\right)_{\beta\gamma} \equiv \nabla^\top\cdot {\cal K}_{.\beta\gamma}+\nabla^\top\cdot {\cal K}_{\beta.\gamma} ,$$
with the condition $\nabla^\top\cdot {\cal K}_{\beta.\gamma}=-\nabla^\top\cdot {\cal K}_{\gamma.\beta} $. By this condition, $ \left( \partial_3^\top\cdot {\cal K}\right)_{\beta\gamma}$ is a rank-$2$ anti-symmetric tensor field. The gauge invariance can be checked by using the two following identities:
$$Q^{(1)}_{3} \nabla^\top_\alpha \Lambda_{\beta\gamma}= \nabla^\top_\alpha Q^{(1)}_{2A} \Lambda_{\beta\gamma},\;\;\;\; \left( \partial_3^\top.\nabla^\top \Lambda\right)_{\beta\gamma}=-\left(Q^{(1)}_{2A}+6\right)\Lambda_{\beta\gamma},$$
where $Q^{(1)}_{2A}$ is the Casimir operator on the rank-$2$ anti-symmetric tensor field:
$$Q^{(1)}_{2A}\Lambda_{\beta\gamma}= Q_0 \Lambda_{\beta\gamma}-2\Lambda_{\beta\gamma}+2x_\gamma \partial^\top\cdot\Lambda_{\beta .}+2x_\beta \partial^\top\cdot\Lambda_{.\gamma}=\left(Q_0 -2\right)\Lambda_{\beta\gamma}.$$

The field equation (\ref{ran3mieq}) can be obtained from the following action integral:
$$ S_2[{\cal K}]=\int d\mu(x){\cal K}^{\alpha_1\alpha_2\alpha_3}
\left[\left(Q^{(1)}_3+6\right) {\cal K}_{\alpha_1\alpha_2\alpha_3}+  \nabla^\top_{\alpha_1} \left( \partial_3^\top\cdot {\cal K}\right)_{\alpha_2\alpha_3} \right]$$
$$=\int d\mu(x){\cal K}^{\alpha_1\alpha_2\alpha_3}\left[ -\nabla^{\top\alpha}  \nabla^\top_\alpha {\cal K}_{\alpha_1\alpha_2\alpha_3}+\nabla^\top_{\alpha_1}\nabla^{\top\alpha}  {\cal K}_{\alpha\alpha_2\alpha_3}+\nabla^\top_{\alpha_1} \nabla^{\top\alpha} {\cal K}_{\alpha_2\alpha\alpha_3}\right.$$ \b \label{gias2r31} \left. - {\cal K}_{\alpha_1\alpha_2\alpha_3}+2{\cal K}_{\alpha_2\alpha_3\alpha_1}+2{\cal K}_{\alpha_3\alpha_1\alpha_2}\right]. \e
By using the commutation relation:
$$[\nabla^{\top}_{\alpha}, \nabla^{\top}_{\beta}]{\cal K}_{\alpha_1\alpha_2\alpha_3}=\nabla^{\top}_{\alpha}\nabla^{\top}_{\beta}{\cal K}_{\alpha_1\alpha_2\alpha_3}-\nabla^{\top}_{\beta}\nabla^{\top}_{\alpha}{\cal K}_{\alpha_1\alpha_2\alpha_3}
$$ $$ =-\Big(\theta_{\alpha\alpha_1}{\cal K}_{\beta\alpha_2\alpha_3}-\theta_{\beta\alpha_1}{\cal K}_{\alpha\alpha_2\alpha_3}+\theta_{\alpha\alpha_2}{\cal K}_{\alpha_1\beta\alpha_3}
-\theta_{\beta\alpha_2}{\cal K}_{\alpha_1\alpha\alpha_3}+\theta_{\alpha\alpha_3}{\cal K}_{\alpha_1\alpha_2\beta}-\theta_{\beta\alpha_3}{\cal K}_{\alpha_1\alpha_2\alpha}\Big)$$
\begin{equation}
\equiv \left(R^{dS}\right)^\lambda_{\;\;\alpha_1\alpha\beta}{\cal K}_{\lambda\alpha_2\alpha_3}+\left(R^{dS}\right)^\lambda_{\;\;\alpha_2\alpha\beta}{\cal K}_{\alpha_1\lambda\alpha_3}+\left(R^{dS}\right)^\lambda_{\;\;\alpha_3\alpha\beta}{\cal K}_{\alpha_1\alpha_2\lambda},
\end{equation}
the action (\ref{gias2r31}) can be written in the following form:
$$ S_2[{\cal K}]=\int d\mu(x){\cal K}^{\alpha_1\alpha_2\alpha_3}\left[ -\nabla^{\top\alpha}  \left(\nabla^\top_\alpha {\cal K}_{\alpha_1\alpha_2\alpha_3}-\nabla^\top_{\alpha_1} {\cal K}_{\alpha\alpha_2\alpha_3}-\nabla^\top_{\alpha_1} {\cal K}_{\alpha_2\alpha\alpha_3}\right)\right.$$ \b \label{gias2r312} \left. +3 {\cal K}_{\alpha_1\alpha_2\alpha_3}-{\cal K}_{\alpha_2\alpha_3\alpha_1}+{\cal K}_{\alpha_3\alpha_1\alpha_2}\right].\e
The dS background curvature tensor in ambient space formalism is 
\begin{equation} \label{dscurv}
\left(R^{dS}\right)_{\gamma\delta\alpha \beta}  \equiv \theta_{\alpha\gamma}\theta_{\beta\delta}-\theta_{\alpha\delta}\theta_{\beta\gamma} .
\end{equation}
We have $\left(R^{dS}\right)^{\gamma}_{\;\;\beta\gamma\delta}= \left(R^{dS}\right)_{\beta\delta}=3\theta_{\beta\delta}$ and then $ \left(R^{dS}\right)=12.$

Similar to the Minkowski space, the gauge fixing terms can be added to the Lagrangian for obtaining the gauge fixing field equation:
\b \label{gffer3m}\left(Q^{(1)}_3+6\right) {\cal K}_{\alpha\beta\gamma}+ c\nabla^\top_\alpha \partial_3^\top\cdot {\cal K}_{.\beta\gamma}=0,\e 
where $c$ is gauge fixing parameter. In section VII-G, we prove that this gauge potential can be written in the sum of two fields with definite symmetry \cite{fr84}, a totally antisymmetric part ${\cal K}_{\alpha_1\alpha_2\alpha_3}^A$ and a mixed symmetric part ${\cal K}_{\alpha_1\alpha_2\alpha_3}^M$.
The mixed symmetric part can be associated with the tensor field with helicity $\pm 2$. This part can propagate on the dS light cone and it should be conformal invariant and transform by an indecomposable representation of dS group. 

\subsubsection{Gauge covariant derivative}

Let us try to reformulate the above discussions on the gauge theory in dS ambient space formalism for the dS gauge group $SO(1,4)$. For simplicity, we change the notation, as in the previous subsection VI.A, and define the dS group generators by:
$$L_{\alpha\beta}=M_{\alpha\beta}+S_{\alpha\beta}\equiv X_A, \;\; A=1,2,...,10,$$ afterwards, the results will be presented in the suitable notation, {\it i.e.} ambient space formalism. Let us write the commutation relation as:
$$[X_A,X_B]=f_{BA}^{\;\;\;\;\;C}X_C.$$
In this case, we need $10$ gauge vector fields or the connection ${\cal K}_\alpha^{\;\;A}$. Since these gauge vector fields exist on the dS hyperboloid, they should be transverse with respect to the first index $x\cdot {\cal K}_{.}^{\;\;A}\equiv x^\alpha{\cal K}_{\alpha}^{\;\;A}=0$. Additionally, one counts $40$ degrees of freedom for these gauge vector fields and recognizes them as the connection coefficients in the general relativity framework. In the ambient space notation the gauge-covariant derivative can be defined as
$$ D^{\Gamma}_\beta=\nabla^\top_\beta+i {\cal K}_\beta^{\;\;A} X_A,$$
where $\nabla^\top_\beta$ is defined in equation (\ref{dscdrt}). 

Now, one can repeat the construction of the gauge gravity in its canonical manner. Under a local infinitesimal gauge transformation generated by $\epsilon^A(x)X_A$, one has the following gauge transformation:
\b \label{gatra3mif} \delta_{\epsilon}{\cal K}_\alpha^{\;\;A}=D^{{\cal K}}_\alpha\epsilon^A=\nabla^\top_\alpha \epsilon^A+ f_{CB}^{\;\;\;\;A}{\cal K}_\alpha^{\;\;C}\epsilon^B.\e
The curvature ${\cal R}$, with values in the Lie algebra of the dS group, is defined by:
$$  {\cal R}(D^{{\cal K}}_\alpha,D^{{\cal K}}_\beta)=-[D^{{\cal K}}_\alpha,D^{{\cal K}}_\beta]=R_{\alpha\beta}^{\;\;\;\; A}X_A.$$
The curvature is divided in two parts $R_{\alpha\beta}^{\;\;\;\; A}=(R^{dS})_{\alpha\beta}^{\;\;\;\; A}+(R^{{\cal K}})_{\alpha\beta}^{\;\;\;\; A}$. The first part is a dS background curvature (\ref{dscurv}) and the second part is the curvature of the gauge potential ${\cal K}$:
 \b \label{cuso14}(R^{\cal K})_{\alpha\beta}^{\;\;\;\; A}=\nabla^\top_\alpha {\cal K}_\beta^{\;\;A}-\nabla^\top_\beta {\cal K}_\alpha^{\;\;A}+{\cal K}_\beta^{\;\;B}{\cal K}_\alpha^{\;\;C}f_{BC}^{\;\;\;\;A}.\e
In the ambient space formalism, the gauge invariant action or Lagrangian for this curvature, may be defined as \cite{wei2}:
\b \label{gias2r31gg} S_2[{\cal K}]=\int  d\mu(x)\left(R_{\alpha\beta}^{\;\;\;\;\;A}Q_{AB}R^{\alpha\beta B}\right).\e
$Q_{AB}$ is a numerical constant. 

Since in the general relativity the difference of two connections is a tensor then with the ten vector fields ${\cal K}_\alpha^{\;\;A}$, we try to construct a tensor field ${\cal K}_\alpha^{\;\;A}\equiv {\cal K}_\alpha^{\;\;\beta\gamma}=-{\cal K}_\alpha^{\;\;\gamma\beta}$. For obtaining a well defined tensor on dS hyperboloid the following subsidiary conditions on the tensor field ${\cal K}_\alpha^{\;\;\beta\gamma}$ must be imposed:
\b \label{subcon14} x_\beta{\cal K}_\alpha^{\;\;\beta\gamma}=0,\;\;\;\;\;x_\gamma{\cal K}_\alpha^{\;\;\beta\gamma}=0.\e 
These conditions (\ref{subcon14}) also preserve the transversality condition for the gauge-covariant derivative. In this case, the combination of the $10$ gauge vector fields ${\cal K}_\alpha^{\;\;A}$ can be considered as a tensor field of rank-$3$ with ${\cal K}_\alpha^{\;\;\beta\gamma}=-{\cal K}_\alpha^{\;\;\gamma\beta}$ on the dS hyperboloid, with $24$ degrees of freedom in the ambient space formalism.

\subsection{Vector-spinor gauge theory}

Now, we consider the vector-spinor gauge field $\Psi_\alpha(x)$ ($j=p=\frac{3}{2}$). From the group theoretical method, using the Casimir operator (\ref{casimirj}) the gauge invariant vector-spinor field equation can be given in the following form \cite{paenta,fatata,azam}:
\b \label{gifes32} \left(Q_{\frac{3}{2}}^{(1)}+\frac{5}{2}\right) \Psi_\alpha+ \nabla^\top_{ \alpha}\partial^\top\cdot\Psi=0.\e
By using the identities
\b \label{cd32}
 Q^{(1)}_{\frac{3}{2}}\nabla_{\alpha}^\top\psi=\nabla_{\alpha}^\top Q^{(1)}_{\frac{1}{2}}\psi,\;\;\; \partial^\top \cdot \nabla^\top\psi=- \left(Q_{\frac{1}{2}}^{(1)}+\frac{5}{2}\right)\psi, \e
one can show the field equation (\ref{gifes32}) is invariant under the gauge transformation \cite{azam,fatata}
$$\Psi_\alpha \longrightarrow \Psi^g_\alpha=\Psi_\alpha+\nabla^\top_{\alpha } \psi.$$
$\psi$ is an arbitrary spinor field and $\nabla_{\alpha}^\top\psi=\left(\partial^\top_\alpha+\gamma^\top_\alpha\not
x\right)\psi$. The field equation (\ref{gifes32}) can be derived from the action:
\begin{equation} \label{as32g}
S_{\frac{3}{2}}[\Psi, \tilde{\Psi}]=\int d\mu(x)\left[\left(\tilde{\nabla}^\top_\alpha\tilde{ \Psi}_\beta-\tilde{\nabla}^\top_\beta\tilde{ \Psi}_\alpha\right)\left( \nabla^{\top\alpha} \Psi^\beta-\nabla^{\top\beta} \Psi^\alpha\right)\right],\end{equation} 
where $\tilde{\Psi}_\alpha=\Psi^\dag_\alpha \gamma^0$ and 
$\tilde{\nabla}^\top_\alpha \tilde{\Psi}_\beta=\partial^\top_\alpha \tilde{ \Psi}_\beta- x_\beta\tilde{\Psi}_\alpha $
is defined in section IV-D. The vector-spinor Lagrangian in equation (\ref{as32g})is invariant under the two types of gauge invariant \cite{paenta},
$$ \Psi_\alpha \longrightarrow \Psi^g_\alpha=\Psi_\alpha+\nabla^\top_{\alpha } \psi=\Psi_\alpha+\partial^\top_{\alpha } \psi+\gamma^\top_\alpha \not x \psi,$$ $$\tilde{\Psi}_\alpha \longrightarrow \tilde{\Psi}^g_\alpha=\tilde{\Psi}_\alpha+\tilde{\nabla}^\top_{\alpha } \tilde{\psi}=\tilde{\Psi}_\alpha+\partial^\top_{\alpha } \tilde{\psi},$$ which will be considered in section VII-E.

What kind of gauge symmetrical group can be associated to this gauge transformation? Now we try to obtain the action (\ref{as32g}) by using the gauge theory. The gauge potential in the present case is a spinor field which satisfies the Grassmann algebra. Correspondingly, the involved symmetry group includes spinorial generators (generators with the anti-commutation relations). Assuming that there are $N$ gauge vector-spinor fields ($\Psi_\alpha^{\;\;n}$, with $n=1,..,N$), one can define the gauge-covariant derivative, similar to the other previous cases, as
$$ D^\Psi_\beta =\nabla_{\beta}^\top +i\left(\Psi_{\beta}^{\;\;n} \right)^\dag \gamma^0 {\cal Q}_n. $$
$\nabla_{\beta}^\top$ is the transverse-covariant derivative on tensor-spinor field (\ref{cdsa}). The generators ${\cal Q}_A$ are spinor fields, satisfying some anti-commutation relations. The super-algebra in the dS ambient space formalism naturally appear. 

A brief discussion of the simple case $N = 1$, would be utterly instructive. The gauge-covariant derivative for this case is:  
$$ D^\Psi_\beta =\nabla_{\beta}^\top +i \left(\Psi_{\beta}\right)^\dag \gamma^0 {\cal Q}=\nabla_{\beta}^\top+i\left(-\bar{\Psi}_{\beta}\gamma^4\right)^{i} {\cal Q}_i, $$
where $i=1,...,4$ is the spinorial index. Since $\Psi_{\beta}^\dag \gamma^0\Psi^\beta$ is a scalar field \cite{bagamota} then one needs to add a spinor generator ${\cal Q}$. The difference between this case and the other cases ($SU(N), SO(1,4), SO(2,4)$) is that the super-algebra between the Grassmanian generators is not closed. The product of two Grassmanian numbers becomes a usual number. So, for obtaining a closed super-algebra, one should couple these generators with the dS group generators $L_{\alpha\beta}$. The $N=1$ super-algebra in the dS ambient space formalism is already calculated \cite{parota}:
\b \{{\cal Q}_i,{\cal Q}_j\}=\left(S^{(\frac{1}{2})}_{\alpha\beta}\gamma^4
\gamma^2\right)_{ij}L^{\alpha\beta},\e
\b
[{\cal Q}_i,L_{\alpha\beta}]=\left(S^{(\frac{1}{2})}_{\alpha\beta}{\cal Q} \right)_i, \;\; \;\;[\tilde
{\cal Q}_i,L_{\alpha\beta}]=-\left(\tilde {\cal Q} S^{(\frac{1}{2})}_{\alpha\beta}\right)_i,\e
\b
[L_{\alpha\beta}, L_{\gamma\delta}] =
-i(\eta_{\alpha\gamma}L_{\beta\delta}+\eta_{\beta\delta}
L_{\alpha\gamma}-\eta_{\alpha\delta}L_{\beta\gamma}-\eta_{\beta\gamma}
L_{\alpha\delta}),\e 
where $\tilde{{\cal Q}}_i=\left({\cal Q}^t \gamma^4
C\right)_i=\bar{{\cal Q}}_i$, and ${\cal Q}^t$ is the transpose of ${\cal Q}$. The charge conjugation $C$ is defined in section IV-B. It
can be shown that $\tilde{{\cal Q}}\gamma^4 {\cal Q}$ is a scalar field under
the dS transformation \cite{morrota}. Then for defining the gauge-covariant derivative, one use the above $N=1$ super-algebra. Consequently, the gauge fields are ${\cal H}_\alpha^{\;\;{\cal A}}\equiv \left( {\cal K}_\alpha^{\;\;\beta\gamma}, \tilde{\Psi}_\alpha^{\;\;i}\right)$, along with the generators are $ Z_{\cal A} \equiv  \left( L_{\alpha\beta}, {\cal Q}_i\right)$: 
\b  D^{\cal H}_\beta =\nabla^\top_\beta +i {\cal H}_{\beta}^{\;\; {\cal A}} Z_{\cal A}. \e
One can rewrite this $N=1$ super-algebra as the following form:
$$[Z_{\cal A},Z_{\cal B}\}={\cal C}_{{\cal B}{\cal A}}^{\;\;\;\;\;{\cal C}}Z_{\cal C}.$$
$[Z_{\cal A},Z_{\cal B}\}$ is a commutation or an anti-commutation relation. Under a local infinitesimal gauge transformation generated by $\epsilon^{\cal A}(x)Z_{\cal A} $, one has
$$\delta_\epsilon {\cal H}_\beta^{\;\;{\cal A}}=D^{\cal H}_\beta \epsilon^{\cal A} =\nabla^\top_\beta \epsilon^{\cal A} + {\cal C}_{{\cal B}{\cal C}}^{\;\;\;\;\;{\cal A}}{\cal H}_\beta^{\;\;{\cal C}} \epsilon^{\cal B}.$$
The curvature ${\cal R}$, with values in this super-algebra is defined by:
$${\cal R}(D^{\cal H}_\alpha,D^{\cal H}_\beta)=-[D^{\cal H}_\alpha,D^{\cal H}_\beta\}=R_{\alpha\beta}^{\;\;\;\;{\cal A}}Z_{\cal A},$$
where $R_{\alpha\beta}^{\;\;\;\; {\cal A}}=(R^{dS})_{\alpha\beta}^{\;\;\;\; A}+(R^{{\cal H}})_{\alpha\beta}^{\;\;\;\; {\cal A}}$. The dS background curvature for vector-spinor field is:
$$  [\nabla^\top_\alpha , \nabla^\top_\beta]\Psi_\gamma=- \left[\left(\theta_{\alpha\gamma}-x_\gamma\gamma^\top_\alpha \not x\right)\theta_{\beta}^{\;\;\delta}-\left(\theta_{\beta\gamma}-x_\gamma\gamma^\top_\beta \not x\right)\theta_{\alpha}^{\;\;\delta}\right]\Psi_\delta \equiv - ({\bf R}^{dS})^{\delta}_{\;\;\gamma\alpha\beta} \Psi_\delta,$$ 
and we have $({\bf R}^{dS})^{\gamma}_{\;\;\beta\gamma\delta}= ({\bf R}^{dS})_{\beta\delta}=3\left(\theta_{\beta\delta}-x_\beta\gamma^\top_\delta\not x\right)$ and $ ({\bf R}^{dS})=12\1$.
For the tensorial part the curvature $(R^{{\cal H}})_{\alpha\beta}^{\;\;\;\; {\cal A}}$ becomes exactly the curvature for gauge dS group (\ref{cuso14}) and for spinorial part the curvature is:
$$ \left(R^\Psi\right)_{\alpha\beta}^{\;\;\;\;\; i}=\nabla^\top_\alpha \Psi_\beta^{\;\;i}-\nabla^\top_\beta \Psi_\alpha^{\;\;i}+{\cal H}_\beta^{\;\;{\cal B}}{\cal H}_\alpha^{\;\;{\cal C}}{\cal C}_{{\cal B}{\cal C}}^{\;\;\;\;\;i},$$
where the gauge potential ${\cal K}_\alpha^{\;\;\beta\gamma}$ appears in the non-linear terms. 

The super-gauge invariant action or Lagrangian in the dS ambient space formalism for these gauge fields, $\Psi_\alpha^{\;\;i}$ and ${\cal K}_\alpha^{\;\;\beta\gamma}$, is \cite{paenta,mama,van,frts}:
$$ S_{\frac{3}{2}}[\Psi,{\cal K}]=\int d\mu(x) \left(\tilde{R}^\Psi\right)_{\alpha\beta }\left(R^\Psi\right)^{\alpha\beta }\;\;\;\;\;\;\;\;\;\;$$ \b \label{gias321}=\int d\mu(x)\left[\left(\tilde{\nabla}^\top_\alpha\tilde{ \Psi}_\beta-\tilde{\nabla}^\top_\beta\tilde{ \Psi}_\alpha\right)\left( \nabla^{\top\alpha} \Psi^\beta-\nabla^{\top\beta} \Psi^\alpha\right)+O\left(\Psi^3,\Psi^2{\cal K},\Psi {\cal K}^2\right)\right],\e where $$ \tilde{\left(R^\Psi\right)}_{\alpha\beta}^{\;\;\;\;\; i}=\tilde{ \nabla}^\top_\alpha \tilde{\Psi}_\beta^{\;\;i}-\tilde{\nabla}^\top_\beta \tilde{\Psi}_\alpha^{\;\;i}+{\cal H}_\beta^{\;\;{\cal B}}{\cal H}_\alpha^{\;\;{\cal C}}{\cal C}_{{\cal B}{\cal C}}^{\;\;\;\;\;i}.$$
The first part of the action integral is the action (\ref{as32g}) and the linear field equation is exactly the gauge invariant field equation (\ref{gifes32}). One cannot establish the gauge-covariant Lagrangian only with a vector-spinor field $\Psi_\alpha$, since the spinor generator ${\cal Q}$ does not have a closed super-algebra and one should couple the vector-spinor field with a rank-$3$ tensor field ${\cal K}_{\alpha\beta\gamma}$. Consequently, such gauge potential
cannot be considered as a new force but it may be considered as a part of the gravitational field.

The gauge fixing terms are added to the Lagrangian and in the linear approximation the gauge fixing field equation becomes:
\b \label{gffer32m}  \left(Q_{\frac{3}{2}}^{(1)}+\frac{5}{2}\right) \Psi_\alpha+ c\nabla^\top_{\alpha}\partial^\top\cdot\Psi=0.\e 
Now, we have the gauge transformations and the gauge invariant field equations for massless fields with spin $1$, $\frac{3}{2}$ and $2$ in the linear approximation. In the next section, our survey continues with the study of the massless quantum field theory and the calculation of its two-point functions.


\setcounter{equation}{0}
\section{Massless quantum field theory}

In this section, the massless quantum field operators are studied. There are two conditions for defining a massless field in the dS space: 
\begin{itemize}
\item
$(1)$ the massless field operator correspond to $j=p=0$ and $j=p=\frac{1}{2}$ must transform by the UIR of the dS group. It corresponds to the massless field of the Minkowskian space in the null curvature limit, 
\item
$(1)'$ the massless field operator correspond to $j=p=1,\frac{3}{2}$ and $2$ must transform as an indecomposable representation of dS group, in which its central part imitates the UIR of the discrete series ($\Pi^+_{j,j} \oplus \Pi^-_{j,j}$) and, in the null curvature limit, this massless field operator corresponds to the massless field of the Minkowskian space-time,  
\item
$(2)$ the massless field must propagate on the dS light cone or, in other words, the propagator must be conformal invariant, implying the existence of an extension of the UIR of the dS group to the conformal group $SO(2,4)$. 
\end{itemize}
For scalar field $(j=0)$, one recognizes two important possibilities; the massless conformally coupled scalar field $p=0$, and the massless minimally coupled scalar field $p=2$. The first case satisfies the two above conditions  $(1)$ and $(2)$ \cite{brmo}, where the second case does not satisfy the above conditions, so it does not transform as an UIR of the dS group and therefore, to validate the covariant quantization procedure, one needs to introduce an indecomposable representation of dS group. Previously, the quantization in the Krein space was presented \cite{gagata}. The minimally coupled scalar field is an auxiliary field, which appears in the indecomposable representation of the vector field and spin-$2$ field. 

The spinor field $j=p=\frac{1}{2}$ satisfies the above conditions $(1)$ and $(2)$ \cite{bagamota}. This massless spinor field was previously introduced in \cite{tak,bagamota} and can simply be considered as the limiting case of the massive spinor field by replacing $(\nu\longrightarrow 0)$ \cite{ta97,bagamota}. The massless field with $j=p\geq 3$ cannot be visualized in the ambient space formalism since the homogeneity degree of plane wave becomes positive (\ref{hdllp}), (\ref{hdmjd}) and the plane wave is infinite for the large values of $x$ (\ref{pws2i}). Additionally, the mass parameter associated with these fields has an imaginary value. 

Three important cases are considered here, namely, $j=p=1, \frac{3}{2}$ and $2$. But for these cases, one encounters the gauge invariance which means the field operator does not satisfy the divergenceless condition and fixing the gauge becomes mandatory. For $j=p=1, \frac{3}{2}$, one can construct the field operators that satisfy the above conditions $(1)'$ and $(2)$ \cite{gagarota,fatata,fahuqere}. For $j=p=2$, a rank-$2$ symmetric tensor field can be quantized but the theory is not the conformal invariant. Therefore, to preserve the dS and the conformal covariants simultaneously for $j=p=2$ case, one must use a rank-$3$ mixed symmetry tensor field \cite{tata}. In what follows, a brief review of quantization of these massless fields will be presented. 

\subsection{Massless conformally coupled scalar field}

Since the various spin massless fields and the auxiliary fields in the discrete series can be constructed in terms of the massless conformally coupled scalar field, then the field operator, the quantum states and the two-point function of conformally coupled scalar field are explicitly constructed in this subsection. The massless conformally coupled scalar field satisfies the following field equation \cite{brmo,ta97}: 
$$ \left(Q_0^{(1)}-2\right) \phi_c(x)=0, \mbox {or} \;\;\left(\square_H+2H^2\right) \phi_c(x)=0.$$
This field corresponds to the complementary series representation of the dS group with $j=p=0$ (\ref{comserp}) which is unitary equivalent with the representation $j=0,\;\;p=1$ (\ref{comserpueq}). This representation was constructed on ${\bf u}-$space or the three-sphere $S^3$, utterly similar to the principal series representation for the scalar case $j=0,\;\;p=\frac{1}{2}+i\nu$  \cite{dix,tak}. The associated mass is $m^2_{b,1}=2H^2$ and the homogeneous degrees of this field are
$ \lambda=-1, -2$. This field satisfies the conditions $(1)$ and $(2)$ for the massless fields. This field can be obtained by replacing the parameter $\nu$ in the principal series representation for massive scalar field by $\nu=\pm \frac{i}{2}$ $(p=0,1)$. In this limit, one can simply write the quantum field operator and the two-point function. Therefore the massless minimally coupled scalar field operator is:
\b \label{scaqfopml} \phi_c(x)=\int_{S^3}d\mu({\bf u})\left[a({\bf \tilde{u}},0;0,1){\cal U}(x;{\bf u},0;0,0)+a^\dag({\bf u},0;0,0){\cal V}(x;{\bf u},0;0,0 )\right],\e
where the coefficients ${\cal U}$ and ${\cal V}$ can be written in terms of $\xi_u$:
\b \label{tscsf}  {\cal U}(x;{\bf u},0;0,0)=\sqrt{ c_0 }(x\cdot\xi_u)^{-2}, \;\;\; \;\;{\cal V}(x;{\bf u},0;0,0)=\sqrt{ c_0 } (x\cdot\xi_u)^{-1}.\e
$c_0$ is the normalization constant. But this field operator cannot be defined properly on dS space-time. Therefore the quantum field operator is presented in the complex dS space-time $X_H^{(c)}$ \cite{brgamo,brmo}
$$\Phi_c(z)=\sqrt{ c_0 }\int_{S^3}d\mu({\bf u})\left[a({\bf\tilde{u}},0;0,1){\cal U}(z;{\bf u},0;0,0 )+a^\dag({\bf u},0;0,0){\cal V}(z;{\bf u},0;0,0 )\right],$$ 
and consequently, the quantum field operator in this notation is given by:
$$ \phi_c(x)=\sqrt{ c_0 }\int_{S^3}   d\mu({\bf u}) \left\lbrace\; a({\bf\tilde{u}},0;0,1)[(x\cdot\xi_u)_+^{-2}
        +e^{-i\pi(-2)}(x\cdot\xi_u)_-^{-2}]\right.\;\;\;\;\;\;\;\;\;\;\;\;\;\;\;\;\;\;\;\;\;\;\;\;\;\;\;\;\;\;$$
     \b\label{foinam}\;\;\;\;\;\;\;\;\;\;\;\;\;\;\;\;\;\;\;\; \;\;\;\;\;\;\;\;\;\;\;\;\;\;\;\;\;\;\;\;\;\;\ \left. +a^\dag({\bf u},0;0,0)[(x\cdot\xi_u)_+^{-1}
        +e^{i\pi(-1)}(x\cdot\xi_u)_-^{-1}]\; \right\rbrace .\e
It is interesting to note that one can construct the field operator on the closed unit ball $B$ or ${\bf q}$-space:
$$ \phi_c(x)=\sqrt{ c'_0 }\int_{B}   d\mu({\bf q}) \left\lbrace\; a({\bf\tilde{q}},0;0,1)[(x\cdot\xi_B)_+^{-2}
        +e^{-i\pi(-2)}(x\cdot\xi_B)_-^{-2}]\right.\;\;\;\;\;\;\;\;\;\;\;\;\;\;\;\;\;\;\;\;\;\;\;\;\;\;\;\;\;\;\;$$
     \b\label{foinamb} \;\;\;\;\;\;\;\;\;\;\;\;\;\;\;\;\;\;\;\;\;\;\;\;\;\;\;\;\;\;\;\;\;\;\;\;\;\;\;\;\;\;\;\ \left. +a^\dag({\bf q},0;0,0)[(x\cdot\xi_B)_+^{-1}
        +e^{i\pi(-1)}(x\cdot\xi_B)_-^{-1}]\; \right\rbrace ,\e
since $\xi_B\equiv \xi_u$ and $d\mu({\bf q})=2\pi^2r^3dr d\mu({\bf u})$.  
These two field operators are equivalent up to a normalization constant which can be fixed by the Hadamard conditions. The field operator (\ref{foinamb}) can be used for construction the other massless fields in the dS ambient space formalism. 

The analytic two-point function for scalar field is \cite{brgamo,brmo}:
\b  \label{tpfscinint}  W_c(z_1,z_2)=\left<\Omega|\Phi(z_1)\Phi(z_2)|\Omega\right>=c_0\int_{S^3}d\mu({\bf u}) (z_1\cdot\xi_u)^{-2}(z_2.\cdot\xi_u)^{-1},\e 
and $c_0=c_{0,\nu}$ is obtain by replacing $\nu=\pm\frac{i}{2}$ in (\ref{ncsv}). One can easily calculate (\ref{tpfscinint}) in terms of the generalized Legendre function (\ref{stpci}):
$$ W_c(z_1,z_2)=\frac{-iH^2}{2^4\pi^2} P_{-1}^{(5)}(H^2 z_1\cdot z_2)=
\frac{H^2}{8\pi^2}\frac{-1}{1-{\cal Z}(z_1,z_2)}. $$
The vacuum state $|\Omega>$ for this normalization constant is exactly the Bunch-Davies vacuum state \cite{chta}. The Wightman two-point function ${\cal W}_c(x_1,x_2)$ is the boundary value (in the sense of its interpretation as a distribution function, according to the theorem A.2 in \cite{brmo}) of the function $W_c(z_1, z_2)$ which is analytic in the domain ${\cal T}_{12}$ of $M_H^{(c)}\times M_H^{(c)}$ (\ref{tuboid}). The boundary value is defined for
$z_1 =x_1+iy_1\in {\cal T}^-$ and
$z_2=x_2+iy_2\in {\cal T}^+$ by
   $$ {\cal Z}(z_1,z_2)={\cal Z}(x_1,x_2)-i\tau\epsilon(x_1^0,x_2^0),$$
where $y_1=(-\tau,0,0,0,0)\in V^-$, $y_2=(\tau,0,0,0,0)\in V^+$ (\ref{v+-}) and $\tau
\rightarrow 0$. Then, one obtains \cite{brmo,ta97,chta}:
  $$ {\cal W}_c(x_1,x_2)=\frac{-H^2}{8\pi^2}\lim_{\tau \rightarrow 0}\frac{1}{1-{\cal
Z}(x_1,x_2)+i\tau\epsilon(x_1^0,x_2^0)}$$ \b \label{stpci2}
=\frac{-H^2}{8\pi^2}\left[
P\frac{1}{1-{\cal Z}(x_1,x_2)}
 -i\pi\epsilon(x_1^0,x_2^0)\delta(1-{\cal Z}(x_1,x_2))\right],\e
where the symbol $P$ is the principal part. ${\cal Z}$ is the geodesic distance between two points $x$ and $y$ on the dS hyperboloid:
$${\cal Z}(x_1,x_2)=-H^2 x_1\cdot x_2=1+\frac{H^2}{2} (x_1-x_2)^2,  $$
and   \b \epsilon (x_1^0-x_2^0)=\left\{\begin{array}{clcr} 1&x_1^0>x_2^0
 \\
  0&x_1^0=x_2^0\\  -1&x_1^0<x_2^0\\    \end{array} \right. .\e
  
Finally, one can show that the two solutions (\ref{tscsf}) are equivalent in the intrinsic coordinate system. In the $SO(4)$-orbital basis $\left(\xi^\alpha_u=(\xi^0, \xi^0\; {\bf u})\right)$, the relation between the dS plane waves  and the partial waves ( intrinsic coordinate solution) is given by \cite{ta97,gagarota,gasiyo}:
\b \label{plapar}
\left( x\cdot\xi_{u} \right)^{\lambda} =  2\pi^2  \left(\xi^0\right)^{\lambda} \sum_{Llm} \Phi_{Llm}^{\lambda}(X){\cal Y}^{\ast}_{Llm}(u)  ,
\e
where $\Re \lambda<0$ and
$$
 \Phi_{Llm}^{\lambda} (X) = i^{L-\lambda}\, e^{-i(L+\lambda +3)\rho}(2 \cos{\rho})^{\lambda + 3} \frac{\Gamma(L-\lambda)}{(L+1)!\Gamma(-\lambda)} $$ \b
\phantom{\Phi_{Llm}^{\lambda} (x) =}{}  \times {}_2F_1\big(\lambda + 2, L+\lambda +3; L+2;-e^{- 2i\rho}\big)  {\cal Y}_{Llm}(v).\e
$ {\cal Y}_{Llm}$ stands for the hyperspherical harmonics. The conformal global coordinates $$ x^\alpha=(H^{-}\tan \rho, (H\cos\rho)^{-1}{\bf v})$$ are being used, where  ${\bf v}=(v^4, \vec v)$ is a quaternion with the norm of $1$. Using the following relation
$$ {}_2F_1\big(a, b; c;z\big)=(1-z)^{c-a-b}{}_2F_1\big(c-a, c-b; c;z\big)$$
one can show that the two solutions $(x\cdot \xi_u)^{-1}$ and $(x\cdot \xi_u)^{-2}$ in the intrinsic global coordinate are equivalent up to a normalization constant. These two solutions can correspond to the two unitary equivalent representations $U^{(0,0)}(g)$ (\ref{comserp}) and $U^{(0,1)}(g)$ (\ref{comserpueq}).

\subsection{Massless minimally coupled scalar field}

The values $j = 0$ and $p = 2$, which define the massless minimally coupled scalar field, are not permissible, according to the classification of the UIR of the dS group. The corresponding mass parameter is $m^2_{b,2}=0$. The field equation can be written in the following form:
$$ Q_0^{(1)} \phi_m(x)=0,\;\;\; \mbox {or} \;\;\;\square_H \phi_m(x)=0.$$ 
This field equation is similar to the $\xi$-space field equation (\ref{xiesegau}) for $j=p$, and is invariant under the transformation 
$$ \phi'_m(x)=\phi_m(x)+ \mbox{const.}\;.$$ 
The homogeneous degrees of this field are:
$ \lambda=0, -3$. The constant solution $(\lambda_1=0, \;(x\cdot\xi)^0=$constant), poses the zero mode problem. With just one solution $(\lambda_2=-3,\; (x\cdot\xi)^{-3} )$, one cannot establish a proper covariant quantum field operator on the Hilbert space-constructed on an UIR of the dS group  \cite{al,gareta,alfo}. One cannot obtain a massless minimally coupled scalar field and its two-point function, by replacing the parameter $\nu$ of a massive field in the principal series representation with the $\nu= \pm\frac{3i}{2}$, since such replacement would cause a singularity in the normalization constant $c_{0,\nu}$, in the sense that such field operator is not in a correspondence with an UIR of the dS group. Nevertheless, one can associate a massless minimally coupled scalar field with an indecomposable representation of the dS group \cite{gareta}. Using these two following identities 
$$Q_0^{(1)} \partial^\top_\alpha\phi(x)- \partial^\top_\alpha Q_0^{(1)} \phi(x)=2\partial^\top_\alpha \phi(x)+2x_\alpha Q_0^{(1)}\phi(x),$$
$$Q_0^{(1)} x_\alpha\phi(x)- x_\alpha Q_0^{(1)}\phi(x) =-2\partial^\top_\alpha\phi(x)-4x_\alpha\phi(x),$$
with $\phi$ as an arbitrary scalar field, one can prove the existence of a magic relation between the minimally coupled and the conformally coupled scalar fields in the dS ambient space formalism:
\b \label{msfincsf}  \phi_m(x)= N\left[Z\cdot\partial^\top + 2 Z\cdot x\right]\phi_c(x). \e  
$N$ is a normalization constant which can be fixed, using the local Hadamard condition, and $Z^\alpha$ is a constant polarization five-vector. The quantum field operator is defined by:
        $$ U(g)\Phi_m(z,Z)U(g)^{-1}=\Phi_m(\Lambda z,\Lambda Z),$$
where $U(g)$ is a representation of the dS group. Such representation can be constructed as the product of two representations of the dS group: the scalar complementary series representation $j=0$ and $p=0$, and a five-dimensional trivial representation with respect to $Z_\alpha^{(l)}$ \cite{gaha}. For a thorough investigation regarding the five existing polarization states $l = 1, 2, 3, 4, 5$, the reader may refer to \cite{gaha}. This subject will not be pursued here.

Concerning the polarization five-vector $Z_\alpha^{(l)}$, the quantum field operator can be defined properly from the quantum field operator of conformally coupled scalar field:
$$ \Phi_m(z)=\sqrt{ c_0 }N\sum_{l=1}^5\left[Z^{(l)}\cdot\partial^\top + 2 Z^{(l)}\cdot z\right] \int_{S^3}   d\mu({\bf u}) \left\lbrace\; a(
{\bf \tilde{u}},0;0,1)(z\cdot\xi_u)^{-2}
       +a^{\dag}({\bf u},0;0,0)(z\cdot\xi_u)^{-1}
        \; \right\rbrace $$
    $$    =\sqrt{ c_0 }N \sum_{l=1}^5\int_{S^3}   d\mu({\bf u}) \left\lbrace\; a(
{\bf \tilde{u}},0;0,1)\left[-2(Z^{(l)}\cdot\xi_u^\top)(z\cdot\xi_u)^{-3} + 2 (Z^{(l)}\cdot z)(z\cdot\xi_u)^{-2}\right] \right.$$ \b \label{mcsfico}
      \left. +a^{\dag}(
{\bf u},0;0,0)\left[-(Z^{(l)}\cdot\xi_u^\top)(z\cdot\xi_u)^{-2} + 2 (Z^{(l)}\cdot z)(z\cdot\xi_u)^{-1}\right]
        \; \right\rbrace\equiv \Phi_m(z,Z).\e      

The analytic two-point function can be defined on the vacuum state of the conformally coupled scalar field or Bunch-Davies vacuum state:
$$ W_{m}^H(z,z') =<\Omega| \Phi_m(z)\Phi_m(z')|\Omega>$$ $$= \sum_{l=1}^5 \sum_{l'=1}^5\left[Z^{(l)}\cdot\partial^\top + 2 Z^{(l)}\cdot z\right]\left[Z^{(l')}\cdot\partial'^\top + 2 Z^{(l')}\cdot z'\right] W_c(z,z').$$
The explicit form of this function depends on the representation $U(g)$. As a simple case, one can choose:
$$\sum_{l=1}^5 \sum_{l'=1}^5Z^{(l)}_\alpha Z^{(l')}_\beta=\eta_{\alpha\beta},$$ which will be concluded to the following analytic two-point function \cite{khrota}
\b\label{tpfmcsh1}  W_{m}^H(z,z')\equiv \left[\partial^\top\cdot\partial'^\top +2z\cdot\partial'^\top + 2 z' \cdot \partial^\top+ 4 z\cdot z'\right] W_c(z,z'),\e
with $W_c$ being the analytic two-point function of conformally coupled scalar field (\ref{stpci}). By using the following relations $${\cal Z}=-H^2z\cdot z',\;\;\;\frac{\partial}{\partial z^\alpha}=-H^2z'_\alpha\frac{d}{d{\cal Z}}, \;\;\;\; \partial^\top_\alpha=\left(z_\alpha {\cal Z}-z'
_\alpha\right)\frac{d}{d{\cal Z}},$$ 
$$ \partial^\top\cdot\partial'^\top=\left(-3+{\cal Z}^2\right)\frac{d}{d {\cal Z}}+{\cal Z}\left(1-{\cal Z}^2\right)\frac{d^2}{d {\cal Z}^2},\;\; z'\cdot\partial^\top=\left(1-{\cal Z}^2\right)\frac{d}{d{\cal Z}},$$
one can show that this two-point function also satisfies the minimally coupled scalar field equation for the variables $z$ and $z'$.
In conclusion, the analytic two-point function (\ref{tpfmcsh1}) is free of any infrared divergences. The two-point function in the real dS space is the boundary value of the analytic two-point function $ W_{m}^H(z,z')$ \cite{khrota}
 \b \label{mth} {\cal W}_{m}^H(x,x') = \sum_{l=1}^5 \sum_{l'=1}^5\left[Z^{(l)}\cdot\partial^\top + 2 Z^{(l)}\cdot x\right]\left[Z^{(l')}\cdot\partial'^\top + 2 Z^{(l')}\cdot x'\right]{\cal W}_c(x,x').\e

For the sole purpose of a covariant quantization of such fields, an entirely new method of quantization, "the Krein space quantization" was developed, based on the definition of the field operator in the intrinsic coordinates, preserving the dS invariance and commencing transformations as a specific indecomposable representation of the dS group \cite{gareta}. Nevertheless, this method has been ignored here, since it breaks the analyticity and uses the intrinsic coordinates contrary to the present formalism which has been established in the ambient space. The reader is encouraged to refer to the previously published papers for a detailed study \cite{ta97,gareta}. Nevertheless, for the sake of an interesting comparison, the two-point function of the massless minimally coupled scalar field in the Krein space quantization is presented \citep{ta01}:
 \b \label{tfsmin} {\cal W}_{m}^K(x,x')=\frac{iH^2}{8\pi} \epsilon (x^0-x'^0)[\delta(1-{\cal
 Z}(x,x'))-\theta ({\cal Z}(x,x')-1)], \e
where $\theta$ is the Heaviside step function. This expression is dS invariant, i.e. coordinate independent and also free of any infra-red divergence. Unfortunately, since the propagation is placed inside the dS light-cone, the appearance of constant term in the two-point function (Heaviside step function), breaks the conformal invariance.

\subsection{Massless spinor field}

The massless spinor field is defined by $j=p=\frac{1}{2}$ and transforms as the UIR of the discrete series $T^{0,\frac{1}{2};\frac{1}{2}}$ and $T^{\frac{1}{2},0;\frac{1}{2}}$ \cite{tak} (or $\Pi^\pm_{\frac{1}{2},\frac{1}{2}}$ in Dixmier notation \cite{dix}). Its corresponding mass parameter and eigenvalue of the Casimir operator are $m^2_{f,\frac{1}{2}}=2H^2$ and $ <Q^{(1)}_{\frac{1}{2},\frac{1}{2}}>=\frac{3}{2}$, respectively. This field satisfies the conditions (1) and (2) as well as the following field equations:
      \b \label{s12mles} \left(Q^{(1)}_0-2\right)
     \psi(x)=0, \;\; \mbox{and} \;\;\left(\not x \not \partial^\top -2\right)\psi(x)=0.\e
The equations (\ref{s12mles}) are invariant under the following gauge transformation
 \b \psi \longrightarrow \psi'=\psi+ H\not x \psi.\e
Through definition of the spinor fields: \cite{gu,gule}
$$ \psi_L(x)=\frac{1+H\not x}{2}\psi(x),\;\;\; \psi_R(x)=\frac{1-H\not x}{2}\psi(x),$$
which are also independent solutions for the field equations (\ref{s12mles}), one obtains $H\not x\psi_L(x)=\psi_R(x)$. The massless spinor field can be simply extracted, replacing the parameter $\nu$ in the quantum field operator and the two-point function for a massive spinor field in the principal series representation with the value of $\nu=0$.

Using the following identity in dS ambient space formalism:
$$ \left(\not x \not \partial^\top-2\right)\left(-\not x \not \partial^\top+1\right)=\left(-\not x \not \partial^\top+1\right)\left(\not x \not \partial^\top-2\right)= Q^{(1)}_0-2,$$
one can write the massless spinor field in terms of the massless conformally coupled scalar field
\b \label{sinsc} \psi(x)= \left(-\not x \not \partial^\top+1\right){\cal U} \phi_c(x),\e
where ${\cal U}$ is a constant spinor, it can be simply fixed in the null curvature limit \cite{bagamota}. The charged spinor field operator can be defined as:
\b \label{spinqfopmasles} \psi(x)=\int_{B}d\mu({\bf q})\sum_{m}\left[a({\bf \tilde{q}},m;\frac{1}{2},\frac{1}{2})U(x;{\bf q},m;\frac{1}{2},\frac{1}{2} )+a^{c\dag}({\bf q},m;\frac{1}{2},\frac{1}{2})V(x;{\bf q},m;\frac{1}{2},\frac{1}{2} )\right],\e
with $m=-\frac{1}{2},\frac{1}{2}$ and $\psi, \; U$ and $V$ being four-component spinors. The homogeneous degrees of the spinor field are $\lambda= -2, \; -1$  (\ref{hdmjd}):
$$ U(x;{\bf q},m;\frac{1}{2},\frac{1}{2} )=(x\cdot\xi_B)^{-2}u(x,{\bf q},m;\frac{1}{2},\frac{1}{2}),$$ $$  V(x;{\bf q},m;\frac{1}{2},\frac{1}{2} )=(x\cdot\xi_B)^{-1}v(x,{\bf q},m;\frac{1}{2},\frac{1}{2}) . $$ 
The analytic two-point function for a spinor field, equation (\ref{tscon}), was previously calculated in \cite{bagamota}:
$$  S_c(z_1,z_2)=\frac{1}{16\pi^2} D_{\frac{1}{2}}(z_2)
                \gamma^4 P_{-1}^{(5)}(H^2z_1.z_2)=\frac{iH^2}{2\pi^2}\frac{(\not z_1-\not z_2)\gamma^4}{[(z_1-z_2)^2]^2}, $$
where $D_{\frac{1}{2}}(z_2)=-\not z_2
\not \partial^\top_{z_2}+1$.
The Wightman spinor two-point function can be written in terms of the Wightman two-point function for a conformally coupled scalar field (\ref{stpci2}): 
   \b {\cal S}_c(x,y)=\frac{H^2}{16\pi^2}D_{\frac{1}{2}}(y)\gamma^4\left[ P \frac{1}{1-{\cal
        Z}(x,y)}-i\pi \epsilon (x^0-y^0)\delta(1-{\cal Z}(x,y))\right].\e

\subsection{Massless vector field}

The massless vector field corresponds to $j=p=1$ with the corresponding eigenvalue of Casimir operator being $<Q_{1,1}^{(1)}>=0$. The associated mass parameter is $m^2_{b,1}=2H^2$. This field corresponds to the discrete series representation $\Pi^\pm_{1,1}$ and the field equation
$$ Q_1^{(1)} K(x)=0,\;\;\;\mbox{or}\;\;\; \left(Q_0^{(1)}-2\right) K(x)=0.$$
Nevertheless, by using the conditions (i-vii) from section II-B, the above field equation cannot be solved. In other words, the quantum states $|{\bf q};1,1>$  cannot be fixed, in the ${\bf q}$-space, by solving the field equation  (\ref{difequj=p}). One needs to drop the divergenceless condition in order the field equation becomes gauge invariant. In the previous section, utilizing the gauge principle, the gauge invariant field equation, in the $x$-space, was obtained (\ref{qifev}):
\b \label{augievec1} Q_1^{(1)} K(x)+cD_1^\top \partial\cdot K(x)=0,\e 
where $D_1^\top=H^{-2}\partial^\top$. For $c=1$, this equation is invariant under the gauge transformation: 
\b \label{gtvf} K(x)\Longrightarrow K^g_\alpha(x)=K_\alpha(x)+D_{1\alpha}^\top \Omega(x),\e 
with $\Omega(x)$ as an arbitrary scalar field. In this case the field equation (\ref{augievec1}) replaces the condition (i) of the section II-D and disables the condition (iv). The gauge invariance can be checked simply by using the two following identities:
$$ Q_1^{(1)} D_1^\top \Omega(x)= D_1^\top Q_0^{(1)} \Omega(x),\;\;\; \partial\cdot D_{1\alpha}^\top \Omega(x)=-Q_0^{(1)} \Omega(x) .$$

By taking the divergence of the equation (\ref{augievec1}), one obtains \cite{gagarota}:
$$ (1-c) Q_0 \partial \cdot K=0.$$ The $\partial.K=\Phi$ is a scalar state, in which for the case $c\neq 1$ yields $Q_0 \Phi=0$. Similarly the divergence of equation (\ref{gtvf}) with $\partial \cdot K=0$, one obtains  $Q_0 \Omega=0$. Combining these two states with the physical states results in the Gupta-Bleuler triplet \cite{gagarota}. Let us now define the Gupta-Bleuler triplet $V_g \subset V \subset V_c$ as the transporter of the indecomposable structure for the representation of the de Sitter group, which appears in our problem.
$V_c/V$ is the scalar state $\Phi$ which is a minimally coupled scalar field. $V_g$ is the gauge state $\Omega$ which is also a minimally coupled scalar field. $V/V_g$ are the physical states $\Pi^+_{1,1}\oplus  \Pi^-_{1,1}$. Then the gauge fixing field equation (\ref{augievec1}), the gauge transformation (\ref{gtvf}) and  the UIR of the discrete series $\Pi^\pm_{1,1}$ will thoroughly define the covariant massless vector field in the dS space-time (see \cite{gagarota}).

The massless vector field, similar to the other massless fields, can be written in terms of the massless conformally coupled scalar field in the ambient space formalism \cite{gagarota}:
\begin{equation}\label{sol}
K_\alpha(x)= \left[Z
_\alpha^\top-\frac{c}{2(1-c)}D_{1\alpha}\left(H^2x\cdot Z
+Z\cdot\partial^\top
\right)+\frac{2-3c}{1-c}H^2D_{1\alpha}\left[Q_0^{(1)}\right]^{-1}x\cdot Z
\right]\phi_c.
\end{equation}
The field operator can be introduced as:
\b \label{vectqfopmasles} K_\alpha(x)=\int_{B}d\mu({\bf q})\sum_{m} \left[a({\bf \tilde{q}},m;1,0){\cal U}_\alpha(x;{\bf q},m;1,1)+a^{\dag}({\bf q},m;1,1){\cal V}_\alpha(x;{\bf q},m;1,1 )\right],\e
where $m=0,1,2,3$. $K_\alpha(x)$ transforms by an indecomposable representation of the dS group. The homogeneous degrees of this field are $\lambda=-1,\;\;-2$ and $ {\cal U}$ and ${\cal V}$ may be written in the following form: 
$$ {\cal U}_\alpha(x;{\bf q},m;1,1 )=(x.\cdot\xi_B)^{-2}u_\alpha(x,{\bf q},m;1,\nu),$$ $$  {\cal V}_\alpha(x;{\bf q},m;1,1 )=(x\cdot\xi_B)^{-1}v_\alpha(x,{\bf q},m;1,1) . $$ 
The coefficients $u_\alpha$ and $v_\alpha$ are obtained by solving the field equation (\ref{augievec1}) but, at the moment, we are not interested in their explicit forms. The reader may follow up this subject in \cite{gagarota}. The analytic two-point function for such massless vector field is
$$ W_{\alpha\alpha'}(z,z')=<\Omega|K_{\alpha}(z)K_{\alpha'}(z')|\Omega>.$$ 
Due to the previous calculations, \cite{gagarota}, it has been shown that:
$$  W_{\alpha \alpha'}(z,z')=\theta_{\alpha }\cdot\theta'_{\alpha' }
W_c(z,z')-\frac{c}{2(1-c)}H^{-2}
\partial^\top_{\alpha}\left[\theta'_{\alpha' }\cdot
\partial^\top  +H^2z\cdot\theta'_{\alpha'
}\right]W_c (z,z') 
$$ \b  +\frac{2-3c}{1-c}
\partial^\top_{\alpha} Q_0^{-1}z\cdot\theta'_{\alpha' }W_c(z,z').
\e
$W_c$ is the two-point function of a massless conformally coupled scalar field (\ref{stpci}). It is proved that the
values 
\b \label{minichoi}  c_s= \frac{2}{2s+1}, \;\; c=\frac{2}{3} \e
correspond to the minimal (or optimal) choice, without any logarithmic singularities \cite{gaha,ta97,gagarota,taro12}

\subsection{Massless vector-spinor field}

The vector-spinor field, associated with the UIR of the discrete series representation $\Pi^\pm_{\frac{3}{2},\frac{3}{2}}$, corresponds to $j=p=\frac{3}{2}$ with the corresponding eigenvalue of Casimir operator being $<Q^{(1)}_{\frac{3}{2},\frac{3}{2}}>=-\frac{5}{2}$. This field satisfies the second order field equation:
$$ \left(Q^{(1)}_{\frac{3}{2}}+\frac{5}{2}\right) \Psi_\alpha(x)=0, \;\; \mbox{or}\;\; \left(Q^{(1)}_0+\frac{i}{2}\gamma_\alpha\gamma_\beta M^{\alpha\beta}-3\right)\Psi_\alpha=0.$$ There are two possibilities for the relevant first order field equation. By using the identity (\ref{spinide}), one obtains:
$$  \left(\not x\not \partial^\top-3\right)\Psi^m_\alpha(x)=0, \;\;\;\;\;\;\;\;\;\; Q^{(1)}_0\Psi^m_\alpha(x)=0,$$
$$  \left(\not x\not \partial^\top-1\right)\Psi^c_\alpha(x)=0, \;\;\;\; \left(Q^{(1)}_0-2\right)\Psi^c_\alpha(x)=0.$$
The mass parameter associated with the first one, $(\Psi^m)$, is $m^2_{\frac{3}{2},\frac{3}{2}}=0$ with its homogeneous degrees being $\lambda=0,\;\;-3$. For the second field, $(\Psi^c)$, the associated mass parameter and the homogeneous degrees of this field are $m^2_{\frac{3}{2},\frac{3}{2}}=2H^2$ and $\lambda=-1,\;\;-2$, respectively. One can describe $\Psi^m$ in terms of $\Psi^c$, using the identity (\ref{msfincsf}). 

By the exclusive use of the conditions (i-vii) from section II-D and because of the appearance of the gauge invariance, one cannot find proper solutions for these field equations. Similar to the case of the vector fields, by using the gauge principle which was obtained in the previous section (\ref{gifes32}), a second order gauge invariant field equation can be obtained:
\b \label{gsvfe} (Q^{(1)}_{\frac{3}{2}}+\frac{5}{2}) \Psi_\alpha(x)+\nabla^\top_{\alpha} \partial^\top \cdot\Psi(x)=0,\e
where $
\nabla^\top_{\alpha}=\partial^\top_\alpha+\gamma^\top_\alpha\not
x$.  One can show
that this equation is invariant under the gauge transformation: \b \label{gsvf}
\Psi_\alpha(x)\rightarrow
\Psi^g_\alpha(x)=\Psi_\alpha(x)+\nabla^\top_{\alpha}\zeta ,\e with
$\zeta(x)$ as an arbitrary spinor field. To provide the gauge
invariance, the following identities are used:
$$Q^{(1)}_{\frac{3}{2}}\nabla^\top_{\alpha}=\nabla^\top_{\alpha}Q^{(1)}_{\frac{1}{2}}\:\:\:,\:\:\:\partial^\top\cdot
\nabla^\top\zeta=-\left(Q^{(1)}_{\frac{1}{2}}+
\frac{5}{2}\right)\zeta.$$ 
Let us introduce a gauge fixing parameter c. The
wave equation now reads as \b \label{s32fegasec}
\left(Q_\frac{3}{2}^{(1)}+\frac{5}{2}\right)\Psi(x)+c\nabla^\top_{\alpha}
\partial^\top\cdot\Psi(x)=0,\e whereas the role of $c$ is just to fix the gauge spinor field $\zeta$.

Up to the first order, there are two gauge invariant field equations. The first one is \cite{fatata}:
\b  \label{s32fega1} (\not x\not \partial^\top-3)\Psi(x) -x_\alpha \not x \not \Psi(x)+ \partial^\top_\alpha \not x \not \Psi(x)=0,\e
which is invariant under the following gauge transformation:
\b 
\Psi_\alpha(x)\rightarrow
\Psi^g_\alpha(x)=\Psi_\alpha(x)+\nabla^\top_{\alpha}\zeta .\e
 The second field equation reads as \cite{fatata}
\b  \label{s32fega2} (\not x\not \partial^\top-1)\Psi(x) -x_\alpha \not x \not \Psi(x)-\nabla^\top_{\alpha} \not x \not \Psi(x)=0.\e
This field equation is invariant under the following gauge transformation:
\b  \Psi_\alpha(x)\rightarrow
\Psi^g_\alpha(x)=\Psi_\alpha(x)+\partial^\top_\alpha\zeta .\e
 
By introducing the massless vector-spinor field operator, one can think of two possibilities; the first choice is to take the equations (\ref{gsvfe}) and (\ref{s32fega1}), and the second, taking the equations (\ref{gsvfe}) and (\ref{s32fega2}). The structure of the Gupta-Bleuler triplet for these two cases are different but the central part or physical states are the same. Similar to the vector field, the pure gauge state $V_g$ and the gauge dependent state $V_c$ can be defined for each case. These two states, plus the physical states, construct the Gupta-Bleuler triplet. Let us now define the Gupta-Bleuler triplet as $V_g \subset V \subset V_c$. $V_c/V$ is the spinor state $\psi_1$ which is a spinor field. $V_g$ is the gauge state $\psi_2$ which is also a spinor field. $V/V_g$ are the physical states $\Pi^+_{\frac{3}{2},\frac{3}{2}}\oplus  \Pi^-_{\frac{3}{2},\frac{3}{2}}$. Then the the gauge fixing field equations, the gauge transformation and the UIR of discrete series $\Pi^\pm_{\frac{3}{2},\frac{3}{2}}$ completely determine the covariant massless vector-spinor field in the dS space.

The vector-spinor field operator can be defined generally as:
\b \label{spinvqfopmasles} \Psi_\alpha(x)=\int_{B}d\mu({\bf q})\sum_{m}\left[a({\bf \tilde{q}},m;\frac{3}{2},-\frac{1}{2})U_\alpha(x;{\bf q},m;\frac{3}{2},\frac{3}{2} )+a^{c\dag}({\bf q},m;\frac{3}{2},\frac{3}{2})V_\alpha(x;{\bf q},m;\frac{3}{2},\frac{3}{2} )\right],\e
where $\Psi_\alpha, \;\;U_\alpha$ and $V_\alpha$ are four-component spinors and we can write: 
$$ U_\alpha(x;{\bf q},m;\frac{3}{2},\frac{3}{2} )=(x\cdot\xi_B)^{-2}u_\alpha(x, {\bf q},m;\frac{3}{2},\frac{3}{2}),$$ $$  V_\alpha(x;{\bf q},m;\frac{3}{2},\frac{3}{2} )=(x\cdot\xi_B)^{-1}v_\alpha(x,{\bf q},m;\frac{3}{2},\frac{3}{2}) . $$ The four-component spinors $u_\alpha$ and $v_\alpha$ can be obtained by imposing the condition that the vector-spinor field must satisfy simultaneously the first and second order field equations which is not important for us here.

The analytic two-point function for vector-spinor field can be written in the following form:
\begin{equation} S_{\alpha \alpha'}(x,x')=
D_{\alpha \alpha'}(x, \partial^\top ;x', {\partial'}^\top;\frac{3}{2})
S_c(x,x'),\label{ts32ml}\end{equation} where
$S_c$ is the massless spinor analytic two-point function (\ref{tscon}) and the explicit form of $D_{\alpha \alpha'}(x, \partial^\top ;x', {\partial'}^\top;\frac{3}{2})$ depends on the structure of the indecomposable representation which is not our concern here \cite{paenta}.

\subsection{Massless spin-$2$ field}

The massless spin-2 field ($j=p=2$) in the dS ambient space formalism may be presented by a rank-2 symmetric tensor field ${\cal H}_{\alpha\beta}$ or a rank-$3$ mixed symmetric tensor field ${\cal K}^M_{\alpha\beta\gamma}$ which  describes the free theories \cite{frhe}. It can be possible to define a map between these two schemes \cite{frhe}. The aim of the present section is to introduce the rank-$ 2$ symmetric tensor field and the rank-$3$ mixed-symmetric tensor field.

\subsubsection{Rank-$2$ symmetric tensor field}

A massless spin-$2$ rank-$2$ symmetric traceless tensor field associates with the discrete series representation $\Pi^\pm_{2,2}$ and satisfies the field equation
$$ \left( Q_2^{(1)}+6\right){\cal H}_{\alpha\beta}=0,\;\;\;\mbox{or}\;\;\; Q^{(1)}_0{\cal H}_{\alpha\beta}=0.$$
The corresponding mass parameter is $m^2_{b,2}=0$. One cannot construct this field from the massive spin-$2$ rank-$2$ symmetric tensor field, only by imposing the limit $\nu=\frac{3}{2}i$. The two-point function (\ref{tpftm2}) in this limit becomes singular, due to the appearance of the gauge invariance.

The homogeneous degrees of the tensor field ${\cal H}_{\alpha\beta}$, are: $ \lambda=0, -3$ (\ref{hdllp}). The solution $\lambda=0$ poses the zero mode problem as well as the appearance of infrared divergence in the process of quantization, just like the case of the minimally coupled scalar field \cite{gareta}. A massless spin-$ 2$ rank-$ 2$ symmetric tensor field can be written in terms of a minimally coupled scalar field $\phi_m$ and polarization tensor ${\cal D}$ \cite{taro12}: 
 $$ {\cal H}_{\alpha \beta}(x)={\cal D}_{\alpha \beta}(x, \partial) \phi_m(x).$$
For calculating the polarization states, the ambient space formalism is needed. In present paper, using the identity (\ref{msfincsf}), a covariant quantization of the minimally coupled scalar fields in the ambient space formalism is being introduced in terms of the conformally coupled scalar fields. 

The procedure of the construction of the Gupta-Bleuler triplet is similar to the massless vector field. Here we briefly review the construction. The gauge fixing field equation 
\b \label{festgi2} \left( Q_2^{(1)}+6\right){\cal H}+cD^\top_2\partial_2^\top. {\cal H}=0,\e 
along with the gauge transformation (\ref{gfst3}) and the UIR of discrete series $\Pi^\pm_{2,2}$ exactly define the covariant massless tensor field in the dS space (see \cite{ta09,taro12}). Applying a divergence operator on the equation (\ref{festgi2}), one obtains \cite{ta09}:
$$ (1-c) Q_1^{(1)} \partial\cdot {\cal H}=0.$$ 
The $\partial\cdot {\cal H}=K_1$ is a vector field in which for $c\neq 1$ yields $Q_1^{(1)} K_1=0$. Imposing the divergenceless condition and taking as ${\cal H}=D_2 K_2$, one obtains  $Q_1^{(1)} K_2=0$. These two vector fields in addition to the physical states construct the Gupta-Bleuler triplet \cite{ta09}.
Let us now define the Gupta-Bleuler triplet $V_g \subset V \subset V_c$ carrying the indecomposable representation of the dS group which appears in our problem. $V_c/V$ is the vector state which is the massless vector field $K_1$. $V_g$ is the gauge state which is also the massless vector field  $K_2$. The space $V/V_g$ contains the physical states $\Pi^+_{2,2}\oplus  \Pi^-_{2,2}$. 

The explicit form of the tensor field depends on the gauge fixing parameter $c$ and for the value $c=\frac{2}{5}$ (\ref{minichoi}) the tensor field becomes \cite{taro12} $$ {\cal H}_{\alpha \beta}(x)={\cal D}_{\alpha \beta}(x, \partial) \phi_m(x),$$
where
$$ {\cal D}(x,\partial,Z_1,Z_2)=\left(-\frac{2}{3}\theta Z_1\cdot+{\cal
S} Z^\top_1+\frac{1}{9 }D^\top_2 (H^2 xZ_1\cdot-Z_1\cdot \partial^\top
+\frac{2}{3}H^2 D^\top_1 Z_1\cdot)\right)$$ \b\left( Z^\top_{2}-\frac{1}{2}
D^\top_{1}(Z_2\cdot
\partial^\top+2H^2x\cdot Z_2)\right),\e
with $Z_1$ and $Z_2$ as two constant five-vectors. $\phi_m(x)$ is a minimally coupled scalar field. By using the identity (\ref{msfincsf}), the tensor field can be written in terms of the conformally coupled scalar field:
\b \label{s2solution} {\cal H}_{\alpha \beta}(x)={\cal D}_{\alpha \beta}(x, \partial)\left[Z_3\cdot\partial^\top + 2 Z_3\cdot x\right]\phi_c(x). \e 

The quantum field operator can be written in the following form:
\b  {\cal H}_{\alpha\beta}(x)=\int_{B}d\mu({\bf q})\sum_{m}\left[a({\bf \tilde{q}},m;2,-1){\cal U}_{\alpha\beta}(x;{\bf q},m;2,2 )+a^\dag({\bf q},m;2,2){\cal V}_{\alpha\beta}(x;{\bf q},m;2,2)\right].\e In terms of the minimally coupled scalar field operator (\ref{mcsfico}), the coefficients ${\cal U}$ and ${\cal V}$ become:
$$ {\cal U}_{\alpha\beta}(x;{\bf q},m;2,2 )=(x\cdot\xi_B)^{-2}u_{\alpha\beta}(x, {\bf q},m;2,2),$$ $$ {\cal V}_{\alpha\beta}(x;{\bf q},m;2,2)=(x\cdot\xi_B)^{-1}v_{\alpha\beta}(x,{\bf q},m;2,2). $$
The coefficients $u_{\alpha\beta}$ and $v_{\alpha\beta}$ can be calculated explicitly by using the equation (\ref{s2solution}).
 
The two-point function can be written in terms of the two-point function of minimally coupled scalar field and a polarization tensor. For $c=\frac{2}{5}$, one has \cite{taro12,ta99,ta09} \b {\cal
W}_{\alpha\beta \alpha'\beta'}(x,x')=\Delta_{\alpha\beta
\alpha'\beta'} (x,x'){\cal W}_{m}(x,x'), \e 
where
$$ \Delta(x,x')=-\frac{2}{3}\Sigma_1'\theta
\theta'\cdot\left(\theta\cdot\theta'
    -{\frac{1}{2}}D^\top_{1}\left[2H^2 x\cdot\theta'+\theta'\cdot\partial^\top\right]\right)$$
$$ +\Sigma_1\Sigma_1'\theta\cdot\theta'\left(\theta\cdot\theta'
    -{\frac{1}{2}}D^\top_{1}\left[2H^2 x\cdot\theta'+\theta'\cdot\partial^\top\right]\right)$$ 
   \b +\frac{H^2}{9}\Sigma_1'D^\top_2 \left(\frac{2}{3}D^\top_1\theta'\cdot + x\theta'\cdot -H^{-2}{\theta'\cdot\partial^\top}\right)\left(\theta\cdot\theta'
    -{\frac{1}{2}}D^\top_{1}\left[2H^2 x\cdot\theta'+\theta'\cdot\partial^\top\right]\right).\e 
 ${\cal W}_{m}$ is the two-point function of the
minimally coupled scalar field. In terms of conformally coupled scalar field it is (\ref{tpfmcsh1}),
and in the Krein space quantization it becomes \ref{tfsmin}).
 It is clear that these two-point functions are dS covariant and also free of any infrared divergences. Note that such tensor field breaks the conformal transformation, and hence, in order to preserve the conformal transformation, the spin-$2$ field must be described by a rank-$3$ mixed symmetric tensor field.

\subsubsection{Rank-3 mixed symmetric tensor field}

The rank-$3$ tensor field on the dS hyperboloid with the conditions $x\cdot{\cal K}=0$ and ${\cal K}_{\alpha_1\alpha_2\alpha_3}=-{\cal K}_{\alpha_1\alpha_3\alpha_2}$ satisfies the field equation (\ref{ran3mieq}). 
To associate this field with an UIR of the dS group, the following subsidiary condition must be imposed:
$ \partial.{\cal K}=0, \;\; \; {\cal K}'=0$,
then the field equation becomes:
$$\left( Q^{(1)}_3+6\right){\cal K}_{\alpha_1\alpha_2\alpha_3}=0,$$
which, in terms of $Q_0$, it is
\b \label{fer31} Q_0 {\cal K}_{\alpha_1\alpha_2\alpha_3}+2{\cal K}_{\alpha_1\alpha_2\alpha_3}+2 {\cal K}_{\alpha_2\alpha_3\alpha_1}+2{\cal K}_{\alpha_3\alpha_1\alpha_2}=0.\e
Because of the existence of the two last terms in the field equation, the rank-$3$ tensor field ${\cal K}$ cannot associates with the UIR of the dS group $\Pi^\pm_{2,2}$. The tensor field ${\cal K}$ can be written in the sum of two fields with definite symmetry \cite{fr84}:
\b \label{r3msas} {\cal K}_{\alpha_1\alpha_2\alpha_3}=\frac{1}{3}\left({\cal K}_{\alpha_1\alpha_2\alpha_3}^M+{\cal K}_{\alpha_1\alpha_2\alpha_3}^A\right), \e
$$ {\cal K}_{\alpha_1\alpha_2\alpha_3}^M=2{\cal K}_{\alpha_1\alpha_2\alpha_3}-{\cal K}_{\alpha_2\alpha_3\alpha_1}-{\cal K}_{\alpha_3\alpha_1\alpha_2},$$
$$ {\cal K}_{\alpha_1\alpha_2\alpha_3}^A={\cal K}_{\alpha_1\alpha_2\alpha_3}+{\cal K}_{\alpha_2\alpha_3\alpha_1}+{\cal K}_{\alpha_3\alpha_1\alpha_2}.$$
${\cal K}_{\alpha_1\alpha_2\alpha_3}^A$ is a totally antisymmetric tensor field with $4$ degrees of freedom, and ${\cal K}_{\alpha_1\alpha_2\alpha_3}^M$ is a mixed symmetric tensor field with $20$ degrees of freedom:
$$ {\cal K}_{\alpha_1\alpha_2\alpha_3}^M+{\cal K}_{\alpha_2\alpha_3\alpha_1}^M+{\cal K}_{\alpha_3\alpha_1\alpha_2}^M=0.$$
The rank-$3$ mixed symmetric tensor field ${\cal K}_{\alpha_1\alpha_2\alpha_3}^M$ on the dS hyperboloid can be considered as a spin-$2$ field in dS space. Such a field may be considered as a part of the gravitational field which its quantum field operator transforms by an indecomposable representation of the dS group. Its physical states correspond to the UIR of the discrete series representations $T^{(0,2;2)}$ and $T^{(2,0;2)}$ (or $\Pi^\pm_{2,2}$ in Dixmier notation).

On the other hand, decomposing a rank- 3 tensor field as (\ref{r3msas}), one can rewrite the field equation (\ref{fer31}) in the following form:
$$ Q_0 {\cal K}_{\alpha_1\alpha_2\alpha_3}^M+ \left(Q_0+6\right) {\cal K}_{\alpha_1\alpha_2\alpha_3}^A=0,$$ 
and consequently, both parts vanish, separately. The mixed symmetric part can be associated to the UIR of the dS group $\Pi^\pm_{2,2}$, since the associated mass parameter is $m_b^2=0$ and the homogeneous degrees of this part are $\lambda=0,-3$
 $$Q_0 {\cal K}_{\alpha_1\alpha_2\alpha_3}^M=0, \;\;\; {\cal K}_{\alpha_1\alpha_2\alpha_3}^M(x)={\cal D}_{\alpha_1\alpha_2\alpha_3}^M(x,\partial)(x\cdot\xi)^\lambda .$$
The totally antisymmetric part cannot associate to any UIR of the dS group and accompanies an imaginary mass $m^2_b=-6H^2$. One of the homogeneous degrees is positive $\lambda=\frac{-3}{2}\pm \frac{1}{2}\sqrt{33}$, 
$$(Q_0 +6){\cal K}_{\alpha_1\alpha_2\alpha_3}^A=0, \;\;\; {\cal K}_{\alpha_1\alpha_2\alpha_3}^A(x)={\cal D}_{\alpha_1\alpha_2\alpha_3}^A(x,\partial)(x.\xi)^\lambda .$$ This part can be eliminated by imposing some conditions which preserve the gauge invariance \cite{fr84}. 

In the previous section the gauge invariant field equations for the massless spin-$ 2$ rank-$ 3$ tensor field in the $x$-space was carried on (\ref{ran3mieq}):
$$ \left(Q^{(1)}_3+6\right) {\cal K}_{\alpha_1\alpha_2\alpha_3}+ \nabla_{\alpha_1}^\top\left( \partial_3^\top.{\cal K}\right)_{\alpha_2\alpha_3}=0.$$ 
The field equation is invariant under the gauge transformation (\ref{gts2r3}):
$$ {\cal K}_{\alpha_1\alpha_2\alpha_3}\longrightarrow {\cal K}_{\alpha_1\alpha_2\alpha_3}^g={\cal K}_{\alpha_1\alpha_2\alpha_3}+ \nabla^\top_{\alpha_1} A_{\alpha_2\alpha_3}.$$
By using the equation (\ref{r3msas}), one can obtain the field  equation and gauge transformation for the mixed symmetric part:
\b \label{gfr3mfv} \left(Q^{(1)}_3+6\right) {\cal K}^M_{\alpha_1\alpha_2\alpha_3}+ \left[\nabla_{\alpha_1}^\top \left(\partial_3^\top.{\cal K}\right)_{\alpha_2\alpha_3}\right]^M=0,\e 
where
$$\left[\nabla^\top_{\alpha_1} A_{\alpha_2\alpha_3}\right]^M=2\nabla^\top_{\alpha_1} A_{\alpha_2\alpha_3}-\nabla^\top_{\alpha_2} A_{\alpha_3\alpha_1}-\nabla^\top_{\alpha_3} A_{\alpha_1\alpha_2}.$$
The gauge transformation becomes:
\b \label{gtr3mfv} {\cal K}^M_{\alpha_1\alpha_2\alpha_3}\longrightarrow {\cal K}^{Mg}_{\alpha_1\alpha_2\alpha_3}={\cal K}^M_{\alpha_1\alpha_2\alpha_3}+ \left[\nabla^\top_{\alpha_1} A_{\alpha_2\alpha_3}\right]^M.\e
The gauge fixing field equation can be obtained similar to the previous cases:
\b \label{gife23t} \left(Q^{(1)}_3+6\right) {\cal K}^M_{\alpha_1\alpha_2\alpha_3}+c \left[\nabla_{\alpha_1}^\top \left(\partial^\top_3\cdot{\cal K}\right)_{\alpha_2\alpha_3}\right]^M=0.\e 

Since the divergencelessness condition is dropped out, the field operator does not transform by an UIR of the dS group and should transform according to an indecomposable representation of the dS group. For the construction of this representation or the Goupta-Bleuler triplet, one needs the gauge transformation (\ref{gtr3mfv}), the gauge fixing equation (\ref{gife23t}) and
the UIR's  $T^{(0,2;2)}$ and $T^{(2,0;2)}$, (or $\Pi^\pm_{2,2}$ in Dixmir notation) of the dS group. The physical states, which correspond to $\Pi^\pm_{2,2}$, transform simultaneously under the conformal group representations ${\cal C}^{(\pm 3,0,2)}$ and ${\cal C}^{(\pm 3,2,0)}$, which will be considered in the next section.
 
Taking the divergence of the equation (\ref{gife23t}), one gains the field equation of the gauge dependent states
$ \left(\partial^\top_3\cdot{\cal K}^M\right)_{\alpha_2\alpha_3}\neq0.$ These states $\left(\partial^\top_3\cdot{\cal K}^M\right)_{\alpha_2\alpha_3}=(A_1)_{\alpha_2\alpha_3}$ are forming a rank-$2$ anti-symmetric tensor field which can be fixed if $c\neq 1$. Imposing the divergenceless condition on the tensor field and visualizing it as ${\cal K}^M_{\alpha_1\alpha_2\alpha_3}=\left[\nabla^\top_{\alpha_1} (A_2)_{\alpha_2\alpha_3}\right]^M$, one obtains the pure gauge state $(A_2)_{\alpha_2\alpha_3}$. These two tensor fields $A_1$ and $A_2$, and the physical states construct the Gupta-Bleuler triplet: $V_g \subset V \subset V_c$.  $V_c/V$ is the space of the solutions as the rank-$2$ antisymmetric tensor field  $A_1$. $V_g$ is the gauge state $A_2$. $V/V_g$ contains the physical states $\Pi^+_{2,2}\oplus  \Pi^-_{2,2}$. 
 
The quantum field operator can be written as the following form:
\b \label{qfop2g} {\cal K}_{\alpha\beta\gamma}(x)=\int_{B}d\mu({\bf q})\sum_{m}\left[a({\bf \tilde{q}},m;2,-1){\cal U}_{\alpha\beta\gamma}(x;{\bf q},m;2,2 )+a^\dag({\bf q},m;2,2){\cal V}_{\alpha\beta\gamma}(x;{\bf q},m;2,2 )\right].\e
By using the homogeneity conditions, one obtains:
$${\cal U}_{\alpha\beta\gamma}(x;{\bf q},m;2,2 )=(x\cdot\xi_B)^{-2} u_{\alpha\beta\gamma}(x,{\bf q},m;2,2 ),$$
$${\cal V}_{\alpha\beta\gamma}(x;{\bf q},m;2,2 )=(x\cdot\xi_B)^{-1} v_{\alpha\beta\gamma}(x,{\bf q},m;2,2 ),$$ 
and also, solving the field equation and imposing the auxiliary conditions, the coefficients $u_{\alpha\beta\gamma}$ and $v_{\alpha\beta\gamma}$ can be established.
 
The analytic two-point function for the Massless spin-$ 2$ rank-$3$ mixed-symmetric tensor field is
$$ W_{\alpha\beta\gamma\alpha'\beta'\gamma'}(z,z')=<\Omega|{\cal K}_{\alpha\beta\gamma}(z)K_{\alpha'\beta'\gamma'}(z')|\Omega>$$
\b \label{tpfvm} =\int_{B}d\mu({\bf q})(z\cdot\xi_B)^{-2}(z'\cdot\xi_B)^{-1}\sum_m u_{\alpha\beta\gamma}(z,{\bf q},m;2,2)v_{\alpha'\beta'\gamma'}(z',{\bf q},m;2,2). \e 
Similar to the other cases, this two-point function can be written in the following form: 
$$  W_{\alpha\beta\gamma\alpha'\beta'\gamma'}(z,z') =D_{\alpha\beta\gamma\alpha'\beta'\gamma'}(z,\partial;z',\partial')W_c (z,z'). $$
$W_c$ is the analytic two-point function of a massless conformally coupled scalar field. The explicit forms of the polarization tensor will be calculated in a forthcoming paper.


\setcounter{equation}{0}
\section{Conformal transformation} 

Considering the global conformal transformation on the Dirac $6$-cone formalism, one notices that the physical states of massless field operators transform according to the UIR of the $SO(2, 4)$ group. We show that the massless fields in the dS ambient space formalism are exactly the same as the massless fields in the Dirac $6$-cone formalism.

Afterwards, the local conformal transformation group, $SO(2, 4)$, can be considered as one of the basis of the gauge gravity model. The gauge potential can be considered as the conformal gravity in the dS ambient space formalism.

\subsection{Dirac $6$-cone formalism}

Undesirably, the conformal group acts non-linearly on the Minkowski coordinates. Obviating this problem, Dirac proposed a manifestly conformal covariant formulation in which the Minkowski coordinates are replaced by a set of suitable coordinates, in order to provide the possibility of linear action of the conformal group. The result was a theory, established as a $5$-dimensional hyper-cone in a $6$-dimensional flat space, named Dirac's six-cone. This approach to conformal symmetry was first used by Dirac, to construct the manifestly conformal invariant field equations for spinor and vector fields in $(1+3)$-dimensional space-time \cite{dirac}, and afterwards has been developed by Mack and Salam \cite{masa}.

Dirac's six-cone or Dirac's projection cone is defined by \b
y^2\equiv (y^0)^{2}-\vec y^{2}-(y^4)^2+(y^5)^{2}=\eta_{ab} y^a y^b=0 ,\;\;
\eta_{ab}=\mbox{diag}(1,-1,-1,-1,-1,1),\e where $ \;y^{a} \in
\R^{6};$ $a,b=0,1,2,3,4,5$ and  $ \vec y \equiv(y^{1},y^{2},y^{3})$. Reduction to four dimensions is achieved by projection after fixing
the degrees of homogeneity of all the fields. Wave equations,
subsidiary conditions, etc., should be expressed in terms of
operators that are defined intrinsically on the cone. These are
well-defined operators that map tensor fields on tensor fields
with the same rank on the cone $y^2=0$. So, the out coming
equations which are obtained by this method, are conformally
invariant.

The tensor fields $\Psi$ on the Dirac $6$-cone are homogeneous functions of variable $y^a$ and are transverse \cite{dirac}:
 $$y^a \frac{\partial}{\partial y^a }\Psi^{cd..}=\sigma \Psi^{cd..},\;\;\;  y_a\Psi^{ab...}=0.$$
In order to project the coordinates on the cone $y^2=0$ to the
$(4+1)$-dimensional dS space, one can choose the following relations:
\b\label{cfdspro} \left\{\ba{rcl}
x^{\alpha}&=&(y^5)^{-1}y^\alpha,\\
x^5&=&y^5.\ea\right.\e Note that $x^5$ becomes superfluous when one
deals with the projective cone. The choice $x^5=H^{-1}$, precisely results in the dS hyperboloid structure, with the tensor field, previously defined in the Dirac $6$-cone formalism, turning to be the tensor field on the dS space-time. For example, we consider rank-$2$ symmetric tensor field $\Psi^{ab}(y)$ on the cone and then it can be projected on dS hyperboloid. This field can be separated as fallows: 
$$ \Psi^{ab}(y)\equiv\left(\Psi^{\alpha\beta}(y),\;\;\;\Psi^{\alpha 5}(y), \;\;\;\Psi^{55}(y) \right).$$
By using the equation (\ref{cfdspro} ) and the transversality condition, these fields can be projected on dS hyperboloid as:
$$ \label{we}
{\cal K}_{\alpha\beta}(x)=\theta_\alpha^\delta\theta_\beta^\gamma\Psi_{\delta\gamma}=\Psi_{\alpha\beta}
+x_\alpha\Psi_{.\beta}\cdot x+x_\beta\Psi_{.\alpha}\cdot x+x_\alpha x_\beta x\cdot\Psi_{..}\cdot x\;, $$
$$ K_\alpha(x)= \theta_\alpha^\delta\Psi_{\delta 5}=\Psi_{\alpha 5}+ x_\alpha \Psi_{. 5}\cdot x,\;\;\;\;\; \phi(x)=\Psi_{55}.$$ 
After performing some algebraic calculation and using the transversality and homogeneity conditions: $$ x^\alpha( x\cdot\partial) {\cal K}_{\alpha\beta}=0,\;\; \; x^\alpha( x\cdot\partial) K_{\alpha}=0,$$ one can show that we obtain ${\cal K}_{\alpha\beta}(x)=\Psi_{\alpha\beta}$ and $K_\alpha(x)=\Psi_{\alpha 5}$ where they are massless fields in dS space-time. That means, there exists an exact correspondence between the tensor fields on the cone and the massless fields in dS hyperboloid.

The same statement holds for tensor (-spinor) fields of other ranks. These fields are conformal invariant, since they are coming from the cone. In the following section, by representing the discrete series of the conformal group to enlighten the subject of the massless fields, the correspondence of these tensor fields with the discrete series representations of the dS group has been illustrated.

\subsection{Discrete UIR of the conformal group}

The conformal group $SO(2,4)$ acts on the cone as:
$$ y'^a=\Lambda^a_b y^b, \;\;\;\Lambda \in SO(2,4) , \;\; \det \Lambda=1,\;\; \Lambda \eta \Lambda^t= \eta \Longrightarrow y\cdot y=0=y'\cdot y' $$ Defining a $4\times 4$ matrix, $Y$, as:
$$ Y=\left( \begin{array}{clcr} y^0+iy^5 & \;\;\;\;{\bf p} \\ {\bf \bar p} & y^0-iy^5 \\    \end{array} \right),$$ where ${\bf p}=(y^4, \vec y)$ is a quaternion, one can show that under a transformation of the conformal group $SU(2,2)$, it transforms as:
$$ Y'=gY\bar g^t, g \in SU(2,2) \Longrightarrow y\cdot y=0=y'\cdot y'.$$
$SU(2,2)$ is the universal covering group of $SO(2,4)$:
 \b SO_0(2,4)  \approx SU(2,2)/ \Z_2 ,\e
which is defined by  \b SU(2,2)=\left\lbrace g=\left(\begin{array}{clcr} {\bf a} &  {\bf b} \\  {\bf c} & {\bf d} \\ \end{array}
        \right),\;\;\mbox{det} g=1 ,\;\;J{\bar g}^t J=g^{-1},
       J=\left(\begin{array}{clcr} \1 &  0 \\  0 & -\1 \\ \end{array}
        \right) \right\rbrace,\e
where ${\bf a},{\bf b},{\bf c}$ and ${\bf d}$ are complex quaternion ${\bf a}={\bf q}_1+i{\bf q}_2$.

In the previous works \cite{masa,mato,ma77,ru72,ru73,anla}, a variety of realizations of the UIR of the conformal group and their corresponded Hilbert spaces have been constructed . Particularly, Graev's realization of the discrete series is important for our proposal here \cite{ru73}. The homogeneous space is by definition a complex $2 \times 2$ matrix, $Z$, in the domain ${\cal D}$ which in turn is defined by the
constraint of positive definiteness \cite{ru72,ru73}:\b \label{shilov} \1-Z^\dag Z>0, \;\;\;  \1-Z Z^\dag >0 .\e
If one defines $Z$ as a quaternion:
$$ Z={\bf q}=\left( \begin{array}{clcr} q^4+iq^3 &\;\; iq^1-q^2 \\ iq^1+q^2 &\;\; q^4-iq^3 \\    \end{array} \right) ,$$
then the condition (\ref{shilov}) holds when $|{\bf q}|<1$. It is exactly the homogeneous space ($\xi^\alpha_B$) in which the discrete series representation of the dS group has been constructed on. 

The discrete series representation of the conformal group and its corresponded Hilbert space on this homogeneous space, has been introduced by Ruhl (for value $E_0>j_1+j_2+3$) \cite{ru72,ru73}
$$ {\cal C}^{(E_0,j_1,j_2)}(g)\left|{\bf q},m_{j_1},m_{j_2};j_1,j_2,E_0\right\rangle=\left[\det({\bf c}{\bf q}+{\bf d})\right]^{-E_0} \times $$ \b \label{dseris2}\sum_{m_{j_1}'m_{j_2}'}D^{(j_1)}_{m_{j_1}m_{j_1}'}\left( {\bf a}^\dag+{\bf q}{\bf b}^\dag\right)D^{(j_2)}_{m_{j_2}'m_{j_2}}\left( {\bf c}{\bf q}+{\bf d}\right) \left|g^{-1}\cdot{\bf q},m_{j_1}',m_{j_2}';j_1,j_2,E_0\right\rangle,  \e
where $g^{-1}=\left( \begin{array}{clcr} {\bf a} & {\bf b} \\ {\bf c} & {\bf d} \\    \end{array} \right) \in SU(2,2)$ and $ g^{-1}\cdot {\bf q}=({\bf a}{\bf q}+{\bf b})({\bf c}{\bf q}+{\bf d})^{-1}.$ $D^{(j_1)}$ and $D^{(j_2)}$ furnish a certain representation of $SU(2)$ group (\ref{su2}) \cite{ru70,ru72,ru73}. The UIR ${\cal C}^{(E_0,j_1,j_2)}(g)$ acts on an infinite dimensional Hilbert space ${\cal H}^{(E_0,j_1,j_2)}_{q}$. The scalar product was defined in \cite{ru73}.  One can easily verify that this representation on the Hilbert space ${\cal H}^{(E_0,j_1,j_2)}_{q}$ satisfies \cite{ru72,ru73}:
$$ {\cal C}^{(E_0,j_1,j_2)}(g){\cal C}^{(E_0,j_1,j_2)}(g')\left|{\bf q},m_{j_1},m_{j_2};j_1,j_2,E_0\right\rangle={\cal C}^{(E_0,j_1,j_2)}(gg')\left|{\bf q},m_{j_1},m_{j_2};j_1,j_2,E_0\right\rangle,$$
$${\cal C}^{(E_0,j_1,j_2)}(g)\left[ {\cal C}^{(E_0,j_1,j_2)}(g)\right]^\dag\left|{\bf q},m_{j_1},m_{j_2};j_1,j_2,E_0\right\rangle =\left|{\bf q},m_{j_1},m_{j_2};j_1,j_2,E_0\right\rangle.$$
There exists an infinite dimensional Hilbert space  ${\cal H}^{(E_0,j_1,j_2)}_{q}$,
$$ \left|{\bf q},m_{j_1},m_{j_2};j_1,j_2,E_0\right\rangle \in {\cal H}^{(E_0,j_1,j_2)}_{q},\;\;\; {\bf q} \in \R^4, |{\bf q}|=r<1, \;\; -j\leq m_j\leq j.$$
Our interest lies only with the values $ j_1j_2=0, \;\; E_0=j_1+j_2+1$, with helicity $ j_1-j_2$ and $2j_1$ and $2j_2$ as non-negative integers (values associated with massless fields). The reader can find a detailed discussion regarding all relevant values of $E_0$, $j_1$ and $j_2$, in correspondence with the UIR of the conformal group, presented by Mack \cite{ma77}.

The representations $ {\cal C}^{(j+1,j,0)}(g)$ and $ {\cal C}^{(j+1,0,j)}(g)$  correspond to the massless representation of the Poincar\'e massless group \cite{babo,micknied,ma77}. They correspond to two helicities of the massless fields. In this case the massless representation $D^{(j)}$ furnishes a certain representation of the little group $ISO(2)$ \cite{wei,ma77}, and therefore, one can write $m_j=m'_j=j$ for $ {\cal C}^{(j+1,j,0)}(g)$ and 
$m_j=m'_j=-j$ for $ {\cal C}^{(j+1,0,j)}(g)$  with the vanishing values for $m_j$ and $m'_j$ in other cases \cite{ma77}.

The massless field equations in the Minkowskian space-time are considered to be conformally invariant, therefore, every massless representation of the Poincar\'e group has only one corresponding representation in the conformal group \cite{babo,anla}. In the dS space, for massless
fields, only two representations in discrete series $T^{(0,j;j)}$ and $T^{(j,0;j)}$ (in Dixmir notation
$\Pi^{\pm}_{j,j}$) have Minkowskian interpretations. The signs
$\pm$ correspond to the two types of helicity for the massless
fields. The representation $\Pi^+_{j,j}$ has a unique extension to a
direct sum of two UIR's ${\cal C}^{(j+1,j,0)}$ and ${\cal C}^{(-j-1,j,0)}$ of
the conformal group $SO_0(2,4)$. Note that  ${\cal C}^{(j+1,j,0)}$ and
${\cal C}^{(-j-1,j,0)}$ correspond to positive and negative energy
representations in the conformal group respectively \cite{babo,anla}.
The concept of energy cannot be defined in the dS space. The UIR of the dS group is restricted to the sum of the massless Poincar\'e UIR's ${\cal P}^{(0,j)}_+$
and ${\cal P}^{(0,j)}_-$ with positive and negative energies respectively. The following diagrams illustrate these connections
$$ \left. \begin{array}{ccccccc}
     &             & {\cal C}^{(j+1,j,0)}
& &{\cal C}^{(j+1,j,0)}   &\hookleftarrow &{\cal P}^{(0,j)}_+\\
 \Pi^+_{j,j} &\hookrightarrow  & \oplus
&\stackrel{H=0}{\longrightarrow} & \oplus  & &\oplus  \\
     &             & {\cal C}^{(-j-1,j,0)}&
& {\cal C}^{(-j-1,j,0)}  &\hookleftarrow &{\cal P}^{(0,j)}_-,\\
    \end{array} \right. $$        
$$ \left. \begin{array}{ccccccc}
     &             & {\cal C}^{(j+1,0,j)}
& &{\cal C}^{(j+1,o,j)}   &\hookleftarrow &{\cal P}^{(0,-j)}_+\\
 \Pi^-_{j,j} &\hookrightarrow  & \oplus
&\stackrel{H=0}{\longrightarrow} & \oplus  & &\oplus  \\
     &             & {\cal C}^{(-j-1,0,j)}&
& {\cal C}^{(-j-1,0,j)}  &\hookleftarrow &{\cal P}^{(0,-j)}_-,\\
    \end{array} \right. $$    
where the arrows $\hookrightarrow $ designate unique extension. $
{\cal P}^{(0,-j)}_{\pm}$ are the massless Poincar\'e UIRs
with positive and negative energies and negative helicity.

\subsection{Conformal gauge gravity}

This subsection contains a generalization of the dS gauge gravity to the conformal group $SO(2, 4)$, using of the Dirac $6$-cone formalism and introducing the following set of the gauge generators:
$$L_{ab}=M_{ab}+S_{ab}\equiv Y_{\cal A}, \;\; {\cal A}=1,2,...,15,\;\; a,b=0,1,..,5.$$ 
In this notation the commutation relation is rewritten as:
$$[Y_{\cal A},Y_{\cal B}]=f_{{\cal B}{\cal A}}^{\;\;\;\;{\cal C}}Y_{\cal C}.$$
Obviously, there are $15$ one-form potentials or gauge vector fields ${\cal G}_\alpha^{\;\;{\cal A}}\equiv {\cal G}_\alpha^{\;\;ab}=-{\cal G}_\alpha^{\;\;ba}$. These gauge potentials satisfy the transversality condition $x^\alpha{\cal G}_\alpha^{\;\;\;\cal A}=0$ and they have $60$ degrees of freedom. In ambient space notation, the conformal gauge-covariant derivative can be defined as
$$ D^{\cal G}_\beta=\nabla^\top_\beta+i {\cal G}_\beta^{\;\;{\cal A}} Y_{\cal A}.$$
The $15$ gauge vector fields ${\cal G}_\alpha^{\;\;ab}$ can be separated into two groups; $10$ gauge vector fields ${\cal G}_\alpha^{\;\;\beta \gamma} \equiv {\cal K}_\alpha^{\;\;\;\beta\gamma}$ (with $40$ degrees of freedom) and $5$ gauge vector fields ${\cal G}_\alpha^{\;\;5\beta }\equiv H_\alpha^{\;\;\;\beta}$ (with $20$ degrees of freedom). By imposing the following subsidiary conditions on these gauge potentials
\b \label{traconf} x_\beta{\cal G}_\alpha^{\;\;\beta\gamma}=0= x_\gamma{\cal G}_\alpha^{\;\;\beta\gamma},\;\; x_\gamma{\cal G}_\alpha^{\;\;5\gamma}=0,\e 
the gauge fields ${\cal G}_\alpha^{\;\;\beta\gamma}$ may be considered as a tensor field of rank-$3$ with $24$ degrees of freedom. The gauge vector potential ${\cal G}_\alpha^{\;\;5\gamma}\equiv H_\alpha^
{\;\;\gamma}$ may be also considered as a tensor field of rank-$2$ with $16$ degrees of freedom on dS hyperboloid.

Now, we can repeat the conformal gauge gravity construction in its conventional way. Under a local infinitesimal gauge transformation generated by $\epsilon^{\cal A}(x)Y_{\cal A}$, we have
$$ \delta_{\epsilon} D^{\cal G}_\alpha=[D^{\cal G}_\alpha, \epsilon^{\cal A}(x)Y_{\cal A}]=(D^{\cal G}_\alpha\epsilon^{\cal A})Y_{\cal A},$$
so that
$$ \delta_{\epsilon}{\cal G}_\alpha^{\;\;{\cal A}}=D^{\cal G}_\alpha\epsilon^{\cal A}=\nabla^\top_\alpha \epsilon^{\cal A}+ f_{{\cal C}{\cal B}}^{\;\;\;\;{\cal A}}{\cal G}_\alpha^{\;\;{\cal C}}\epsilon^{\cal B}.$$
The curvature ${\cal C}$, with values in the Lie algebra of conformal group $SO(2,4)$, is defined by:
$${\cal C}(D^{\cal G}_\alpha,D^{\cal G}_\beta)=-[D^{\cal G}_\alpha,D^{\cal G}_\beta]=C_{\alpha\beta}^{\;\;\;\; {\cal A}}Y_{\cal A}.$$

The $SO(2,4)$ gauge invariant action or Lagrangian in dS background for the gauge field ${\cal G}_\alpha^{\;\;{\cal A}}$ is \cite{wei2}:
\b \label{cagi}  S_c[{\cal G}]=\int d\mu(x) \left(  C_{\alpha\beta}^{\;\;\;\;{\cal A}}{\cal Q}_{{\cal A}{\cal B}}C^{\alpha\beta{\cal B}}\right), \e
where ${\cal Q}_{{\cal A}{\cal B}}$ are the numerical constants. One can impose the subsidiary conditions on ${\cal Q}_{{\cal A}{\cal B}}$ to achieve the maximal irreducibility of the gauge multiple ${\cal G}_\alpha^{\;\;\beta \gamma} \equiv {\cal K}_\alpha^{\;\;\;\beta\gamma}$ and ${\cal G}_\alpha^{\;\;5\beta }\equiv H_\alpha^{\;\;\;\beta}$ on the dS space \cite{frts}. The effect of these definitions splits the action into two parts:
$$ S_c[H,{\cal K},H{\cal K}]=\int d\mu(x) {\cal L}(H,{\cal K},H{\cal K})\equiv S_{c_2}[{\cal K},H{\cal K}]+S_{c_1}[H,H{\cal K}].$$ The linear part of the field equation, which obtain from the action $S_{c_2}$, is in direct correspondence to the field equation of the spin-$2$ rank-$3$ field in the dS space (\ref{ran3mieq}).
The field equation, which is obtained from the local conformal invariant action (\ref{cagi}), is a second order field equation. Similar to the Minkowski space, the gauge fixing terms can be added to the conformal Lagrangian. Using the gauge fixing field equation, the gauge transformation and the UIR of the conformal group ${\cal C}^{(3,2,0)} \oplus {\cal C}^{(-3,2,0)}$ and ${\cal C}^{(3,0,2)} \oplus {\cal C}^{(-3,0,2)}$, one can construct the Gupta-Bleuler triplet for conformal gauge gravity. This rather complicated procedure will be the subject of a forthcoming paper.


\section{Conclusion and outlook}

In this paper, quantum field theory (including the gauge theory) is reformulated in ambient space formalism. This formalism allows us to construct quantum field operator in a rigorous mathematical framework based on complexified pseudo-Riemannian manifold and the group representation theory. The field equations for various spin fields are obtained by using the second order Casimir operator and dS algebra machinery. The procedure of defining the Lagrangian from the field equations for the free fields and the gauge theory are presented in ambient space formalism. Finally the quantum field operators, the quantum states and the two-point functions are presented for various spin fields.

For simplicity, we recall some interesting results of this formalism: 

\begin{itemize}

\item{Since the action of the dS group on the ambient space coordinate $x^\alpha$ is linear, QFT in this formalism has a very simple form, in comparison with its form in the intrinsic coordinate $X^\mu$ where the action of the dS group is non-linear.}

\item{Using the dS plane waves in ambient space formalism, the construction of quantum field theory in dS space is very similar to its counterpart in the Minkowski space-time.}

\item{The dS space predicts the existence of a maximum length for an observable (or equivalently, an ''event horizon'') in the $x$-space. The uncertainty principle results in to the existence of a minimum size in the $\xi$-space. We know that the total volume of this parameter in Hilbert space or $\xi$-space, is finite. As a direct consequence, the total number of quantum states in the Hilbert space turns out to be finite physically \cite{ta13}. This is the one of the most important results of this formalism.}

\item{The massless fields with $s=j=p\geq 3$ cannot propagate on the dS space in our formalism. The massless fields with spins $j=0,\frac{1}{2}, 1,\frac{3}{2},2$ and $\frac{5}{2}$ may be exist in our universe.}

\item{A novel $N=1$ de Sitter superalgebra can be attained by the use of spinor field and charge conjugation in the ambient space notation \cite{morrota}.}

\item{Gauge vector fields $K_\alpha(x)$ with $j=p=1$ can be considered as the potential of the electromagnetic, weak and strong nuclear forces. The spin-$2$ gauge field ${\cal K}^M_{\alpha\beta\gamma}(x)$ in the gauge gravity model may be considered as a part of gravitational field. The case $j=p=\frac{3}{2}$ corresponds to a vector-spinor gauge potential $\Psi_\alpha(x)$ with anti commutation relations. The consequence of this formalism is
the natural appearance of super-gravity.}

\item{The two spinor generators would not generate a closed super-algebra. Therefore these spinor generators must engage their couplings with the dS group generators and the vector-spinor gauge potential $\Psi_\alpha(x)$ with the gauge potentials ${\cal K}^M_{\alpha\beta\gamma}(x)$. Then the gravitational field may be considered as: classical gravitational field $\theta_{\alpha\beta}$ or dS background, the spin-$2$ gauge potential ${\cal K}^M_{\alpha\beta\gamma}(x)$ and the vector-spinor gauge potential $\Psi_\alpha(x)$.}

\item{The spin-$2$ gauge potential ${\cal K}^M_{\alpha\beta\gamma}(x)$ can be quantized and their quantum field operators transform according to the indecomposable representations of dS group. The physical states transform simultaneously as the UIR of dS and conformal groups.}

\item{One can construct the Hilbert space and then the Fock space for the universe including the gauge potentials ${\cal K}^M_{\alpha\beta\gamma}(x)$ and $\Psi_\alpha(x)$. The total number of quantum states for our universe turns out to be finite.}

\item{Previously, the Krein space quantization (Hilbert space $\oplus$ anti-Hilbert space) has been used for the extraction of the covariant quantization of minimally coupled scalar fields that suppresses the infrared divergences but breaks the analyticity \cite{gareta}. To remedy this later drawback, we have used the identity (\ref{msfincsf}) in ambient space formalism. This innovative formalism permits one to quantize the massless minimally coupled scalar field in terms of a massless conformally coupled scalar field. It means that the quantum field operator (\ref{mcsfico}) and the two-point function (\ref{mth}) of minimally scalar field can be constructed on the Bunch-Davies vacuum state. Then the problem of infrared divergence of the linear gravity in the dS space is completely solved on the Bunch-Davies vacuum. The theory is also analytic.}

\item{The quantum massless tensor (-spinor) field can be reformulated in terms of a polarization tensor (-spinor) and a massless conformally coupled scalar field . Its analytic two-point function can also be written in terms of the analytic two-point function of conformally coupled scalar field and a polarization tensor (-spinor).}

\item{In the Dirac six-cone formalism the conformal invariance of the theory becomes manifest. One of the interesting properties of Dirac six-cone formalism is that the tensor and spinor fields on the cone can be simply mapped to the massless fields on de Sitter space-time in the ambient space formalism. It means that there is an exact correspondence between massless conformally fields on the dS ambient space formalism with tensor (or spinor) fields on the Dirac six-cone formalism.}

\end{itemize}

\vspace{0.5cm} 
\noindent {\bf{Acknowlegements}}:  I would like to thank J.P. Gazeau, J. Iliopoulos, J. Renaud and S. Rouhani for their helpful discussions, over some parts of this work, throughout the years and  M. Enayati, M.R. Masomi Nia, M. Tabrizi, and M.R. Tanhayi for their useful suggestions to improve the paper.

\begin{appendix}

\setcounter{equation}{0}
\section{Relation with intrinsic coordinates}

In order to compare the work of the other authors in the intrinsic coordinates \cite{goklsu,giha,stva,bomast,mari,ardunitr,wit,spstvo,mast,bafi,bafipa,pave,pave2,kakalitr,
novawe,nom,dav,mipa,bama} with the ambient space formalism in this paper the relation between these two formalism, is presented here. These relation are discussed for the tensor fields, the spinor fields and the two-point functions. We recall that the intrinsic coordinates is presented by $X^\mu,\;\; \mu=0,1,2,3,$ and the ambient space formalism by $x^\alpha, \;\; \alpha=0,1,2,3,4$. 

\subsection{Tensor fields}

Here we consider only the spin-$2$ rank-$2$ symmetric tensor field and one can simply generalized this method to the other tensor fields. We
use the fact that the intrinsic field $h_{\mu\nu}(X)$ is
locally determined by the transverse tensor field \cite{ta97,gata,gagata}
$\K_{\alpha\beta}(x)$ through \b\lbl{passage}
h_{\mu\nu}(X)=\frac{\partial x^{\alpha}}{\partial
X^{\mu}}\frac{\partial x^{\beta}}{\partial
X^{\nu}}\K_{\alpha\beta}(x(X)). \e  In the same way one can show
that the transverse projector $\theta$ is the only symmetric and
transverse tensor that is linked to the dS metric $g_{\mu\nu}$:
$$g^{dS}_{\mu\nu}=\frac{\partial x^{\alpha}}{\partial
X^{\mu}}\frac{\partial x^{\beta}} {\partial
X^{\nu}}\,\theta_{\alpha\beta}.$$ Covariant derivatives acting on
a symmetric second rank tensor are transformed according to \b
\nabla_{\rho}\nabla_{\lambda}h_{\mu\nu}= \frac{\partial
x^{\alpha}}{\partial X^{\rho}}\frac{\partial x^{\beta}}{\partial
X^{\lambda}}\frac{\partial x^{\gamma}}{\partial
X^{\mu}}\frac{\partial x^{\delta}}{\partial
X^{\nu}}\mbox{Trpr}
\partial^\top_\alpha \mbox{Trpr}\partial^\top_\beta \K_{\gamma\delta},
\e where $\nabla_\mu$ is the covariant derivative in the intrinsic coordinates. The transverse projection (Trpr) defined by $$
(\mbox{Trpr}\K)_{\alpha\beta}=\theta^\gamma_\alpha
\theta^\delta_\beta \K_{\gamma\delta},$$ guarantees
transversality in each index, e.g. \cite{gagata}
$$\nabla_{\rho}\nabla_{\lambda}h_{\mu\nu}= \frac{\partial
x^{\alpha}}{\partial X^{\rho}}\frac{\partial x^{\beta}}{\partial
X^{\lambda}}\frac{\partial x^{\gamma}}{\partial
X^{\mu}}\frac{\partial x^{\delta}}{\partial X^{\nu}}\nabla_\alpha^\top \nabla_\beta^\top {\cal K}_{\gamma\delta} $$ $$= \frac{\partial
x^{\alpha}}{\partial X^{\rho}}\frac{\partial x^{\beta}}{\partial
X^{\lambda}}\frac{\partial x^{\gamma}}{\partial
X^{\mu}}\frac{\partial x^{\delta}}{\partial X^{\nu}}\left[
\partial^\top_\alpha\left(\partial^\top_\beta \K_{\gamma\delta}-
x_{\gamma}\K_{\beta\delta}- x_{\delta}\K_{\beta\gamma}\right)
- x_\beta\left(\partial^\top_\alpha \K_{\gamma\delta}-
x_{\gamma}\K_{\alpha\delta}- x_{\delta}\K_{\alpha\gamma}
\right)\right.$$ \b
-x_\gamma\left(\partial^\top_\beta \K_{\alpha\delta}-
x_{\alpha}\K_{\beta\delta}- x_{\delta}\K_{\beta\alpha}\right) \left.
-x_\delta\left(\partial^\top_\beta \K_{\gamma\alpha}-
x_{\gamma}\K_{\beta\alpha}- x_{\alpha}\K_{\beta\gamma}\right)\right].\e 
By contraction of the covariant derivatives, {\it i.e.}
$\nabla_{\rho}\nabla^{\rho}$, the d'Alambertian operator becomes:
\b \Box_{H}
h_{\mu\nu}=g^{\lambda\rho}\nabla_{\lambda}\nabla_{\rho}h_{\mu\nu}=\frac{\partial
x^{\alpha}}{\partial X^{\mu}}\frac{\partial x^{\beta}}{\partial
X^{\nu}}\left(\left[ \partial^\top_\gamma \partial^{\top\gamma}
-2\right]\K_{\alpha\beta}-2\Sigma_1x_\alpha(  \partial^\top\cdot\K)_\beta
+2x_\alpha x_\beta \K'\right),\e 
and one can define the other necessary relations accordingly.

\subsection{Spinor fields}

The Dirac equation in the general curved space-time is \cite{morrota}
\b \label{dsdiecu}
\left(\gamma^{\mu}(X)\nabla_{\mu}-m\right)\Psi(X)=0=(\bar{\gamma}^{a}\nabla_{a}-m)\Psi(X),\e
where
$$\{\gamma^{\mu}(X),\gamma^{\nu}(X)\}=2g^{\mu\nu},\;\{\bar{\gamma}^{a},\bar{\gamma}^{b}\}=
2\eta^{ab} ,\;\; \mu,a=0,1,2,3.$$  Here $\nabla_a$ is the spinor
covariant derivative
$$ \nabla_{a}\Psi(X)={e^{\mu}}_{a}(\partial_{\mu}+
\frac{i}{2}{e^{c}}_{\nu}\nabla_{\mu}e^{b\nu}\Sigma_{cb})\Psi(X),$$
with ${e_{\mu}}^{a}$ as the local vierbein, ${e_{\mu}}^{a}
{e_{\nu}}^{b}\eta_{ab}=g_{\mu\nu}$, and $
\Sigma_{cb}=\frac{i}{4}[\bar{\gamma}_{c},\bar{\gamma}_{b}]$ as the
spinor representation of the generators of the Lorentz
transformation. At this point, it seems relevant to briefly recall the relation of the dS-Dirac equation in two formalism (\ref{dsdiracfe}) and (\ref{dsdiecu}), extracted by G\"ursey and Lee \cite{gule}. They have introduced a set of coordinates $\{y^{\alpha}\}\equiv (y^\mu,
y^4=H^{-1})$ related to the $\{x^{\alpha}\}$'s by $$ x^{\alpha}=
\bigl( Hy^{4} \bigl) f^{\alpha}(y^{0},y^{1},y^{2},y^{3}), $$ where
arbitrary functions $f^{\alpha}$ satisfies
$f^{\alpha}f_{\alpha}=-H^{-2}$. Matrices $
\beta^{\alpha}\equiv \left( \frac{\partial y^{\alpha}}{\partial
x^{\beta}} \right) \gamma^{\beta}$ can be introduced trough the anticommutation
properties $$ \{ \beta^{\mu},\beta^{\nu}\}=2g^{\mu \nu},\; \{
\beta^{\mu},\beta^{4}\}=0. $$ with $g^{\mu
\nu}=\frac{\partial y^{\mu}}{\partial
x^{\alpha}}\frac{\partial y^{\nu}}{\partial x^{\beta}}\eta^{\alpha\beta}$,
$\mu,\nu=0,\dots ,3$. The spinor field in the ambient space notation can be written in terms of a new spinor field $\chi$ by: $\psi= (1\pm i\beta^{4})\chi$. The spinor field $ \chi
$ satisfies the G\"ursey-Lee equation
\begin{equation}
\left( \beta^{\mu} \frac{\partial}{\partial y^{\mu}} -2H\,
\beta^{4}-m \right) \chi(y) =0, \label{GurLee}
\end{equation}
for $m=H\nu$. Through the selection of a local vierbein ${e_{\mu}}^{a}$ and setting the gamma function to be $\gamma^{\mu}(X)\equiv
{e^{\mu}}_{a}\gamma^{a}$, one can find a transformation $V$ so $ \gamma^{\mu}(X)=V\beta^{\mu}(y)V^{-1}$. Then, under the transformation
$V$, the G\"ursey-Lee equation (\ref{GurLee}) becomes an exact replica of the equation (\ref{dsdiracfe}) with $\Psi(X)=V\chi(y).$ It is interesting to
note that the matrix $\beta^4=\gamma_\alpha x^\alpha=\not x$ is related
to the constant matrix $\gamma^4$ by \cite{gule}:
$$\gamma^{4}=V\beta^{4}V^{-1}=V\not x V^{-1}.$$ 
Now, we can write the relation between the spinor field in two
steps, at first introducing $$ \psi(x)=V^{-1}\left[\frac{a}{2}\left(1+ i\gamma^4\right)+\frac{b}{2} \left(1- i\gamma^4\right)\right]\Psi(X),$$ 
where $a$ and $b$ are normalization constants and second by calculating the matrix that transforms the $\psi(x) $ to $\Psi(X)$ as follows:
\b \label{samre} \Psi(X)=\left[\frac{1}{2a}\left(1+ i\gamma^4\right)+\frac{1}{2b} \left(1- i\gamma^4\right)\right]V\psi(x).\e
One may directly conclude the relation between the tensor-spinor fields in the two formalism by the equation (\ref{samre}).

\subsection{Two-point function}

Following Allen and Jacobson in \cite{alja} we will
write the two-point functions in de Sitter space (maximally
symmetric) in terms of bi-tensors. These are functions of two
points $(x,x')$ which behave like tensors under coordinate
transformations at either point. As shown in \cite{alja}, any maximally symmetric bi-tensor can be expressed as a sum of products of three  basic
tensors. The coefficients in this expansion are functions of the
geodesic distance $\sigma(x,x')$, that is the distance along the
geodesic connecting the points $x$ and $x'$ (note that $\sigma(x,x')$
can be defined by unique analytic extension also when no geodesic
connects $x$ and $x'$). In this sense,  these fundamental tensors
form a complete set.  They can be obtained by differentiating the
geodesic distance \cite{gagata}:
$$n_\mu = \nabla_\mu \sigma(x, x')\;\;\;,\;\;\; n_{\mu'} = \nabla_{\mu'} \sigma(x,
x'), $$ and the parallel propagator
$$g_{\mu\nu'}=-c^{-1}({\cal{Z}})\nabla_{\mu}n_{\nu'}+n_\mu n_{\nu'}.$$
The basic bi-tensors in
ambient space notation are found through
$$ {\partial}^\top_\alpha \sigma(x,x')\;\;\;,\;\;\;{\partial}'^\top_{\beta'}
\sigma(x,x')\;\;\;,\;\;\;\theta_\alpha .\theta'_{\beta'},$$
restricted to the hyperboloid by
$$ {\cal{T}}_{\mu\nu'}(X,X')=\frac{\partial x^\alpha}{\partial
X^\mu}\frac{\partial x'^{\beta'}}{\partial
X'^{\nu'}}T_{\alpha\beta'}(x,x').$$

For space-like geodesic we have $ {\cal{Z}}=-H^{2}x\cdot x'=\cos(\sigma) $ and one can find
$$n_\mu=\frac{\partial x^\alpha}{\partial X^\mu}\bar{\partial}_\alpha \sigma(x,x')=
\frac{\partial x^\alpha}{\partial X^\mu} \frac{(x' \cdot
\theta_\alpha)}{\sqrt{1-{\cal{Z}}^2}},\;\; n_{\nu'}=\frac{\partial
x'^{\beta'}}{\partial X'^{\nu'}}\bar{\partial}_{\beta'}
\sigma(x,x') =\frac{\partial x'^{\beta'}}{\partial X'^{\nu'}}
\frac{(x\cdot\theta'_{\beta'})}{\sqrt{1-{\cal{Z}}^2}},$$
$$\nabla_\mu n_{\nu'}=\frac{\partial x^\alpha}{\partial
X^\mu}\frac{\partial x'^{\beta'}}{\partial
X'^{\nu'}}\theta^\varrho_\alpha
\theta'^{\gamma'}_{\beta'}\bar{\partial}_\varrho\bar{\partial}_{\gamma'}
\sigma(x, x')=c({\cal{Z}})\left[n_\mu n_{\nu'}{\cal{Z}}-\frac{\partial
x^\alpha}{\partial X^\mu}\frac{\partial x'^{\beta'}}{\partial
X^{\nu'}}\theta_\alpha \cdot\theta'_{\beta'}\right],$$ with $
c^{-1}({\cal{Z}})=-\frac{1}{\sqrt{1-{\cal{Z}}^2}}.$  In the case of the time-like geodesic $
{\cal{Z}}=\cosh (\sigma), $  $ n_\mu $ and $ n_\nu$ are multiplied
by $i$ and $ c({\cal{Z}}) $ becomes
$-\frac{i}{\sqrt{1-{\cal{Z}}^2}}.$ In both cases we have
$$ g_{\mu\nu'}+({\cal{Z}}-1)n_\mu n_{\nu'}=\frac{\partial x^\alpha}{\partial
X^\mu}\frac{\partial x'^{\beta'}}{\partial X'^{\nu'}}\theta_\alpha
\cdot\theta'_{\beta'}.$$ 
By using these equations, the relation between the two-point functions in
the two formalisms can be obtained directly \cite{gata,gagata,gagarota,berotata,derotata,petata}.


\end{appendix}

\end{document}